\documentclass{nature}
\usepackage[utf8]{inputenc}
\usepackage{graphicx}
\usepackage{graphics}
\graphicspath{ {./Figures/} }
\usepackage{duckuments} %
\usepackage{float}
\usepackage{lineno}
\usepackage{comment}
\usepackage{tabularx}
\usepackage{booktabs}
\usepackage{multirow}
\usepackage{amsmath,amssymb}
\usepackage{bbold}
\usepackage[dvipsnames]{xcolor}
\usepackage{mathtools}
\usepackage{braket}

\usepackage{textcomp,gensymb}
\usepackage{prettyref}
\newrefformat{fig}{Fig.~\ref{#1}}
\newrefformat{exfig}{Extended Data Fig.~\ref{#1}}

\usepackage{hyperref}
\hypersetup{
    colorlinks=true,
    linkcolor=blue,
    filecolor=magenta,      
    urlcolor=cyan,
    citecolor=red,
    pdftitle={Twist-Programmable Superconductivity in Spin-Orbit Coupled Bilayer Graphene},
}

\usepackage{caption}
\DeclareCaptionLabelFormat{vertline}{#1 #2 $|$}
\usepackage{enumitem}

\makeatletter \let\saved@includegraphics\includegraphics
\AtBeginDocument{\let\includegraphics\saved@includegraphics} 
\renewenvironment{figure}{\@float{figure}}{\end@float}
\makeatother

\definecolor{orange}{rgb}{1,0.5,0}

\newcommand{\beginsupplement}{%
        \setcounter{table}{0}
        \setcounter{figure}{0}
        \renewcommand{\figurename}{Extended Data Fig.}
        \renewcommand{\tablename}{SI Table}
     }

\newcommand{\vk}{{\boldsymbol{k}}}

\newcommand{\vb}{{\boldsymbol{b}}}

\newcommand{\vs}{{\boldsymbol{s}}}

\renewcommand{\vec}[1]{{\boldsymbol #1}} %

\DeclareMathAlphabet{\mathcald}{U}{dutchcal}{m}{n}
\SetMathAlphabet{\mathcald}{bold}{U}{dutchcal}{b}{n}
\DeclareMathAlphabet{\mathalt}{U}{dutchcal}{b}{n}

\title{Twist-Programmable Superconductivity in Spin-Orbit Coupled Bilayer Graphene}
\author{Yiran Zhang$^{1,2,3\dagger}$, Gal Shavit$^{2,3,4}$, Huiyang Ma$^{5,6}$, Youngjoon Han$^{2,3}$, Kenji Watanabe$^7$, Takashi Taniguchi$^7$, David Hsieh$^{2,3}$, Cyprian Lewandowski$^{5,6}$, Felix von Oppen$^{8}$, Yuval Oreg$^{9}$, and Stevan Nadj-Perge$^{1,2\dagger}$}
\begin{document}

\maketitle

\begin{affiliations}

  \item T. J. Watson Laboratory of Applied Physics, California Institute of
  Technology, 1200 East California Boulevard, Pasadena, California 91125, USA
  \item Institute for Quantum Information and Matter, California Institute of
  Technology, Pasadena, California 91125, USA
  \item Department of Physics, California Institute of Technology, Pasadena,
  California 91125, USA
  \item Walter Burke Institute of Theoretical Physics, California Institute of Technology, Pasadena, California 91125, USA
  \item National High Magnetic Field Laboratory, Tallahassee, FL, USA
  \item Department of Physics, Florida State University, Tallahassee, FL, USA
  \item National Institute for Materials Science, Namiki 1-1, Tsukuba, Ibaraki 305 0044, Japan
  \item Dahlem Center for Complex Quantum Systems and Fachbereich Physik, Freie Universit\"at Berlin, 14195 Berlin, Germany
  \item Department of Condensed Matter Physics, Weizmann Institute of Science, Rehovot, Israel
  \item[$^\dagger$] Correspondence: yzhang7@caltech.edu; s.nadj-perge@caltech.edu
  
\end{affiliations}
\begin{abstract}

The relative twist angle between layers of near-lattice-matched van der
Waals materials is critical for the emergent correlated phenomena associated with moir\' e flat bands\cite{bistritzerMoireBandsTwisted2011, caoCorrelatedInsulatorBehaviour2018, 
caoUnconventionalSuperconductivityMagicangle2018}. However, the concept of angle rotation
control is not exclusive to moir\'e superlattices in which electrons directly experience a twist-angle-dependent periodic 
potential. Instead, it can also be employed to induce programmable
symmetry-breaking perturbations with the goal of stabilizing desired correlated states.
Here, we experimentally demonstrate `moir\'eless' twist-tuning of superconductivity
together with other correlated orders in Bernal bilayer graphene
proximitized by tungsten diselenide. 
The precise alignment between the two materials systematically controls the strength 
of the induced Ising spin-orbit coupling (SOC), profoundly altering the phase diagram. As Ising SOC
is increased, superconductivity onsets at a higher displacement field and
features a higher critical temperature, reaching up to 0.5~K. Within the main superconducting 
dome and in the strong Ising SOC limit, we find an unusual phase transition characterized by a nematic redistribution of 
holes among trigonally warped Fermi pockets and enhanced resilience to in-plane magnetic fields. 
The behavior of the superconducting phase is well captured by our theoretical model, which emphasizes the prominent 
role of interband interactions between Fermi pockets arising due to interaction-enhanced symmetry breaking.
Moreover, we identify two additional superconducting regions, one of which descends from an inter-valley coherent 
normal state and exhibits a Pauli-limit violation ratio exceeding $\mathbf{40}$, 
among the highest for all known superconductors\cite{zhouIsospinMagnetismSpinpolarized2022, ranExtremeMagneticFieldboosted2019, ranNearlyFerromagneticSpintriplet2019, luFullSuperconductingDome2018}. Our results provide essential 
insights into ultra-clean graphene-based superconductors and underscore the 
potential of utilizing moir\' eless-twist engineering across a wide range of van der Waals heterostructures.

\end{abstract}

In a large electrical displacement field, the electronic bands of Bernal bilayer 
graphene (BLG) flatten out around the corners of the Brillouin zone, giving rise 
to various correlated phases that spontaneously break BLG
symmetries\cite{zhouIsospinMagnetismSpinpolarized2022, 
delabarreraCascadeIsospinPhase2022, seilerQuantumCascadeCorrelated2022}. These 
correlated phases naturally compete with each other and, therefore, are highly 
sensitive to explicit symmetry-breaking perturbations. One way to induce such 
perturbations is by placing BLG adjacent to 
transition metal dichalcogenides such as tungsten diselenide (WSe$_2$). Due to a large 
mismatch of the unit cells, WSe$_2$ does not induce a moir\'e potential that is deep 
enough to renormalize the low-energy bands of BLG significantly. Yet, the SOC
perturbations induced by WSe$_2$\cite{wangOriginMagnitudeDesigner2016, gmitraProximityEffectsBilayer2017,khooOndemandSpinOrbit2017,khooTunableQuantumHall2018,islandSpinOrbitdrivenBand2019,wangQuantumHallEffect2019,liTwistangleDependenceProximity2019} alter the phase diagram and remarkably enhance the otherwise exceedingly fragile superconductivity in BLG\cite{zhouIsospinMagnetismSpinpolarized2022}, enabling superconductivity at zero 
magnetic field and boosting the critical 
temperature significantly\cite{zhangEnhancedSuperconductivitySpin2023, holleisNematicityOrbitalDepairing2024, liTunableSuperconductivityElectron2024}. However, 
the nature and the extent of this enhancement, and more generally, the impact of spin-orbit interaction on the correlated phase diagram of BLG and other graphene systems, remain elusive and experimentally underexplored.

\begin{figure}[p]
    \centering
    \includegraphics[width=16cm]{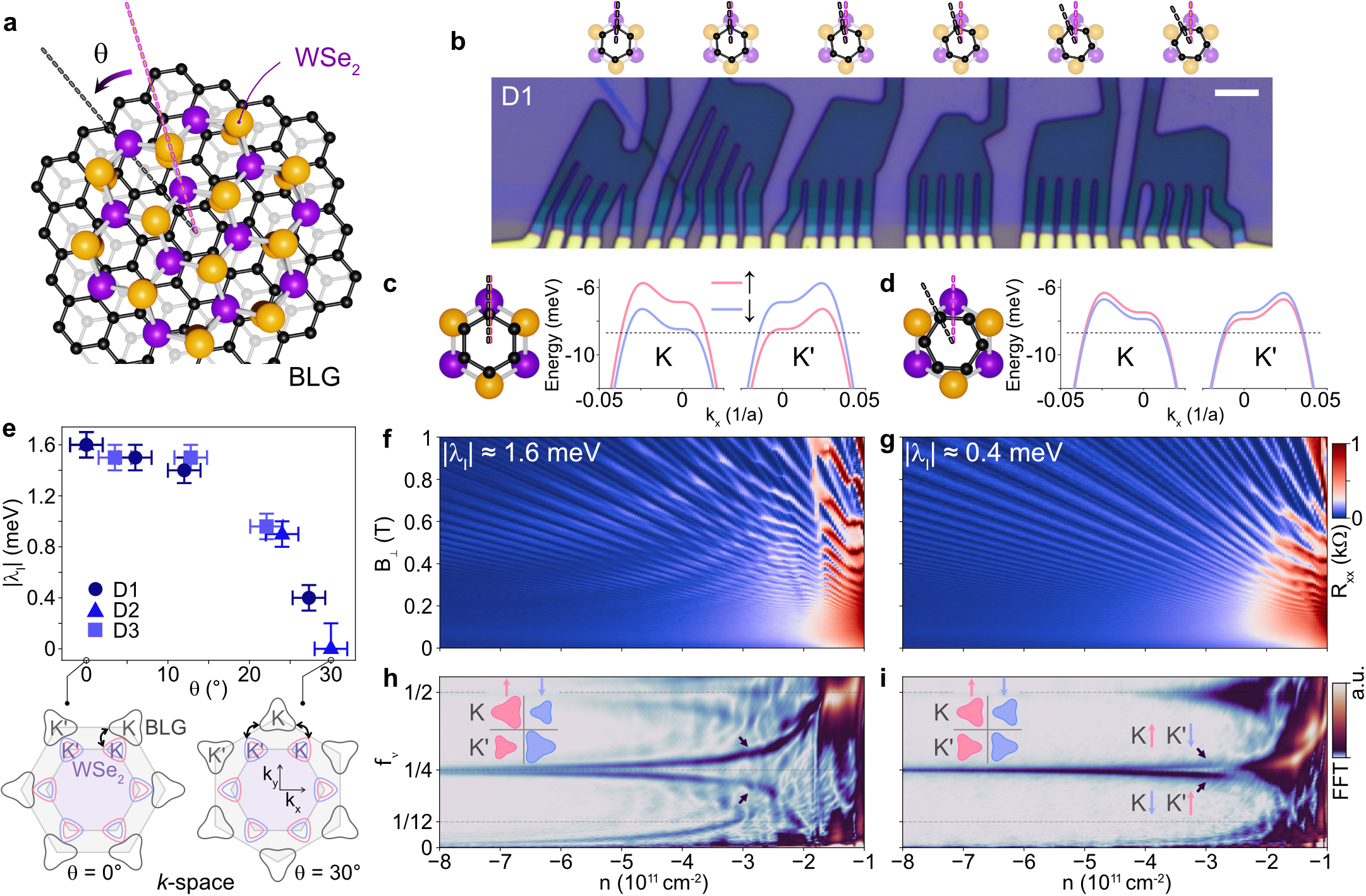}
    \caption{{\bf Programmable Ising SOC by interfacial twisting between BLG and WSe$_2$.} {\bf a},~Schematic showing the twisting of the BLG-WSe$_2$ interface; tuning the interfacial twist angle $\theta$ between the two largely lattice-mismatched materials modifies the Ising SOC strength $|\lambda_{I}|$ and the correlated phase diagram. {\bf b}, Optical image of the device set D1. Crystal axes of graphene and WSe$_2$ are rotated relative to each other with an angle $\theta \sim 0\degree$, $6\degree$, $12\degree$, $18\degree$, $24\degree$, and $30\degree$ for the six devices, respectively. The scale bar corresponds to 5$\mu$m. 
    {\bf c},{\bf d}, Non-interacting valence bands of BLG near the $K$ and $K'$ points of the Brillouin zone at $D/\epsilon_0 =
    0.2~\text{V/nm}$, with proximitized Ising SOC $|\lambda_I| \approx 1.6$~meV ({\bf c}) and $0.4$~meV ({\bf d}), respectively. 
    {\bf e}, Ising SOC strength $|\lambda_{I}|$ versus BLG-WSe$_2$ interfacial twist angle $\theta$; data were extracted from three sets of devices D1-D3. The bottom schematics show the relative rotation between the BLG and WSe$_2$ Brillouin zones. At $\theta \approx 0\degree$, $K$/$K'$ valleys of BLG couple more effectively to one of the two WSe$_2$ valleys, resulting in large induced Ising SOC. 
    In contrast, at $\theta \approx 30\degree$, 
    inter-valley and intra-valley tunneling between WSe$_2$ and BLG have the same amplitude by reflection symmetry so that Ising couplings of opposite sign cancel each other and result in vanishing proximity value.
    {\bf f},{\bf g}, $R_{xx}$ versus out-of-plane magnetic field $B_\perp$ and doping $n$ measured at $D/\epsilon_0 = 0.2~\text{V/nm}$ for devices with $|\lambda_I| \approx 1.6$~meV ({\bf f}) and $0.4$~meV ({\bf g}), respectively. {\bf h},{\bf i}, Fast Fourier transform (FFT) of $R_{xx}(1/B_\perp)$ versus $n$ and $f_\nu$, where $f_\nu$ denotes the quantum oscillation frequency normalized to the Luttinger volume. The arrow-marked FFT splittings reflect the Ising-induced Fermi-surface imbalance within each valley, where larger Ising SOC ({\bf h}) features a larger splitting than small Ising SOC ({\bf i}).}
    \label{fig:fig1}
\end{figure}

Theoretically, the induced SOC is predicted to depend on the relative twist 
angle $\theta$ between WSe$_2$ and graphene\cite{liTwistangleDependenceProximity2019, chouEnhancedSuperconductivityVirtual2022, davidInducedSpinorbitCoupling2019, naimerTwistangleDependentProximity2021, zollnerTwistGatetunableProximity2023} (Fig.~\ref{fig:fig1}a). This dependence, 
however, has not been experimentally studied or utilized with systematic control. 
Here we employ this novel tuning knob to explore how Ising SOC modifies the 
correlated phases and emerging superconductivity in BLG. This approach offers 
several unique opportunities for exploring the properties of ultra-clean and highly 
tunable superconductors: %
($i$) the strength of the induced Ising SOC in BLG can be 
precisely quantified, essential for developing theoretical understanding; 
($ii$) the induced SOC is much less sensitive to twist-angle variations
compared to moir\'e systems, allowing for fine control of SOC; and ($iii$) 
the proximity to WSe$_2$ does not induce additional disorder, making experimental 
insights highly reproducible.  
We investigated a series of BLG-WSe$_2$ devices (Fig.~\ref{fig:fig1}b) fabricated from 
the same BLG and WSe$_2$ crystals, separating the large BLG flake into multiple 
pieces by atomic-force-microscope-actuated 
cutting\cite{liElectrodeFreeAnodicOxidation2018}. 
With the first BLG piece roughly lattice-aligned with WSe$_2$ ($\theta \approx 0\degree$), 
the following ones are sequentially twisted relative to WSe$_2$ with a $\sim6\degree$ 
increment (see Methods and \prettyref{exfig:fabrication} for device fabrication details).

To characterize the SOC in our devices, we first perform high-resolution measurements of Shubnikov–de Haas oscillations (\prettyref{fig:fig1}f,g)
in regions of the $n$-$D$ phase diagram ($n$ is the doping density, $D$ is the electrical displacement field) that are well-described by non-interacting theory.
When a positive $D$ field  $D/\epsilon_0 = 0.2~\text{V/nm}$ is applied, the 
hole-carrier wavefunctions are strongly polarized towards the top graphene layer adjacent to WSe$_2$, which in turn induces Ising SOC in BLG\cite{islandSpinOrbitdrivenBand2019, zhangEnhancedSuperconductivitySpin2023, holleisNematicityOrbitalDepairing2024, liTunableSuperconductivityElectron2024, masseroniSpinorbitProximityMoS2024, seilerLayerselectiveSpinorbitCoupling2024}. %
For this $D$ field, Ising SOC is already maximal, i.e., larger $D$ values do not further increase the Ising SOC strength (see Methods and \prettyref{exfig:Ising_extraction} for further discussion). In this regime,
we observe a clear beating pattern in longitudinal resistance $R_{xx}$ as a function of out-of-plane magnetic field ($B_\perp$) at 
higher doping densities (\prettyref{fig:fig1}f), indicating two close 
oscillation frequencies originating from Fermi pockets of slightly different sizes. 
To quantitatively analyze this Fermi-surface imbalance, we normalize the oscillation frequencies of $R_{xx}(1/B_\perp)$ to the Luttinger volume corresponding to the total doping density. The resulting normalized frequency $f_\nu$ reveals the fraction of the total Fermi surface area enclosed by a cyclotron orbit. Figure \ref{fig:fig1}h,i shows an example comparison of the density-dependent frequencies $f_\nu$ from two devices with twist angles $\theta \approx 0\degree$ and $30 \degree$, respectively. 
In both cases, two frequencies ($f_\nu^{(1)}$ and $f_\nu^{(2)}$) are found satisfying $f_\nu^{(1)} + f_\nu^{(2)} = 1/2$. 
These frequencies can be understood as a splitting from $f_\nu = 1/4$, which signals the broken four-fold spin-valley symmetry, and are a direct measure of how the Ising SOC modifies the 
single-particle band structure (\prettyref{fig:fig1}c,d). Due to Ising SOC, nominally four-fold degenerate bands separate into two pairs of spin-valley locked bands with slightly different Fermi-surface areas (illustrated in insets of \prettyref{fig:fig1}h,i). 

We can now confirm experimentally that the induced Ising SOC is modulated
by the twist angle $\theta$ between WSe$_2$ and BLG. This is evident from 
the more pronounced splitting shown in \prettyref{fig:fig1}h compared to the one in \prettyref{fig:fig1}i. By extracting the sizes of the Fermi pockets corresponding to spin-valley locked bands (see Methods and 
\prettyref{exfig:Ising_extraction}) and comparing them to band-structure calculations, we precisely quantify the Ising SOC strength ($|\lambda_I| \approx 1.6$~meV in \prettyref{fig:fig1}h and $|\lambda_I| \approx 0.4$~meV in \prettyref{fig:fig1}i, respectively). \prettyref{fig:fig1}e summarizes systematic measurements across three sets of moir\'eless twisting BLG-WSe$_2$ devices (D1-D3), all of which demonstrate robust $\theta$-modulated Ising SOC strengths. Our results are consistent with the picture that virtual interlayer tunneling is responsible for the induced SOC. When the lattices of BLG and WSe$_2$ are angle-aligned, i.e., $\theta \approx 0\degree$, the $K$/$K'$ valleys of BLG couple more effectively to one of the two valleys of WSe$_2$ (left schematic in \prettyref{fig:fig1}e), resulting in a large induced Ising SOC. In contrast, for $\theta \approx 30\degree$, the inter-valley and intra-valley tunneling between WSe$_2$ and BLG have the same amplitude due to reflection symmetry (right schematic in \prettyref{fig:fig1}e). The induced Ising SOC in BLG vanishes accordingly\cite{sunDeterminingSpinorbitCoupling2023}. The overall twist-angle dependence and the magnitude of Ising SOC are qualitatively consistent with predictions\cite{liTwistangleDependenceProximity2019, chouEnhancedSuperconductivityVirtual2022, davidInducedSpinorbitCoupling2019, naimerTwistangleDependentProximity2021, zollnerTwistGatetunableProximity2023}.

\begin{figure}[p]
    \centering
    \includegraphics[width=14.5cm]{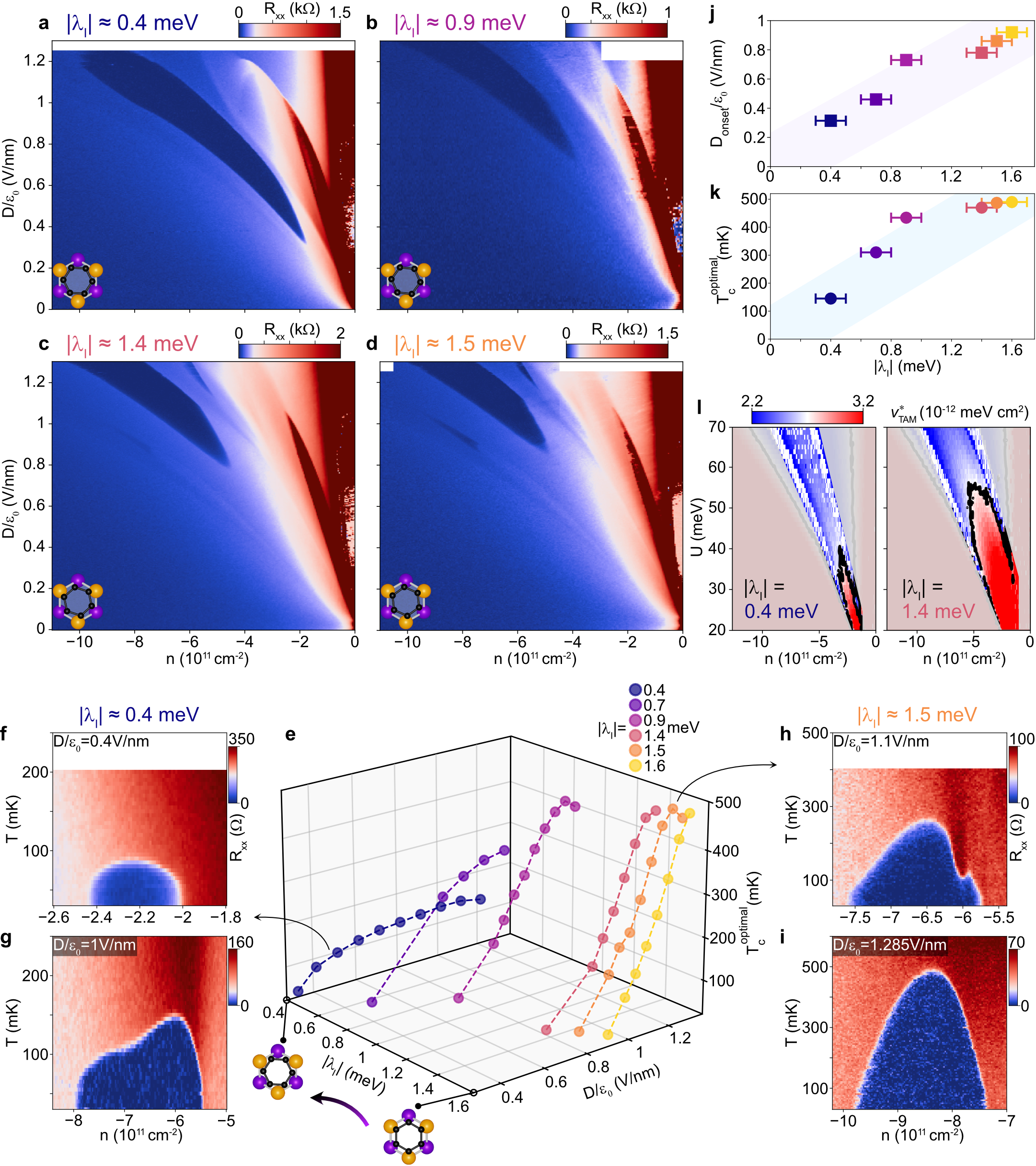}
    \caption{
    {\bf Twist-programmable superconducting phase diagram.} 
    {\bf a}--{\bf d}, $R_{xx}$ versus doping density $n$ and displacement field $D$ for devices with Ising strength $|\lambda_I| \approx 0.4$~meV ({\bf a}), $0.9$~meV ({\bf b}), $1.4$~meV ({\bf c}), and $1.5$~meV ({\bf d}), respectively. {\bf e}, Optimal superconducting critical temperature $T_c^\mathrm{optimal}$ versus $|\lambda_I|$ and $D$. {\bf f},{\bf g}, $R_{xx}$ versus doping  $n$ and temperature for a device with $|\lambda_I| \approx 0.4$~meV, showing superconducting domes at $D/\epsilon_0 = 0.4~\text{V/nm}$ ({\bf f}) and $1~\text{V/nm}$ ({\bf g}), respectively. {\bf h},{\bf i}, $R_{xx}$ versus doping $n$ and temperature for a device with $|\lambda_I| \approx 1.5$~meV, showing superconducting domes at $D/\epsilon_0 = 1.1~\text{V/nm}$ ({\bf h}) and $1.285~\text{V/nm}$ ({\bf i}), respectively. {\bf j},{\bf k}, Displacement field $D_\mathrm{onset}$ at which  superconductivity onsets  ({\bf j}) and optimal critical temperature $T_c^\mathrm{optimal}$ ({\bf k}) versus Ising SOC strength $|\lambda_I|$. {\bf l}, The residual Cooper channel repulsion  $v^*_{\rm TAM}$ versus doping $n$ and interlayer potential difference $U$ for $|\lambda_I| = 0.4$~meV (left) and $1.4$~meV (right), respectively. 
    }
    \label{fig:fig2}
\end{figure}

Using the exquisite twist-angle control of the Ising SOC strength, we explore the SOC-dependent correlated phase diagrams occurring at large $D$ fields. Devices with various 
Ising strengths all show characteristic $R_{xx}$ features that are associated with strong 
correlations 
and superconductivity stabilized at zero magnetic field\cite{zhangEnhancedSuperconductivitySpin2023, holleisNematicityOrbitalDepairing2024, liTunableSuperconductivityElectron2024} (\prettyref{fig:fig2}a-d, 
see \prettyref{exfig:nD diagram_different Ising} for all six $n$-$D$ phase diagrams). 
Importantly, the main superconducting pocket, which emerges from a polarized state 
with a dominant population of two out of the four spin-valley 
flavors\cite{zhangEnhancedSuperconductivitySpin2023, holleisNematicityOrbitalDepairing2024}, 
shows a strong dependence on the Ising SOC strength. 
For low Ising SOC ($|\lambda_I| \approx 0.4$~meV; \prettyref{fig:fig2}a), the superconducting 
region occupies a large $D$ field range, starting from $D/\epsilon_0 \approx 0.3~\text{V/nm}$ 
and extending up to $D/\epsilon_0 \approx 1.25~\text{V/nm}$. For large Ising SOC 
($|\lambda_I| \approx 1.5$~meV; \prettyref{fig:fig2}d), however, superconductivity onsets only 
at $D/\epsilon_0 \approx 0.9~\text{V/nm}$. Overall, the value $D_{\rm onset}$ marking the onset of the 
superconducting pocket grows with increasing $|\lambda_I|$ (\prettyref{fig:fig2}j).

This trend of $D_{\rm onset}$ can be understood as a consequence of \emph{interband} interactions 
between the majority ($K\uparrow$, $K'\downarrow$) and the minority ($K\downarrow$, $K'\uparrow$) spin-valley flavors. 
The difference in hole populations between the majority and minority bands scales with $\left|\lambda_I\right|$ 
and is further enhanced by Coulomb interactions. Consequently, the region in the phase diagram where both bands 
have a large density of states (DOS) near the Fermi level is pushed towards higher values of $D$ and $n$ as Ising SOC is increased. The experimentally observed trend is well reproduced by a simple model 
that takes into account the pairing between majority and minority bands. 
We perform multiband-superconductivity calculations with polarized normal state and find that the residual Cooper-channel repulsion\cite{tolmachevLogarithmicCriterionSuperconductivity1962,morelCalculationSuperconductingState1962} ($v^*_{\rm TAM}$) grows with increasing $\left|\lambda_I\right|$ and decreases at higher $D$ fields (see~\prettyref{fig:fig2}l and Supplementary Information (SI), section~\ref{sec:Importance_Interband} to \ref{sec:multibandSC}). The majority-minority interactions that scatter electron pairs between the pair of bands, greatly enhance the screening of the bare repulsion compared to the single-band case. A larger density imbalance between the bands effectively suppresses this interaction, shifting $D_{\rm onset}$ to higher values. Note that the significant role of interband interactions in BLG-WSe$_2$ is in stark contrast to the case of moir\'e graphene\cite{caoUnconventionalSuperconductivityMagicangle2018,aroraSuperconductivityMetallicTwisted2020,parkTunableStronglyCoupled2021,haoElectricFieldTunable2021,parkRobustSuperconductivityMagicangle2022, zhangPromotionSuperconductivityMagicangle2022}, where superconductivity emerges from a polarized phase in the absence of majority and minority carriers. %

Intriguingly, the superconducting critical temperature $T_c$ also shows a striking dependence on $|\lambda_I|$. 
While $D_\mathrm{onset}$ is smaller and superconductivity persists over a wide range of $D$ fields for small Ising SOC, 
the superconducting critical temperature remains low throughout and saturates at $T_c \approx 150$~mK (\prettyref{fig:fig2}f,g). 
In contrast, for large Ising SOC, superconductivity onsets only at higher $D$ fields, but $T_c$ quickly increases, 
reaching $T_c \approx 500$~mK at the optimal $D$ (\prettyref{fig:fig2}h,i). This is the highest $T_c$ reported for crystalline (untwisted) 
graphene systems. Thus, the optimal critical temperature also shows an increasing trend with $|\lambda_I|$ (\prettyref{fig:fig2}k). 
A detailed three-dimensional map of the optimal critical temperature $T_c^\mathrm{optimal}$ versus $D$ field and $|\lambda_I|$ 
extracted from multiple devices is plotted in \prettyref{fig:fig2}e.
These observations motivate further investigations of the phase diagram with even stronger Ising SOC, e.g., through proximity to other transition 
metal dichalcogenides or the application of pressure\cite{yankowitzTuningSuperconductivityTwisted2019, kedvesStabilizingInvertedPhase2023}.

\begin{figure}[p]
    \centering
    \includegraphics[width=16cm]{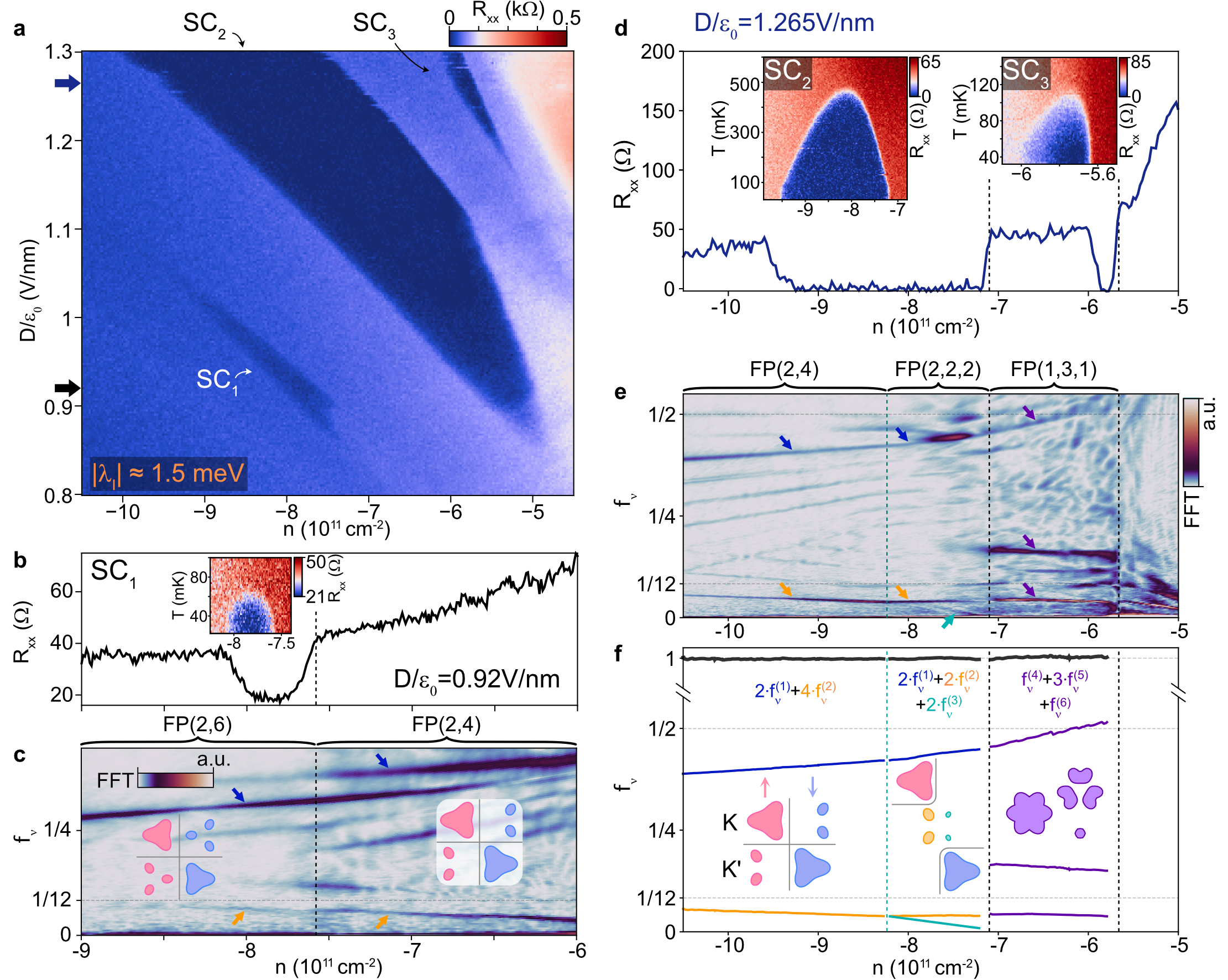}
    \caption{
    {\bf Superconductivity across nematic redistribution and from inter-valley coherence.}
    {\bf a}, $R_{xx}$ versus doping density $n$ and displacement field $D$ for a device with Ising SOC strength $|\lambda_I| \approx 1.5$~meV, focusing around the phase space where the three superconducting regions coexist. {\bf b}, $R_{xx}$ versus $n$ measured at $D/\epsilon_0 = 0.92~\text{V/nm}$. The inset shows $R_{xx}$ versus $n$ and temperature for the superconducting dome SC$_1$. {\bf c}, Frequency-normalized FFT of $R_{xx}(1/B_\perp)$ over the same doping range as in {\bf b}; schematics show the corresponding flavor symmetry-breaking Fermi surfaces. {\bf d}, $R_{xx}$ versus  $n$ measured at $D/\epsilon_0 = 1.265~\text{V/nm}$. Insets show $R_{xx}$ versus $n$ and temperature for the superconducting domes SC$_2$ (left) and SC$_3$ (right), respectively. {\bf e}, Frequency-normalized FFT of $R_{xx}(1/B_\perp)$ over the same doping range as in {\bf d}. The arrows mark the primary FFT peaks, as shown in {\bf f}. The green dashed line marks the continuous transition from $\mathrm{FP}(2,2,2)$ to $\mathrm{FP}(2,4)$; black dashed lines mark first-order flavor symmetry-breaking transitions. {\bf f}, Intensity peaks in $f_\nu$ extracted from {\bf e}. The black solid lines around $f_\nu = 1$ indicate the results from the Luttinger sum rule. Schematics show the possible flavor-polarized phases, from left to right corresponding to spin-valley locked nematic $\mathrm{FP}(2,4)$, nematic $\mathrm{FP}(2,2,2)$ with two sizes (green and orange) of trigonal-warping pockets, and inter-valley coherent $\mathrm{FP}(1,3,1)$.
    } 
    
    \label{fig:fig3}
\end{figure}

The case for investigating devices with large Ising SOC is further emphasized by 
the observation of two additional superconducting pockets in this regime 
($1.4$~meV $\lesssim |\lambda_I| \lesssim 1.6$~meV; see \prettyref{fig:fig3}a, \prettyref{fig:fig2}c,d, and \prettyref{exfig:nD diagram_different Ising}). 
We refer to the observed superconducting regions as SC$_1$, SC$_2$, and SC$_3$, enumerated from higher to lower hole doping, respectively (with the main
superconducting region discussed above being SC$_2$; see \prettyref{exfig:three_SC} for additional temperature and $B_\perp$ characterizations). 
Each superconducting pocket descends from a distinct flavor-symmetry-breaking 
normal state (\prettyref{fig:fig3}b-f) and is terminated by a first-order symmetry-breaking
phase transition (marked by black dashed lines in \prettyref{fig:fig3}b-f) on the low-doping side. Region SC$_1$ features an optimal critical temperature $T_c \approx 60$~mK  (\prettyref{fig:fig3}b inset), while the critical temperatures for SC$_2$ and SC$_3$  are
$T_c \approx 500$~mK and $100$~mK, respectively (\prettyref{fig:fig3}d). 
The normal state of SC$_1$ is the only one that can be directly related to the non-interacting 
band structure.
Quantum oscillations in this regime ($-9 \lesssim n \lesssim -7.6\times 10^{11}~\text{cm}^{-2}$, \prettyref{fig:fig3}c) show two main frequencies (marked by blue and orange arrows)
obeying $2\cdot f_\nu^{(1)} + 6 \cdot f_\nu^{(2)} \approx 1$. This indicates 
two large Fermi pockets from the two majority Ising flavors and six small pockets originating from trigonal warping\cite{mccannElectronicPropertiesBilayer2013} of the two minority Ising 
flavors\cite{holleisNematicityOrbitalDepairing2024} (\prettyref{fig:fig3}c left schematic); we denote the flavor-polarized phase as $\mathrm{FP}(2,6)$ ($\mathrm{FP}(n,m)$ denotes a flavor-polarized phase
with $n$ and $m$ degenerate-sized Fermi pockets, from large to small).

\begin{figure}[p]
    \centering
    \includegraphics[width=14cm]{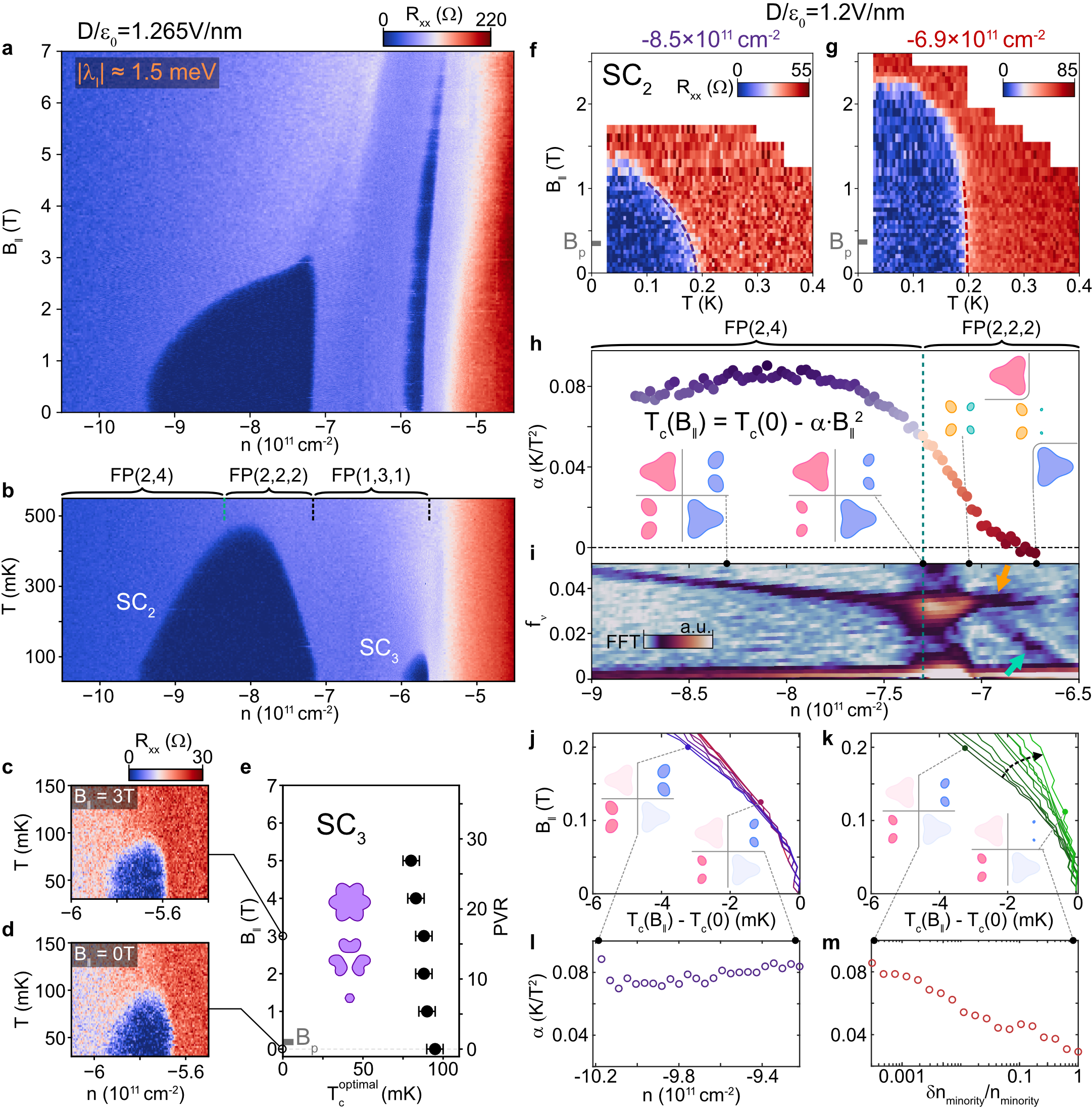}
    \caption{{\bf Ultra-high Pauli-limit violation and nematicity-intertwined $B_\parallel$ depairing.} {\bf a}, $R_{xx}$ versus doping $n$ and in-plane magnetic field $B_{\parallel}$ at $D/\epsilon_0 = 1.265~\text{V/nm}$ for a device with $|\lambda_I| \approx 1.5$~meV. {\bf b}, $R_{xx}$ versus $n$ and temperature at the same $D$. {\bf c},{\bf d}, $R_{xx}$ versus $n$ and temperature for SC$_3$ measured at $B_{\parallel} = 0$~T ({\bf c}) and  $3$~T ({\bf d}), respectively. {\bf e}, 
    Optimal  critical temperature $T_c^\mathrm{optimal}$ of SC$_3$ versus $B_{\parallel}$. The grey bar marks the Pauli limit $B_p$. 
    {\bf f},{\bf g}, $R_{xx}$ versus temperature and $B_{\parallel}$ at $n = -8.5\times 10^{11} ~\text{cm}^{-2}$ ({\bf f}) and $-6.9\times 10^{11} ~\text{cm}^{-2}$ ({\bf g}), respectively for $D/\epsilon_0 = 1.2~\text{V/nm}$. The colored dashed lines are quadratic fitting by $T_{c}(B_{\parallel}) = T_{c}(0)-\alpha \times B_{\parallel}^2$. {\bf h}, Coefficient $\alpha$ versus doping  $n$ within the SC$_2$ dome at $D/\epsilon_0 = 1.2~\text{V/nm}$. {\bf i}, Normalized FFT of $R_{xx}(1/B_\perp)$ over the same $n$ and $D$ range as in {\bf h}, focusing at low frequencies. Green dashed line marks the nematic redistribution of holes from $\mathrm{FP}(2,4)$ to $\mathrm{FP}(2,2,2)$. Schematics in {\bf h} show the Fermi-surface evolution versus $n$, where the smallest trigonal-warping pockets (green) grow rapidly from low to high doping ($-7.3 \lesssim n \lesssim -6.6\times 10^{11}~\text{cm}^{-2}$). {\bf j},{\bf k}, Theoretical $B_\parallel$ depairing with the prominent interband pairing ({\bf j}) and the suppressed case by valley polarization ({\bf k}). {\bf l},{\bf m}, Theoretical $\alpha$ versus $n$ for $\mathrm{FP}(2,4)$ ({\bf i}) and versus minority imbalance $\delta n_\mathrm{minority}/n_\mathrm{minority}$ for $\mathrm{FP}(2,2,2)$ ({\bf m}).} 
    
    \label{fig:fig4}
\end{figure}

Ultrahigh-resolution quantum oscillations reveal the unusual features of the Fermi surfaces in the correlated normal states forming SC$_2$ and SC$_3$. 
In the 
higher-doping region of SC$_2$ ($n \lesssim -8.25\times 10^{11}~\text{cm}^{-2}$, \prettyref{fig:fig3}e), we observe two dominant 
oscillation frequencies  marked by blue ($f_\nu^{(1)}$) and orange ($f_\nu^{(2)}$) lines in \prettyref{fig:fig3}f, satisfying $2 \cdot f_\nu^{(1)} + 4 \cdot f_\nu^{(2)} \approx 1$ (black line).
Thus, the normal state, denoted as $\mathrm{FP}(2,4)$, is a flavor-polarized phase hosting two majority and four minority Fermi pockets. The occupation of two out of 
the three trigonal-warping pockets for both the minority 
spin-valley flavors implies a nematic normal state that breaks the $C_3$ rotational symmetry\cite{holleisNematicityOrbitalDepairing2024} (\prettyref{fig:fig3}f left schematic). 
Remarkably, we observe a different Fermi-pocket configuration in the doping range 
$-8.25 \lesssim n \lesssim -7.1\times 10^{11}~\text{cm}^{-2}$ within the same superconducting pocket. Here, the lowest third frequency $f_\nu^{(3)}$ 
(green line in \prettyref{fig:fig3}f) can be clearly resolved 
(see Methods and \prettyref{exfig:D1p265_Fan} to \ref{exfig:frequency_raw data} 
for frequency extraction). Starting from the same value as $f_\nu^{(2)}$, the value of $f_\nu^{(3)}$ 
rapidly decreases to zero at the low-doping phase boundary, implying that two out 
of the four small Fermi pockets shrink considerably in this 
density range. We denote this phase as $\mathrm{FP}(2,2,2)$ in view of the relation $2 \cdot f_\nu^{(1)} + 2 \cdot f_\nu^{(2)} + 2 \cdot f_\nu^{(3)} \approx 1$. Here, the second and third numbers ($2$ and $2$) imply an additional broken symmetry 
within the trigonal-warping pockets\cite{dongIsospinMomentumpolarizedOrders2023, linSpontaneousMomentumPolarization2023} (orange and green pockets of the middle schematic in \prettyref{fig:fig3}f), signaling 
a nematic redistribution of holes. This remarkable 
continuous transition from $\mathrm{FP}(2,2,2)$ to $\mathrm{FP}(2,4)$ 
within the superconducting dome has a significant impact on the in-plane magnetic field 
response of SC$_2$ (see \prettyref{fig:fig4} and the associated discussion).

A rather exceptional correlated phase denoted $\mathrm{FP}(1,3,1)$ emerges 
upon further decreasing the doping ($-7.1\lesssim n \lesssim -5.7\times 10^{11}~\text{cm}^{-2}$; \prettyref{fig:fig3}e,f). 
The Fermi-surface configuration is reflected in three oscillation frequencies obeying 
$f_\nu^{(4)} + 3 \cdot f_\nu^{(5)} + f_\nu^{(6)} \approx 1$, with $f_\nu^{(4)}$ 
being larger than $1/2$, ensuring that the largest Fermi pocket is non-degenerate 
 (see Methods and 
\prettyref{exfig:QO_FP(1,3,1)_1} to \ref{exfig:QO_FP(1,3,1)_lowfield} for further discussion).  Remarkably, this normal state supports the superconducting region SC$_3$, although all Fermi pockets have odd multiplicities. The combination of superconductivity and odd Fermi-pocket multiplicities strongly points at an inter-valley coherent (IVC) state\cite{nuckollsQuantumTexturesManybody2023a, kimImagingIntervalleyCoherent2023, arpIntervalleyCoherenceIntrinsic2024, chatterjeeIntervalleyCoherentOrder2022, kohSymmetrybrokenMetallicOrders2024,youKohnLuttingerSuperconductivityIntervalley2022, xieFlavorSymmetryBreaking2023} 
(purple schematic of \prettyref{fig:fig3}f). 
This is further 
corroborated by an analysis of the response of the phase boundaries to a $B_\perp$ field\cite{arpIntervalleyCoherenceIntrinsic2024} (see Methods and \prettyref{exfig:phase boundary moving}). The odd multiplicity of all Fermi pockets excludes the possibility 
of conventional $s$-wave pairing, suggesting unconventional superconductivity. Moreover, provided that the pockets are intrinsically superconducting, the number of three mid-size Fermi pockets ($f_\nu^{(5)}$) 
implies time-reversal symmetry breaking regardless of the inter-valley coherent nature\cite{kohSymmetrybrokenMetallicOrders2024, thomsonGatedefinedWiresTwisted2022, kohCorrelatedPhasesSpinorbitcoupled2024, zhumagulovSwappingExchangeSpinorbit2024}. It is interesting that SC$_3$ only develops in a multi-band situation and not from one of the IVC-ordered normal states at lower hole doping with a single or a smaller number of Fermi pockets (see \prettyref{exfig:phase boundary moving}).

All three superconducting pockets show extraordinary resilience to in-plane magnetic field $B_{\parallel}$.
SC$_1$ is characterized by an in-plane critical field $B_{c \parallel} \approx 2.5$~T, significantly higher 
than observed previously\cite{holleisNematicityOrbitalDepairing2024} (\prettyref{exfig:SC1_inplane_field}). SC$_2$ and SC$_3$ show distinct features in the $B_{\parallel}$ response, 
reflecting the highly unusual intertwining with the underlying normal states.
\prettyref{fig:fig4}a,b shows the dependence of $R_{xx}$ on in-plane magnetic field $B_\parallel$ (\prettyref{fig:fig4}a) and  temperature (\prettyref{fig:fig4}b)
measured at $D/\epsilon_0 \approx 1.265~\text{V/nm}$. SC$_2$ occupies a significantly 
larger doping range, and its optimal $T_c$ is roughly five times that of 
SC$_3$ (\prettyref{fig:fig4}b). In comparison, the two superconducting regions show a striking response 
to $B_\parallel$.  While SC$_2$ is fully suppressed by $B_\parallel \approx 3$~T, SC$_3$ 
persists up to $B_\parallel = 7$~T (\prettyref{fig:fig4}a) at the phase boundary.
Crucially, the optimal critical temperature $T_c^\mathrm{optimal}$ of SC$_3$ appears to be insensitive to $B_\parallel$ (\prettyref{fig:fig4}e and \prettyref{exfig:Inplane_T}), with the superconducting domes at $B_\parallel = 0$~T and $B_\parallel = 3$~T being almost the same (\prettyref{fig:fig4}c,d). For a weak-coupling spin-singlet Bardeen–Cooper–Schrieffer (BCS) superconductor, the Pauli limit $B_p$ is related to the zero-magnetic-field critical temperature $T_c(0)$ as 
$B_p = 1.86~\text{T}/\text{K}\times T_c(0)$.  $T_c(0) = 100$~mK for SC$_3$ would produce a Pauli limit $B_p = 0.186$~T.
Thus, the observed in-plane critical field $B_{c \parallel} = 7$~T yields a Pauli-limit 
violation ratio (PVR) $B_{c \parallel}/ B_p\sim 40$, placing SC$_3$ among the superconducting phases with the 
highest Pauli-limit violation ratios\cite{zhouIsospinMagnetismSpinpolarized2022, ranExtremeMagneticFieldboosted2019, ranNearlyFerromagneticSpintriplet2019, luFullSuperconductingDome2018}. Note that the exceedingly large PVR is not 
present in the other two superconducting regions where Fermi pockets of the same size appear in pairs, 
further reflecting the remarkable nature of SC$_3$.

While having a significantly lower PVR, the analysis of SC$_2$ provides further insights into 
the pairing scenarios. SC$_2$ features two doping regions with 
distinct $B_\parallel$ responses (\prettyref{fig:fig4}f,g) that are directly intertwined with the continuous transition from $\mathrm{FP}(2,2,2)$ to $\mathrm{FP}(2,4)$ (\prettyref{fig:fig4}h,i).
Figure~\ref{fig:fig4}f,g shows representative $R_{xx}$ versus temperature and $B_\parallel$ measured in the overdoped and underdoped regions ($n = -8.5$ and $-6.9\times 10^{11}~\text{cm}^{-2}$, respectively) for $D/\epsilon_0 = 1.2~\text{V/nm}$. 
SC$_2$ exhibits the same  $T_c(0) \approx 200$~mK at both doping densities, but the $B_\parallel$ responses are distinct.
The overdoped $T_c$ is quickly suppressed by $B_\parallel$ following a conventional quadratic scaling (\prettyref{fig:fig4}f).
The underdoped $T_c$, however, is insensitive to  $B_\parallel$ for $B_\parallel \leq 1$~T (\prettyref{fig:fig4}g), with the depairing at higher fields $B_\parallel>1.5$~T likely  due to the Fermi-surface changes induced by $B_\parallel$.

To quantify the $B_\parallel$-induced suppression of SC$_2$, we fit $T_c$ versus 
$B_\parallel$ by $T_c(B_\parallel) = T_c(0)-\alpha \cdot B_\parallel^2$, where
$\alpha$ quantifies the pair-breaking tendency of $B_\parallel$. The resulting $\alpha$ shows a striking dependence on doping (\prettyref{fig:fig4}h). At higher doping, $\alpha$ plateaus around 0.08~K/T$^2$. At lower doping, $\alpha$ approaches zero, indicating vanishing sensitivity 
to $B_\parallel$. Importantly, the qualitative change in $\alpha$ (\prettyref{fig:fig4}h) coincides 
with the redistribution of the trigonal-warping pockets
(\prettyref{fig:fig4}i). The region with the plateau ($-8.8 \lesssim n \lesssim -7.3\times 10^{11}~\text{cm}^{-2}$) and the region with the 
rapidly changing $\alpha$ ($-7.3 \lesssim n \lesssim -6.6\times 10^{11}~\text{cm}^{-2}$) correspond to the $\mathrm{FP}(2,4)$ 
and $\mathrm{FP}(2,2,2)$ phases, respectively. Within the $\mathrm{FP}(2,2,2)$, 
both the value of $\alpha$ and the size of the smallest Fermi pockets (green pockets of the schematics in \prettyref{fig:fig4}h) approach zero at the phase boundary ($n \approx -6.6\times 10^{11}~\text{cm}^{-2}$). These observations suggest that the smallest Fermi pockets determine the $B_\parallel$ response (see \prettyref{exfig:Inplane_SC2_D1p265} to \ref{exfig:Inplane_SC2_D1p07} for additional data).

The disparity in the response of the two SC$_2$ regions to $B_\parallel$ invites an analysis of possible microscopic mechanisms.
We propose that this disparity may be attributed to the prominence of majority-minority 
interband interactions, so that the $B_\parallel$ response (\prettyref{fig:fig4}f-i) and the trend observed in $D_{\rm onset}$ (\prettyref{fig:fig2}j) share a common origin. Due to strong interactions, a modest $B_\parallel$  
(compared to $\left|\lambda_I\right|$) may lead to significant spin canting, 
where majority- (minority-) band spins cant towards (away from) the magnetic field direction (see SI, section~\ref{sec:PVRheirarchy}).
In the spin-canted normal state\cite{dongSuperconductivitySpincantingFluctuations2024, kohSymmetrybrokenMetallicOrders2024}, the interband (intraband) Cooper-channel 
interactions are naturally suppressed (enhanced) due to the in-plane spin 
projection of the scattered Copper pairs.
As a consequence, since interband scattering is beneficial to pairing, 
one expects an appreciable decrease in $T_c$ with applied $B_\parallel$ (\prettyref{fig:fig4}j,l).
On the other hand, further symmetry breaking 
in the minority bands (e.g., spin or valley polarization) may critically suppress 
pairing between minority carriers and thus decouple them completely from the 
majority band in the Cooper channel. In such a scenario
provided that 
the minority bands of $\mathrm{FP}(2,2,2)$ are valley-polarized (right schematic in \prettyref{fig:fig4}k), 
one expects the adverse magnetic field effects on the interband interaction 
to gradually disappear, making the superconductor less field-sensitive (\prettyref{fig:fig4}k,m).
This coincides with the experimental trend in \prettyref{fig:fig4}f-i. Note that the same interband $B_\parallel$-suppression mechanism
is consistent with the hierarchy of PVR 
between  SC$_1$, SC$_2$, and SC$_3$ 
(SI, section ~\ref{sec:PVRheirarchy}). Our conclusions are further supported by 
an analysis of conventional depairing mechanisms within the Ising superconductor 
framework\cite{frigeriSuperconductivityInversionSymmetry2004,
luEvidenceTwodimensionalIsing2015,
saitoSuperconductivityProtectedSpinvalley2016,zhangEnhancedSuperconductivitySpin2023},
 which is qualitatively inconsistent with the experimental observations (see Methods and 
 SI, section~\ref{sec: conventional}).

Unprecedented control over the strength of Ising SOC in BLG allowed us to explore its rich set of superconducting regions systematically. Superconductivity occurs for a diverse set of Fermi-pocket configurations, including for Fermi pockets with odd multiplicity pointing at unconventional superconductivity. Remarkably, all the superconductors exhibit distinctive resilience to in-plane magnetic field. A newly discovered inter-valley coherent Fermi-pocket configuration exhibits a PVR value, which reaches one of the highest values for any superconductor to date. All the superconducting regions are multiband superconductors, which we argue to explain differences in their resilience to in-plane fields and their dependence on the displacement field. More generally, the approach of inducing
tunable symmetry-breaking fields via moir\'eless-twist engineering, can be applied to a broad family of van der Waals materials and extended beyond SOC to include magnetism, charge orders, etc. This opens promising avenues towards tailoring desired perturbations and realizing exotic phases of matter on demand.


\noindent {\bf References:}

\bibliographystyle{naturemag} 
\bibliography{main.blg}

\begin{thebibliography}{10}
\expandafter\ifx\csname url\endcsname\relax
  \def\url#1{\texttt{#1}}\fi
\expandafter\ifx\csname urlprefix\endcsname\relax\def\urlprefix{URL }\fi
\providecommand{\bibinfo}[2]{#2}
\providecommand{\eprint}[2][]{\url{#2}}

\bibitem{bistritzerMoireBandsTwisted2011}
\bibinfo{author}{Bistritzer, R.} \& \bibinfo{author}{MacDonald, A.~H.}
\newblock \bibinfo{title}{Moir{\'e} bands in twisted double-layer graphene}.
\newblock \emph{\bibinfo{journal}{PNAS}} \textbf{\bibinfo{volume}{108}},
  \bibinfo{pages}{12233--12237} (\bibinfo{year}{2011}).

\bibitem{caoCorrelatedInsulatorBehaviour2018}
\bibinfo{author}{Cao, Y.} \emph{et~al.}
\newblock \bibinfo{title}{Correlated insulator behaviour at half-filling in
  magic-angle graphene superlattices}.
\newblock \emph{\bibinfo{journal}{Nature}} \textbf{\bibinfo{volume}{556}},
  \bibinfo{pages}{80--84} (\bibinfo{year}{2018}).

\bibitem{caoUnconventionalSuperconductivityMagicangle2018}
\bibinfo{author}{Cao, Y.} \emph{et~al.}
\newblock \bibinfo{title}{Unconventional superconductivity in magic-angle
  graphene superlattices}.
\newblock \emph{\bibinfo{journal}{Nature}} \textbf{\bibinfo{volume}{556}},
  \bibinfo{pages}{43--50} (\bibinfo{year}{2018}).

\bibitem{zhouIsospinMagnetismSpinpolarized2022}
\bibinfo{author}{Zhou, H.} \emph{et~al.}
\newblock \bibinfo{title}{Isospin magnetism and spin-polarized
  superconductivity in {{Bernal}} bilayer graphene}.
\newblock \emph{\bibinfo{journal}{Science}} \textbf{\bibinfo{volume}{375}},
  \bibinfo{pages}{774--778} (\bibinfo{year}{2022}).

\bibitem{ranExtremeMagneticFieldboosted2019}
\bibinfo{author}{Ran, S.} \emph{et~al.}
\newblock \bibinfo{title}{Extreme magnetic field-boosted superconductivity}.
\newblock \emph{\bibinfo{journal}{Nat. Phys.}} \textbf{\bibinfo{volume}{15}},
  \bibinfo{pages}{1250--1254} (\bibinfo{year}{2019}).

\bibitem{ranNearlyFerromagneticSpintriplet2019}
\bibinfo{author}{Ran, S.} \emph{et~al.}
\newblock \bibinfo{title}{Nearly ferromagnetic spin-triplet superconductivity}.
\newblock \emph{\bibinfo{journal}{Science}} \textbf{\bibinfo{volume}{365}},
  \bibinfo{pages}{684--687} (\bibinfo{year}{2019}).

\bibitem{luFullSuperconductingDome2018}
\bibinfo{author}{Lu, J.} \emph{et~al.}
\newblock \bibinfo{title}{Full superconducting dome of strong {{Ising}}
  protection in gated monolayer {{WS2}}}.
\newblock \emph{\bibinfo{journal}{Proceedings of the National Academy of
  Sciences}} \textbf{\bibinfo{volume}{115}}, \bibinfo{pages}{3551--3556}
  (\bibinfo{year}{2018}).

\bibitem{delabarreraCascadeIsospinPhase2022}
\bibinfo{author}{{de la Barrera}, S.~C.} \emph{et~al.}
\newblock \bibinfo{title}{Cascade of isospin phase transitions in
  {{Bernal-stacked}} bilayer graphene at zero magnetic field}.
\newblock \emph{\bibinfo{journal}{Nat. Phys.}} \textbf{\bibinfo{volume}{18}},
  \bibinfo{pages}{771--775} (\bibinfo{year}{2022}).

\bibitem{seilerQuantumCascadeCorrelated2022}
\bibinfo{author}{Seiler, A.~M.} \emph{et~al.}
\newblock \bibinfo{title}{Quantum cascade of correlated phases in trigonally
  warped bilayer graphene}.
\newblock \emph{\bibinfo{journal}{Nature}} \textbf{\bibinfo{volume}{608}},
  \bibinfo{pages}{298--302} (\bibinfo{year}{2022}).

\bibitem{wangOriginMagnitudeDesigner2016}
\bibinfo{author}{Wang, Z.} \emph{et~al.}
\newblock \bibinfo{title}{Origin and magnitude of `designer' spin-orbit
  interaction in graphene on semiconducting transition metal dichalcogenides}.
\newblock \emph{\bibinfo{journal}{Phys. Rev. X}} \textbf{\bibinfo{volume}{6}},
  \bibinfo{pages}{041020} (\bibinfo{year}{2016}).

\bibitem{gmitraProximityEffectsBilayer2017}
\bibinfo{author}{Gmitra, M.} \& \bibinfo{author}{Fabian, J.}
\newblock \bibinfo{title}{Proximity effects in bilayer graphene on monolayer
  {{WSe}}$_{2}$: Field-effect spin valley locking, spin-orbit valve, and spin
  transistor}.
\newblock \emph{\bibinfo{journal}{Phys. Rev. Lett.}}
  \textbf{\bibinfo{volume}{119}}, \bibinfo{pages}{146401}
  (\bibinfo{year}{2017}).

\bibitem{khooOndemandSpinOrbit2017}
\bibinfo{author}{Khoo, J.~Y.}, \bibinfo{author}{Morpurgo, A.~F.} \&
  \bibinfo{author}{Levitov, L.}
\newblock \bibinfo{title}{On-demand spin--orbit interaction from which-layer
  tunability in bilayer graphene}.
\newblock \emph{\bibinfo{journal}{Nano Lett.}} \textbf{\bibinfo{volume}{17}},
  \bibinfo{pages}{7003--7008} (\bibinfo{year}{2017}).

\bibitem{khooTunableQuantumHall2018}
\bibinfo{author}{Khoo, J.~Y.} \& \bibinfo{author}{Levitov, L.}
\newblock \bibinfo{title}{Tunable quantum {{Hall}} edge conduction in bilayer
  graphene through spin-orbit interaction}.
\newblock \emph{\bibinfo{journal}{Phys. Rev. B}} \textbf{\bibinfo{volume}{98}},
  \bibinfo{pages}{115307} (\bibinfo{year}{2018}).

\bibitem{islandSpinOrbitdrivenBand2019}
\bibinfo{author}{Island, J.~O.} \emph{et~al.}
\newblock \bibinfo{title}{Spin--orbit-driven band inversion in bilayer graphene
  by the van der {{Waals}} proximity effect}.
\newblock \emph{\bibinfo{journal}{Nature}} \textbf{\bibinfo{volume}{571}},
  \bibinfo{pages}{85--89} (\bibinfo{year}{2019}).

\bibitem{wangQuantumHallEffect2019}
\bibinfo{author}{Wang, D.} \emph{et~al.}
\newblock \bibinfo{title}{Quantum {{Hall}} effect measurement of spin--orbit
  coupling strengths in ultraclean bilayer graphene/{{WSe2}} heterostructures}.
\newblock \emph{\bibinfo{journal}{Nano Lett.}} \textbf{\bibinfo{volume}{19}},
  \bibinfo{pages}{7028--7034} (\bibinfo{year}{2019}).

\bibitem{liTwistangleDependenceProximity2019}
\bibinfo{author}{Li, Y.} \& \bibinfo{author}{Koshino, M.}
\newblock \bibinfo{title}{Twist-angle dependence of the proximity spin-orbit
  coupling in graphene on transition-metal dichalcogenides}.
\newblock \emph{\bibinfo{journal}{Phys. Rev. B}} \textbf{\bibinfo{volume}{99}},
  \bibinfo{pages}{075438} (\bibinfo{year}{2019}).

\bibitem{zhangEnhancedSuperconductivitySpin2023}
\bibinfo{author}{Zhang, Y.} \emph{et~al.}
\newblock \bibinfo{title}{Enhanced superconductivity in spin--orbit
  proximitized bilayer graphene}.
\newblock \emph{\bibinfo{journal}{Nature}} \textbf{\bibinfo{volume}{613}},
  \bibinfo{pages}{268--273} (\bibinfo{year}{2023}).

\bibitem{holleisNematicityOrbitalDepairing2024}
\bibinfo{author}{Holleis, L.} \emph{et~al.}
\newblock \bibinfo{title}{Nematicity and {{Orbital Depairing}} in
  {{Superconducting Bernal Bilayer Graphene}} with {{Strong Spin Orbit
  Coupling}}}.
\newblock \emph{\bibinfo{journal}{arXiv:2303.00742 [cond-mat]}}
  (\bibinfo{year}{2024}).
\newblock \eprint{2303.00742}.

\bibitem{liTunableSuperconductivityElectron2024}
\bibinfo{author}{Li, C.} \emph{et~al.}
\newblock \bibinfo{title}{Tunable superconductivity in electron- and hole-doped
  {{Bernal}} bilayer graphene}.
\newblock \emph{\bibinfo{journal}{Nature}} \bibinfo{pages}{1--7}
  (\bibinfo{year}{2024}).

\bibitem{chouEnhancedSuperconductivityVirtual2022}
\bibinfo{author}{Chou, Y.-Z.}, \bibinfo{author}{Wu, F.} \&
  \bibinfo{author}{Das~Sarma, S.}
\newblock \bibinfo{title}{Enhanced superconductivity through virtual tunneling
  in {{Bernal}} bilayer graphene coupled to
  \$\{{\textbackslash}mathrm\{\vphantom{\}\}}{{WSe}}\vphantom\{\}\vphantom\{\}\_\{2\}\$}.
\newblock \emph{\bibinfo{journal}{Phys. Rev. B}}
  \textbf{\bibinfo{volume}{106}}, \bibinfo{pages}{L180502}
  (\bibinfo{year}{2022}).

\bibitem{davidInducedSpinorbitCoupling2019}
\bibinfo{author}{David, A.}, \bibinfo{author}{Rakyta, P.},
  \bibinfo{author}{Korm{\'a}nyos, A.} \& \bibinfo{author}{Burkard, G.}
\newblock \bibinfo{title}{Induced spin-orbit coupling in twisted
  graphene--transition metal dichalcogenide heterobilayers: {{Twistronics}}
  meets spintronics}.
\newblock \emph{\bibinfo{journal}{Phys. Rev. B}}
  \textbf{\bibinfo{volume}{100}}, \bibinfo{pages}{085412}
  (\bibinfo{year}{2019}).

\bibitem{naimerTwistangleDependentProximity2021}
\bibinfo{author}{Naimer, T.}, \bibinfo{author}{Zollner, K.},
  \bibinfo{author}{Gmitra, M.} \& \bibinfo{author}{Fabian, J.}
\newblock \bibinfo{title}{Twist-angle dependent proximity induced spin-orbit
  coupling in graphene/transition metal dichalcogenide heterostructures}.
\newblock \emph{\bibinfo{journal}{Phys. Rev. B}}
  \textbf{\bibinfo{volume}{104}}, \bibinfo{pages}{195156}
  (\bibinfo{year}{2021}).

\bibitem{zollnerTwistGatetunableProximity2023}
\bibinfo{author}{Zollner, K.}, \bibinfo{author}{Jo{\~a}o, S.~M.},
  \bibinfo{author}{Nikoli{\'c}, B.~K.} \& \bibinfo{author}{Fabian, J.}
\newblock \bibinfo{title}{Twist- and gate-tunable proximity spin-orbit
  coupling, spin relaxation anisotropy, and charge-to-spin conversion in
  heterostructures of graphene and transition metal dichalcogenides}.
\newblock \emph{\bibinfo{journal}{Phys. Rev. B}}
  \textbf{\bibinfo{volume}{108}}, \bibinfo{pages}{235166}
  (\bibinfo{year}{2023}).

\bibitem{liElectrodeFreeAnodicOxidation2018}
\bibinfo{author}{Li, H.} \emph{et~al.}
\newblock \bibinfo{title}{Electrode-{{Free Anodic Oxidation Nanolithography}}
  of {{Low-Dimensional Materials}}}.
\newblock \emph{\bibinfo{journal}{Nano Lett.}} \textbf{\bibinfo{volume}{18}},
  \bibinfo{pages}{8011--8015} (\bibinfo{year}{2018}).

\bibitem{masseroniSpinorbitProximityMoS2024}
\bibinfo{author}{Masseroni, M.} \emph{et~al.}
\newblock \bibinfo{title}{Spin-orbit proximity in {{MoS}}\$\_2\$/bilayer
  graphene heterostructures}.
\newblock \emph{\bibinfo{journal}{arXiv:2403.17120 [cond-mat]}}
  (\bibinfo{year}{2024}).
\newblock \eprint{2403.17120}.

\bibitem{seilerLayerselectiveSpinorbitCoupling2024}
\bibinfo{author}{Seiler, A.~M.} \emph{et~al.}
\newblock \bibinfo{title}{Layer-selective spin-orbit coupling and strong
  correlation in bilayer graphene}.
\newblock \emph{\bibinfo{journal}{arXiv:2403.17140 [cond-mat]}}
  (\bibinfo{year}{2024}).
\newblock \eprint{2403.17140}.

\bibitem{sunDeterminingSpinorbitCoupling2023}
\bibinfo{author}{Sun, L.} \emph{et~al.}
\newblock \bibinfo{title}{Determining spin-orbit coupling in graphene by
  quasiparticle interference imaging}.
\newblock \emph{\bibinfo{journal}{Nat Commun}} \textbf{\bibinfo{volume}{14}},
  \bibinfo{pages}{3771} (\bibinfo{year}{2023}).

\bibitem{tolmachevLogarithmicCriterionSuperconductivity1962}
\bibinfo{author}{Tolmachev, V.~V.}
\newblock \bibinfo{title}{Logarithmic criterion for superconductivity}.
\newblock \emph{\bibinfo{journal}{SPhD}} \textbf{\bibinfo{volume}{6}},
  \bibinfo{pages}{800} (\bibinfo{year}{1962}).

\bibitem{morelCalculationSuperconductingState1962}
\bibinfo{author}{Morel, P.} \& \bibinfo{author}{Anderson, P.~W.}
\newblock \bibinfo{title}{Calculation of the {{Superconducting State
  Parameters}} with {{Retarded Electron-Phonon Interaction}}}.
\newblock \emph{\bibinfo{journal}{Phys. Rev.}} \textbf{\bibinfo{volume}{125}},
  \bibinfo{pages}{1263--1271} (\bibinfo{year}{1962}).

\bibitem{aroraSuperconductivityMetallicTwisted2020}
\bibinfo{author}{Arora, H.~S.} \emph{et~al.}
\newblock \bibinfo{title}{Superconductivity in metallic twisted bilayer
  graphene stabilized by {{WSe2}}}.
\newblock \emph{\bibinfo{journal}{Nature}} \textbf{\bibinfo{volume}{583}},
  \bibinfo{pages}{379--384} (\bibinfo{year}{2020}).

\bibitem{parkTunableStronglyCoupled2021}
\bibinfo{author}{Park, J.~M.}, \bibinfo{author}{Cao, Y.},
  \bibinfo{author}{Watanabe, K.}, \bibinfo{author}{Taniguchi, T.} \&
  \bibinfo{author}{{Jarillo-Herrero}, P.}
\newblock \bibinfo{title}{Tunable strongly coupled superconductivity in
  magic-angle twisted trilayer graphene}.
\newblock \emph{\bibinfo{journal}{Nature}} \textbf{\bibinfo{volume}{590}},
  \bibinfo{pages}{249--255} (\bibinfo{year}{2021}).

\bibitem{haoElectricFieldTunable2021}
\bibinfo{author}{Hao, Z.} \emph{et~al.}
\newblock \bibinfo{title}{Electric field--tunable superconductivity in
  alternating-twist magic-angle trilayer graphene}.
\newblock \emph{\bibinfo{journal}{Science}} \textbf{\bibinfo{volume}{371}},
  \bibinfo{pages}{1133--1138} (\bibinfo{year}{2021}).

\bibitem{parkRobustSuperconductivityMagicangle2022}
\bibinfo{author}{Park, J.~M.} \emph{et~al.}
\newblock \bibinfo{title}{Robust superconductivity in magic-angle multilayer
  graphene family}.
\newblock \emph{\bibinfo{journal}{Nat. Mater.}} \textbf{\bibinfo{volume}{21}},
  \bibinfo{pages}{877--883} (\bibinfo{year}{2022}).

\bibitem{zhangPromotionSuperconductivityMagicangle2022}
\bibinfo{author}{Zhang, Y.} \emph{et~al.}
\newblock \bibinfo{title}{Promotion of superconductivity in magic-angle
  graphene multilayers}.
\newblock \emph{\bibinfo{journal}{Science}} \textbf{\bibinfo{volume}{377}},
  \bibinfo{pages}{1538--1543} (\bibinfo{year}{2022}).

\bibitem{yankowitzTuningSuperconductivityTwisted2019}
\bibinfo{author}{Yankowitz, M.} \emph{et~al.}
\newblock \bibinfo{title}{Tuning superconductivity in twisted bilayer
  graphene}.
\newblock \emph{\bibinfo{journal}{Science}} \textbf{\bibinfo{volume}{363}},
  \bibinfo{pages}{1059--1064} (\bibinfo{year}{2019}).

\bibitem{kedvesStabilizingInvertedPhase2023}
\bibinfo{author}{Kedves, M.} \emph{et~al.}
\newblock \bibinfo{title}{Stabilizing the {{Inverted Phase}} of a
  {{WSe2}}/{{BLG}}/{{WSe2 Heterostructure}} via {{Hydrostatic Pressure}}}.
\newblock \emph{\bibinfo{journal}{Nano Lett.}} \textbf{\bibinfo{volume}{23}},
  \bibinfo{pages}{9508--9514} (\bibinfo{year}{2023}).

\bibitem{mccannElectronicPropertiesBilayer2013}
\bibinfo{author}{McCann, E.} \& \bibinfo{author}{Koshino, M.}
\newblock \bibinfo{title}{The electronic properties of bilayer graphene}.
\newblock \emph{\bibinfo{journal}{Rep. Prog. Phys.}}
  \textbf{\bibinfo{volume}{76}}, \bibinfo{pages}{056503}
  (\bibinfo{year}{2013}).

\bibitem{dongIsospinMomentumpolarizedOrders2023}
\bibinfo{author}{Dong, Z.}, \bibinfo{author}{Davydova, M.},
  \bibinfo{author}{Ogunnaike, O.} \& \bibinfo{author}{Levitov, L.}
\newblock \bibinfo{title}{Isospin- and momentum-polarized orders in bilayer
  graphene}.
\newblock \emph{\bibinfo{journal}{Phys. Rev. B}}
  \textbf{\bibinfo{volume}{107}}, \bibinfo{pages}{075108}
  (\bibinfo{year}{2023}).

\bibitem{linSpontaneousMomentumPolarization2023}
\bibinfo{author}{Lin, J.-X.} \emph{et~al.}
\newblock \bibinfo{title}{Spontaneous momentum polarization and diodicity in
  {{Bernal}} bilayer graphene}.
\newblock \emph{\bibinfo{journal}{arXiv:2302.04261 [cond-mat]}}
  (\bibinfo{year}{2023}).
\newblock \eprint{2302.04261}.

\bibitem{nuckollsQuantumTexturesManybody2023a}
\bibinfo{author}{Nuckolls, K.~P.} \emph{et~al.}
\newblock \bibinfo{title}{Quantum textures of the many-body wavefunctions in
  magic-angle graphene}.
\newblock \emph{\bibinfo{journal}{Nature}} \textbf{\bibinfo{volume}{620}},
  \bibinfo{pages}{525--532} (\bibinfo{year}{2023}).

\bibitem{kimImagingIntervalleyCoherent2023}
\bibinfo{author}{Kim, H.} \emph{et~al.}
\newblock \bibinfo{title}{Imaging inter-valley coherent order in magic-angle
  twisted trilayer graphene}.
\newblock \emph{\bibinfo{journal}{Nature}} \textbf{\bibinfo{volume}{623}},
  \bibinfo{pages}{942--948} (\bibinfo{year}{2023}).

\bibitem{arpIntervalleyCoherenceIntrinsic2024}
\bibinfo{author}{Arp, T.} \emph{et~al.}
\newblock \bibinfo{title}{Intervalley coherence and intrinsic spin--orbit
  coupling in rhombohedral trilayer graphene}.
\newblock \emph{\bibinfo{journal}{Nat. Phys.}} \bibinfo{pages}{1--8}
  (\bibinfo{year}{2024}).

\bibitem{chatterjeeIntervalleyCoherentOrder2022}
\bibinfo{author}{Chatterjee, S.}, \bibinfo{author}{Wang, T.},
  \bibinfo{author}{Berg, E.} \& \bibinfo{author}{Zaletel, M.~P.}
\newblock \bibinfo{title}{Inter-valley coherent order and isospin fluctuation
  mediated superconductivity in rhombohedral trilayer graphene}.
\newblock \emph{\bibinfo{journal}{Nat Commun}} \textbf{\bibinfo{volume}{13}},
  \bibinfo{pages}{6013} (\bibinfo{year}{2022}).

\bibitem{kohSymmetrybrokenMetallicOrders2024}
\bibinfo{author}{Koh, J.~M.}, \bibinfo{author}{Thomson, A.},
  \bibinfo{author}{Alicea, J.} \& \bibinfo{author}{{Lantagne-Hurtubise},
  {\'E}.}
\newblock \bibinfo{title}{Symmetry-broken metallic orders in spin-orbit-coupled
  {{Bernal}} bilayer graphene}.
\newblock \emph{\bibinfo{journal}{arXiv:2407.09612 [cond-mat]}}
  (\bibinfo{year}{2024}).
\newblock \eprint{2407.09612}.

\bibitem{youKohnLuttingerSuperconductivityIntervalley2022}
\bibinfo{author}{You, Y.-Z.} \& \bibinfo{author}{Vishwanath, A.}
\newblock \bibinfo{title}{Kohn-{{Luttinger}} superconductivity and intervalley
  coherence in rhombohedral trilayer graphene}.
\newblock \emph{\bibinfo{journal}{Phys. Rev. B}}
  \textbf{\bibinfo{volume}{105}}, \bibinfo{pages}{134524}
  (\bibinfo{year}{2022}).

\bibitem{xieFlavorSymmetryBreaking2023}
\bibinfo{author}{Xie, M.} \& \bibinfo{author}{Das~Sarma, S.}
\newblock \bibinfo{title}{Flavor symmetry breaking in spin-orbit coupled
  bilayer graphene}.
\newblock \emph{\bibinfo{journal}{Phys. Rev. B}}
  \textbf{\bibinfo{volume}{107}}, \bibinfo{pages}{L201119}
  (\bibinfo{year}{2023}).

\bibitem{thomsonGatedefinedWiresTwisted2022}
\bibinfo{author}{Thomson, A.}, \bibinfo{author}{Sorensen, I.~M.},
  \bibinfo{author}{{Nadj-Perge}, S.} \& \bibinfo{author}{Alicea, J.}
\newblock \bibinfo{title}{Gate-defined wires in twisted bilayer graphene:
  {{From}} electrical detection of intervalley coherence to internally
  engineered {{Majorana}} modes}.
\newblock \emph{\bibinfo{journal}{Phys. Rev. B}}
  \textbf{\bibinfo{volume}{105}}, \bibinfo{pages}{L081405}
  (\bibinfo{year}{2022}).

\bibitem{kohCorrelatedPhasesSpinorbitcoupled2024}
\bibinfo{author}{Koh, J.~M.}, \bibinfo{author}{Alicea, J.} \&
  \bibinfo{author}{{Lantagne-Hurtubise}, {\'E}.}
\newblock \bibinfo{title}{Correlated phases in spin-orbit-coupled rhombohedral
  trilayer graphene}.
\newblock \emph{\bibinfo{journal}{Phys. Rev. B}}
  \textbf{\bibinfo{volume}{109}}, \bibinfo{pages}{035113}
  (\bibinfo{year}{2024}).

\bibitem{zhumagulovSwappingExchangeSpinorbit2024}
\bibinfo{author}{Zhumagulov, Y.}, \bibinfo{author}{Kochan, D.} \&
  \bibinfo{author}{Fabian, J.}
\newblock \bibinfo{title}{Swapping exchange and spin-orbit induced correlated
  phases in proximitized {{Bernal}} bilayer graphene}.
\newblock \emph{\bibinfo{journal}{Phys. Rev. B}}
  \textbf{\bibinfo{volume}{110}}, \bibinfo{pages}{045427}
  (\bibinfo{year}{2024}).

\bibitem{dongSuperconductivitySpincantingFluctuations2024}
\bibinfo{author}{Dong, Z.}, \bibinfo{author}{{Lantagne-Hurtubise}, {\'E}.} \&
  \bibinfo{author}{Alicea, J.}
\newblock \bibinfo{title}{Superconductivity from spin-canting fluctuations in
  rhombohedral graphene}.
\newblock \emph{\bibinfo{journal}{arXiv:2406.17036 [cond-mat]}}
  (\bibinfo{year}{2024}).
\newblock \eprint{2406.17036}.

\bibitem{frigeriSuperconductivityInversionSymmetry2004}
\bibinfo{author}{Frigeri, P.~A.}, \bibinfo{author}{Agterberg, D.~F.},
  \bibinfo{author}{Koga, A.} \& \bibinfo{author}{Sigrist, M.}
\newblock \bibinfo{title}{Superconductivity without {{Inversion Symmetry}}:
  {{MnSi}} versus {{CePt3Si}}}.
\newblock \emph{\bibinfo{journal}{Phys. Rev. Lett.}}
  \textbf{\bibinfo{volume}{92}}, \bibinfo{pages}{097001}
  (\bibinfo{year}{2004}).

\bibitem{luEvidenceTwodimensionalIsing2015}
\bibinfo{author}{Lu, J.~M.} \emph{et~al.}
\newblock \bibinfo{title}{Evidence for two-dimensional ising superconductivity
  in gated {{MoS}}$_2$}.
\newblock \emph{\bibinfo{journal}{Science}} \textbf{\bibinfo{volume}{350}},
  \bibinfo{pages}{1353--1357} (\bibinfo{year}{2015}).

\bibitem{saitoSuperconductivityProtectedSpinvalley2016}
\bibinfo{author}{Saito, Y.} \emph{et~al.}
\newblock \bibinfo{title}{Superconductivity protected by spin-valley locking in
  ion-gated {{MoS}}$_2$}.
\newblock \emph{\bibinfo{journal}{Nat Phys}} \textbf{\bibinfo{volume}{12}},
  \bibinfo{pages}{144--149} (\bibinfo{year}{2016}).

\bibitem{seylerElectricalControlSecondharmonic2015}
\bibinfo{author}{Seyler, K.~L.} \emph{et~al.}
\newblock \bibinfo{title}{Electrical control of second-harmonic generation in a
  {{WSe2}} monolayer transistor}.
\newblock \emph{\bibinfo{journal}{Nature Nanotech}}
  \textbf{\bibinfo{volume}{10}}, \bibinfo{pages}{407--411}
  (\bibinfo{year}{2015}).

\bibitem{cohenMagneticBreakdownCrystals1961}
\bibinfo{author}{Cohen, M.~H.} \& \bibinfo{author}{Falicov, L.~M.}
\newblock \bibinfo{title}{Magnetic {{Breakdown}} in {{Crystals}}}.
\newblock \emph{\bibinfo{journal}{Phys. Rev. Lett.}}
  \textbf{\bibinfo{volume}{7}}, \bibinfo{pages}{231--233}
  (\bibinfo{year}{1961}).

\bibitem{jungAccurateTightbindingModels2014}
\bibinfo{author}{Jung, J.} \& \bibinfo{author}{MacDonald, A.~H.}
\newblock \bibinfo{title}{Accurate tight-binding models for the $\pi$ bands of
  bilayer graphene}.
\newblock \emph{\bibinfo{journal}{Phys. Rev. B}} \textbf{\bibinfo{volume}{89}},
  \bibinfo{pages}{035405} (\bibinfo{year}{2014}).

\bibitem{shavitInducingSuperconductivityBilayer2023}
\bibinfo{author}{Shavit, G.} \& \bibinfo{author}{Oreg, Y.}
\newblock \bibinfo{title}{Inducing superconductivity in bilayer graphene by
  alleviation of the {{Stoner}} blockade}.
\newblock \emph{\bibinfo{journal}{Phys. Rev. B}}
  \textbf{\bibinfo{volume}{108}}, \bibinfo{pages}{024510}
  (\bibinfo{year}{2023}).

\bibitem{nagaosaQuantumFieldTheory1999}
\bibinfo{author}{Nagaosa, N.}
\newblock \emph{\bibinfo{title}{Quantum {{Field Theory}} in {{Condensed Matter
  Physics}}}} (\bibinfo{publisher}{Springer}, \bibinfo{address}{Berlin,
  Heidelberg}, \bibinfo{year}{1999}).

\bibitem{chouAcousticphononmediatedSuperconductivityBernal2022}
\bibinfo{author}{Chou, Y.-Z.}, \bibinfo{author}{Wu, F.}, \bibinfo{author}{Sau,
  J.~D.} \& \bibinfo{author}{Das~Sarma, S.}
\newblock \bibinfo{title}{Acoustic-phonon-mediated superconductivity in
  {{Bernal}} bilayer graphene}.
\newblock \emph{\bibinfo{journal}{Phys. Rev. B}}
  \textbf{\bibinfo{volume}{105}}, \bibinfo{pages}{L100503}
  (\bibinfo{year}{2022}).

\bibitem{lianTwistedBilayerGraphene2019}
\bibinfo{author}{Lian, B.}, \bibinfo{author}{Wang, Z.} \&
  \bibinfo{author}{Bernevig, B.~A.}
\newblock \bibinfo{title}{Twisted {{Bilayer Graphene}}: {{A Phonon-Driven
  Superconductor}}}.
\newblock \emph{\bibinfo{journal}{Phys. Rev. Lett.}}
  \textbf{\bibinfo{volume}{122}}, \bibinfo{pages}{257002}
  (\bibinfo{year}{2019}).

\bibitem{kheirabadiMagneticRatchetEffect2016}
\bibinfo{author}{Kheirabadi, N.}, \bibinfo{author}{McCann, E.} \&
  \bibinfo{author}{Fal'ko, V.~I.}
\newblock \bibinfo{title}{Magnetic ratchet effect in bilayer graphene}.
\newblock \emph{\bibinfo{journal}{Phys. Rev. B}} \textbf{\bibinfo{volume}{94}},
  \bibinfo{pages}{165404} (\bibinfo{year}{2016}).

\bibitem{zaletelGatetunableStrongFragile2019}
\bibinfo{author}{Zaletel, M.~P.} \& \bibinfo{author}{Khoo, J.~Y.}
\newblock \bibinfo{title}{The gate-tunable strong and fragile topology of
  multilayer-graphene on a transition metal dichalcogenide}.
\newblock \emph{\bibinfo{journal}{arXiv:1901.01294 [cond-mat]}}
  (\bibinfo{year}{2019}).
\newblock \eprint{1901.01294}.

\bibitem{maUpperCriticalInplane2024}
\bibinfo{author}{Ma, H.}, \bibinfo{author}{Chichinadze, D.~V.} \&
  \bibinfo{author}{Lewandowski, C.}
\newblock \bibinfo{title}{Upper critical in-plane magnetic field in
  quasi-{{2D}} layered superconductors}.
\newblock \emph{\bibinfo{journal}{In preparation.}}  (\bibinfo{year}{2024}).

\bibitem{saint-jamesTypeIISuperconductivity1969}
\bibinfo{author}{{Saint-James}, D.}, \bibinfo{author}{Sarma, G.},
  \bibinfo{author}{Thomas, E.~J.} \& \bibinfo{author}{Silverman, P.}
\newblock \emph{\bibinfo{title}{Type {{II}} Superconductivity}}
  (\bibinfo{publisher}{Oxford, New York, Pergamon Press},
  \bibinfo{year}{1969}).

\bibitem{zwicknaglCriticalMagneticField2017}
\bibinfo{author}{Zwicknagl, G.}, \bibinfo{author}{Jahns, S.} \&
  \bibinfo{author}{Fulde, P.}
\newblock \bibinfo{title}{Critical {{Magnetic Field}} of {{Ultra-Thin
  Superconducting Films}} and {{Interfaces}}}.
\newblock \emph{\bibinfo{journal}{J. Phys. Soc. Jpn.}}
  \textbf{\bibinfo{volume}{86}}, \bibinfo{pages}{083701}
  (\bibinfo{year}{2017}).

\end{thebibliography}

\clearpage
\clearpage

\clearpage

\clearpage

\section*{Methods}

\begin{figure}[p]
    \includegraphics[width=12cm]{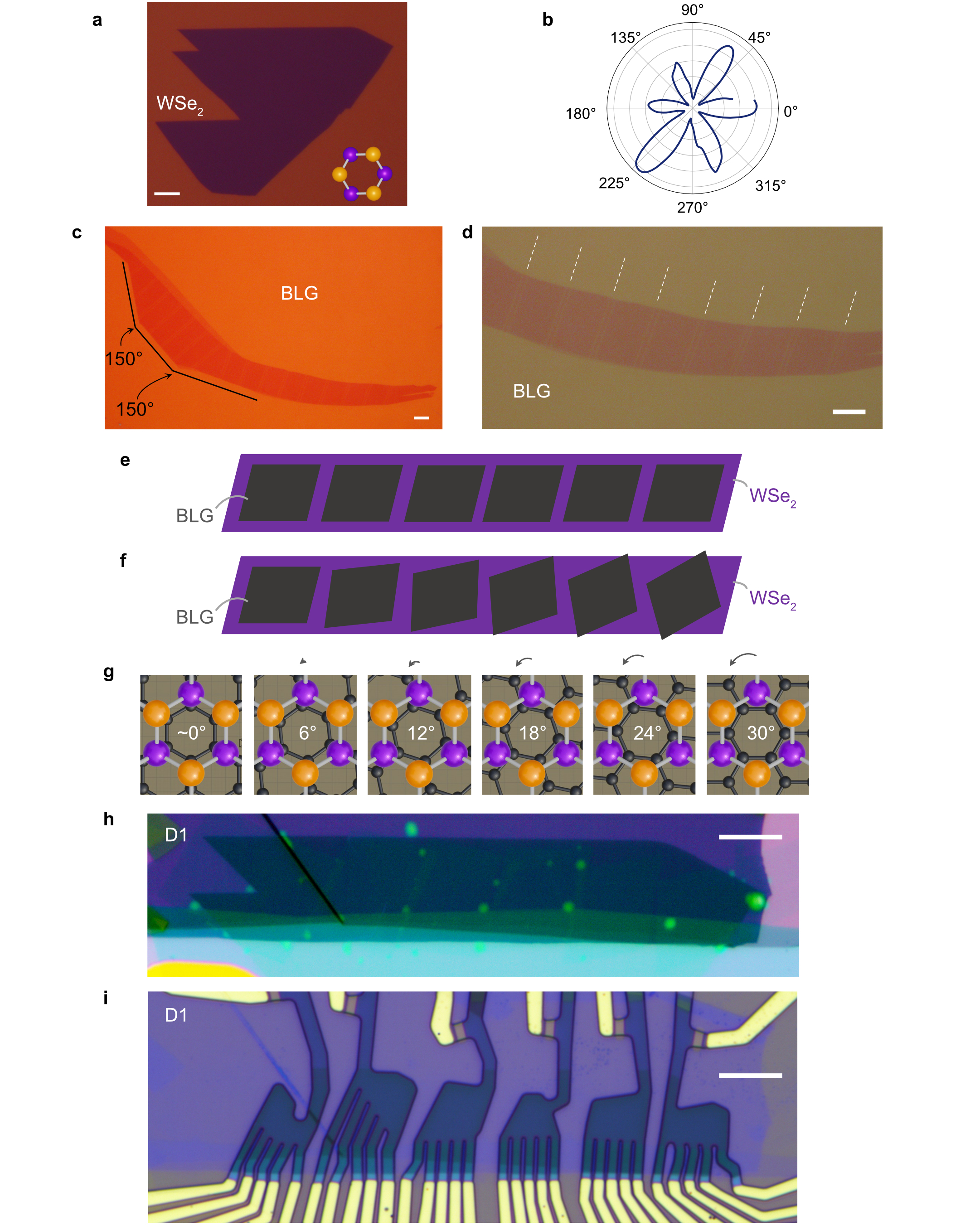}
    \centering
    \caption{
    {\bf Device fabrication for BLG-WSe$_2$ twisting.} {\bf a}, Optical image of a WSe$_2$ crystal. {\bf b}, Second harmonic generation for the WSe$_2$ flake shown in {\bf a}; the polarization of the incident and reflected beams are selected to lie parallel to the scattering plane.
    {\bf c}, Optical image of a large BLG flake. Straight edges form angles $150 \degree$ that are consistent with the three straight edges being along zigzag- or armchair-edge direction. {\bf d}, Zoom-in image of the BLG in {\bf c}, showing small BLG pieces that are separated by atomic-force-microscope-actuated cutting. {\bf e}-{\bf g}, Schematics showing the flake transferring processes for the continuous interfacial twisting. The BLG pieces are sequentially picked up with angles relative to WSe$_2$ in increments of $6 \degree$, from $\sim 0 \degree$ to $30 \degree$. {\bf h}, Optical image of the twisting stack, clearly showing that the BLG pieces form different twist angles relative to the WSe$_2$ crystal. {\bf i}, Optical image of the finished device set D1. All the scale bars correspond to 10$\mu$m.
    }

\label{exfig:fabrication}
\end{figure}

\textbf{Device fabrication:} 
The majority of the fabrication processes follow the standard flake transfer and lithography explained in a previous study\cite{zhangEnhancedSuperconductivitySpin2023}.
Here, we focus on the interfacial twisting between BLG and WSe$_2$. Large flakes of BLG and WSe$_2$ are exfoliated on SiO$_2$/Si chips.
The crystal orientation of WSe$_2$ can be identified by second harmonic generation\cite{seylerElectricalControlSecondharmonic2015} (SHG; \prettyref{exfig:fabrication}b), where the
polarization of the incident and reflected beams are selected to lie  parallel to the scattering plane.
The directions with maximized SHG signal correspond to the in-plane crystal orientations along the W-Se direction (armchair direction). BLG is somewhat trickier. We identify flakes with long straight edges forming angles that are multiple of 30$\degree$, e.g., three edges form two angles of 150$\degree$ in \prettyref{exfig:fabrication}c.
The configuration is consistent with the assignment that the straight edges are along the zigzag- or armchair-edge direction of graphene.
We then cut the large BLG flake into small pieces\cite{liElectrodeFreeAnodicOxidation2018}; \prettyref{exfig:fabrication}d.
First, pick up topmost hBN, top graphite gate, top hBN dielectric, and the large WSe$_2$ flake using propylene carbonate (PC) film on a polydimethylsiloxane (PDMS).
Then, align the straight edge of BLG with the crystal orientation of WSe$_2$ and control the approach of PC/PDMS stamp so that only one BLG piece is picked up.
SiO$_2$/Si chip was manually rotated by an angle $\theta \sim 6\degree$, and a second piece of BLG was picked up but not overlapping with the first one. Repeat the same processes for the remaining BLG pieces (\prettyref{exfig:fabrication}f,g).
Depending on whether the BLG straight edge used for alignment is along the zigzag or armchair direction, the crystal axes of the six BLG pieces are rotated relative to the WSe$_2$ axis by an angle $\theta \sim$ 0$\degree$, 6$\degree$, 12$\degree$, 18$\degree$, 24$\degree$, and 30$\degree$ (armchair direction), or vice versa (zigzag direction). The two configurations  can be distinguished by measuring the Ising SOC strengths of the devices at the two ends. The large (small) Ising device corresponds to $\sim 0 \degree$ ($\sim 30 \degree$) alignment due to the reflection symmetry\cite{liTwistangleDependenceProximity2019, chouEnhancedSuperconductivityVirtual2022, davidInducedSpinorbitCoupling2019, naimerTwistangleDependentProximity2021}. 
A typical finished stack is shown in \prettyref{exfig:fabrication}h; a series of different rotation angles between BLG and WSe$_2$ can be clearly seen from the optical image. The stack went through standard lithographic and etching processes for final device preparation (\prettyref{exfig:fabrication}i).

\textbf{Measurements:} All measurements were performed in a dilution
refrigerator (Oxford Triton) with a base temperature of $\sim30$~mK and an 1T/1T/9T (XYZ) superconducting vector magnet, using
standard low-frequency lock-in amplifier techniques. Unless otherwise
specified, measurements are taken at the base temperature. Frequencies of the
lock-in amplifiers (Stanford Research, models 865a) were kept in the range of
$7-40$~Hz in order to reduce the electronic noise and measure the device's DC
properties. The AC excitation was kept $<5$~nA; most measurements were taken at
$1$~nA to preserve the linearity of the system and to avoid disturbing the
fragile states at low temperatures. Each of the DC fridge lines passes through
cold filters, including 4 Pi filters that filter out a range from $\sim80$~MHz
to $>10$~GHz, as well as a two-pole RC low-pass filter. 

\textbf{Ising SOC:} One way to quantify the WSe$_{2}$-induced Ising SOC $H_I = \frac{1}{2}\lambda_I\tau_z s_z$ ($\lambda_I$ is the Ising SOC strength in the main text) is to probe the octet zeroth Landau level (LL) in
BLG. Note that these LL energies are not sensitive to Rashba
SOC\cite{khooTunableQuantumHall2018}. The sets of two Landau levels that cross at filling
factors $\nu = \pm 3$ have opposite layer polarizations, so that their energy difference (at
zero $D$ field) is given by $\Delta E = E_Z \pm \lambda_I/2$ ($E_Z$ is the
Zeeman gap between spin-up and spin-down LLs). Therefore, the critical field $B_\perp^*$ that makes $\Delta E$ vanish is
$2 E_Z = 2g\mu_B B_\perp^\ast =\lambda_I$. However, this method doesn't work when $\lambda_I$ is negative since the energy-level crossing is not inverted\cite{wangQuantumHallEffect2019}.

\begin{figure}[p]
    \includegraphics[width=13cm]{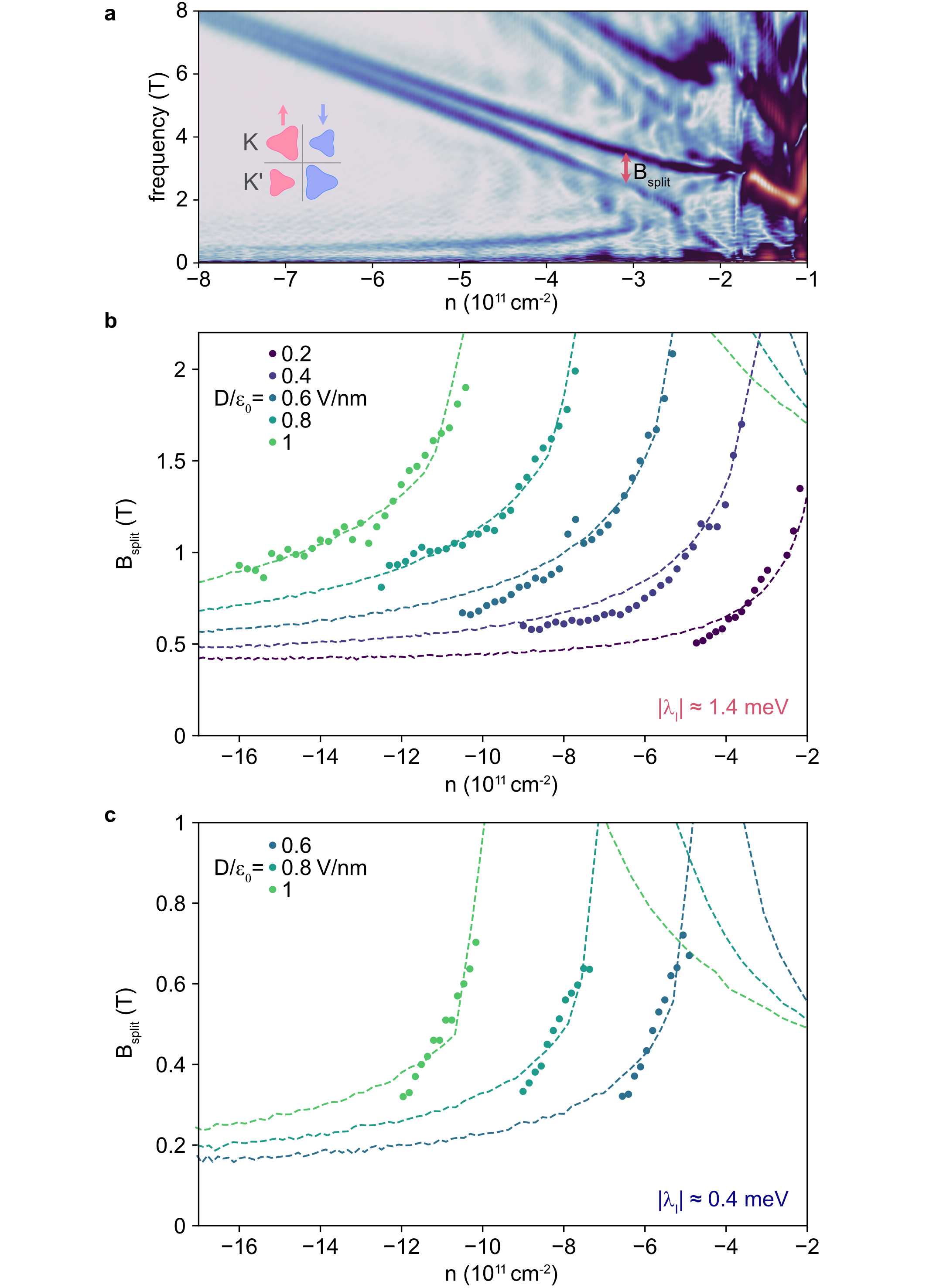}
    \centering
    \caption{
    {\bf Quantifying Ising SOC strength $|\lambda_I|$ by quantum oscillations.} {\bf a}, The same data as the one in \prettyref{fig:fig1}h, but without the frequency normalization to show $B_\mathrm{split}$. {\bf b},{\bf c}, Experimental (dots) doping-dependent frequency splitting around $f_\nu = 1/4$ measured at different $D$ fields for a large Ising device ({\bf b}; $|\lambda_I| \approx 1.4$~meV) and a small Ising device ({\bf c}; $|\lambda_I| \approx 0.4$~meV). The dashed lines are $B_\mathrm{split}$ calculated from single-particle band structure using the corresponding Ising SOC values.
    }

\label{exfig:Ising_extraction}
\end{figure}

\begin{figure}[p]
    \includegraphics[width=15cm]{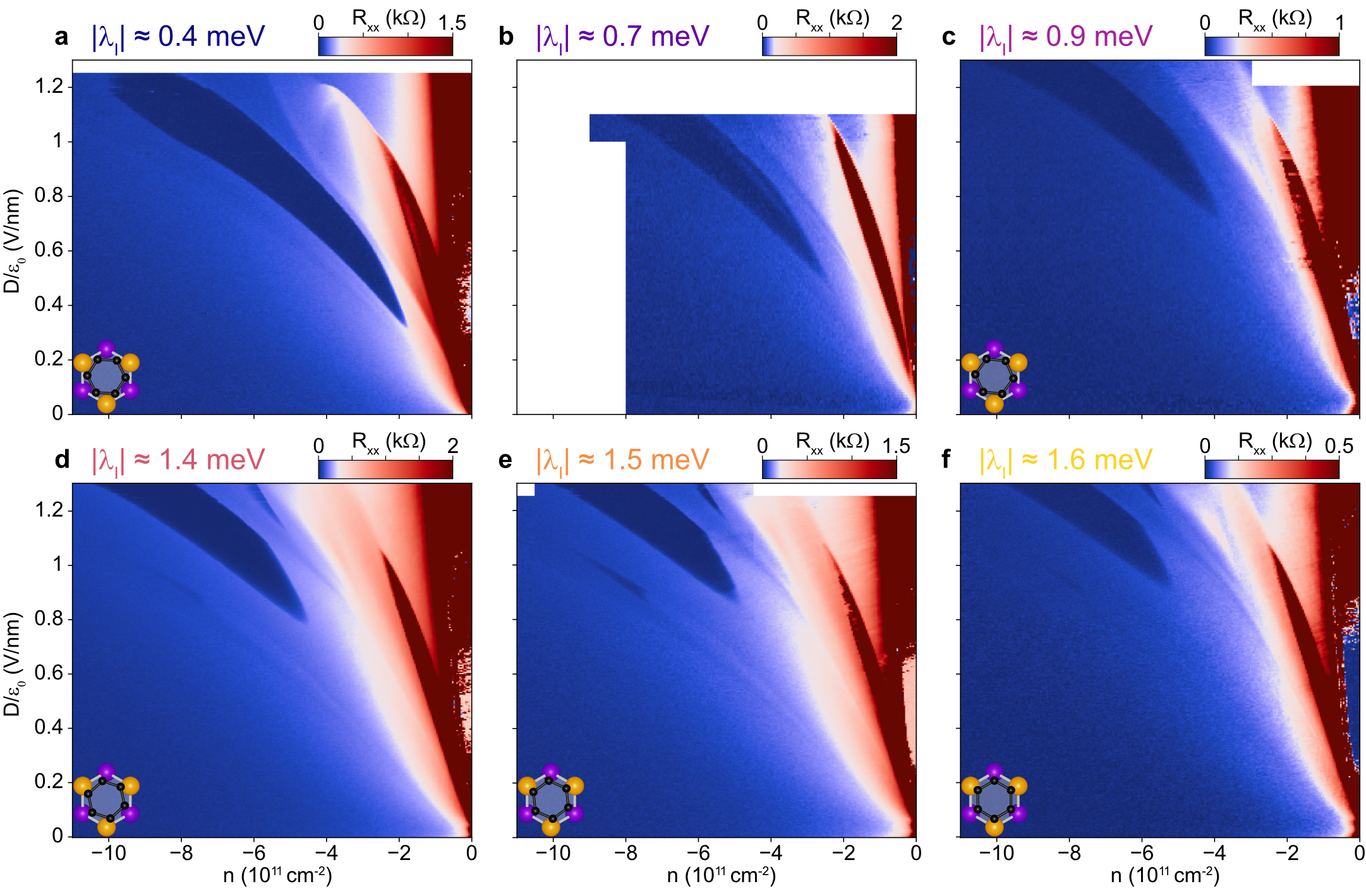}
    \centering
    \caption{{\bf $n$-$D$ phase diagrams for devices with various Ising SOC strengths.} {\bf a}-{\bf f}, $R_{xx}$ versus doping density $n$ and displacement field $D$ for devices with Ising SOC strength $|\lambda_I| \approx 0.4$~meV ({\bf a}), $0.7$~meV ({\bf b}), $0.9$~meV ({\bf c}), $1.4$~meV ({\bf d}), $1.5$~meV ({\bf e}), and $1.6$~meV ({\bf f}), respectively.}

\label{exfig:nD diagram_different Ising}
\end{figure}

\begin{figure}[p]
    \includegraphics[width=14cm]{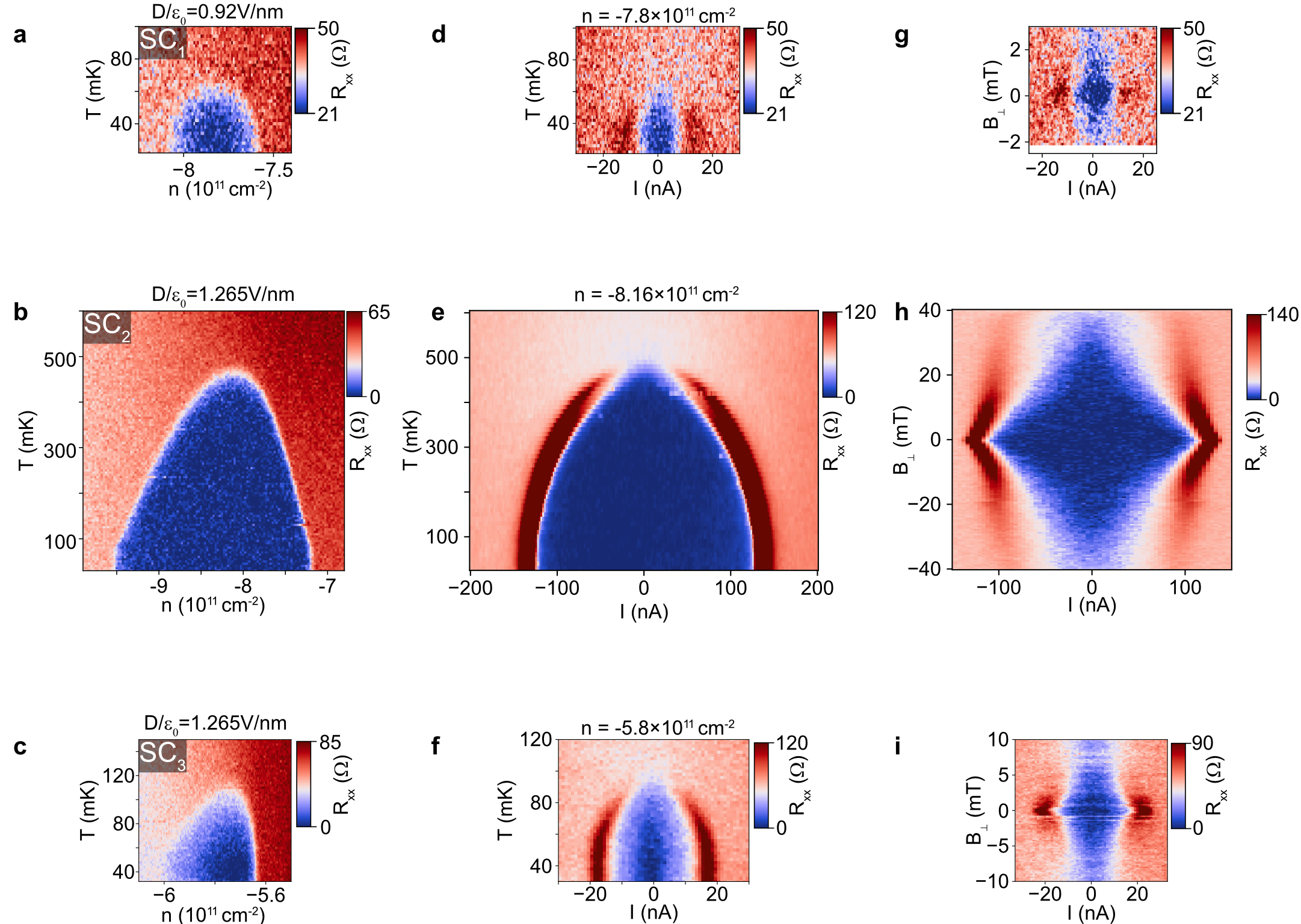}
    \centering
    \caption{{\bf Characterizations of the three superconducting regions SC$_1$, SC$_2$, and SC$_3$.}
    {\bf a}-{\bf c}, Temperature dependence of the three superconducting domes SC$_1$ ({\bf a}), SC$_2$ ({\bf b}), and SC$_3$ ({\bf c}), respectively. {\bf d}-{\bf f}, Critical current versus temperature at the corresponding $D$ and $n$.  {\bf g}-{\bf i}, Critical current disappearing with $B_\perp$ at the same $D$ and $n$ as in {\bf d}-{\bf f}.  
    } 

\label{exfig:three_SC}
\end{figure}

Independently, $|\lambda_I|$ can also be extracted from the doping-dependent FFT
splitting of the quantum oscillations, regardless the sign of $\lambda_I$ (\prettyref{fig:fig1}f-i and \prettyref{exfig:Ising_extraction}a). \prettyref{exfig:Ising_extraction}b,c shows the
doping-dependent FFT splitting $B_\text{split}$ measured at different $D$ fields within the non-interacting phase (schematics in \prettyref{fig:fig1}h,i). 
Ising-type splitting is suppressed with increasing $|n|$, in
contrast to the Rashba-type splitting which increases with increasing $|n|$\cite{zhangEnhancedSuperconductivitySpin2023}. 
The detailed mapping of $B_\text{split}$ as a function of $n$ and $D$ enables comparison to single-particle band structure calculation that quantifies Ising-induced Fermi surface imbalance.
The dashed lines in \prettyref{exfig:Ising_extraction}b,c are calculated frequency splittings for $|\lambda_I| \approx 1.4$~meV (\prettyref{exfig:Ising_extraction}b) and $|\lambda_I| \approx 0.4$~meV (\prettyref{exfig:Ising_extraction}c), respectively.
Both cases roughly match the experimental data. 
The overall trend is that ($i$) at constant $|\lambda_I|$,  higher $D$ features larger $B_\text{split}$ and ($ii$) at constant $D$, higher $|\lambda_I|$ features larger $B_\text{split}$. The observed trends put strong constraints on the estimates of the Ising SOC strength. Note that one single Ising SOC strength provides a good fit to the data at different $D$ fields (from $D/\epsilon_0 = 0.2~\text{V/nm}$ to $1~\text{V/nm}$; \prettyref{exfig:Ising_extraction}b), suggesting that Ising SOC is already maximal at $D/\epsilon_0 = 0.2~\text{V/nm}$ and larger $D$ values do not further increase the Ising SOC strength.
 
\begin{figure}[p]
    \includegraphics[width=14cm]{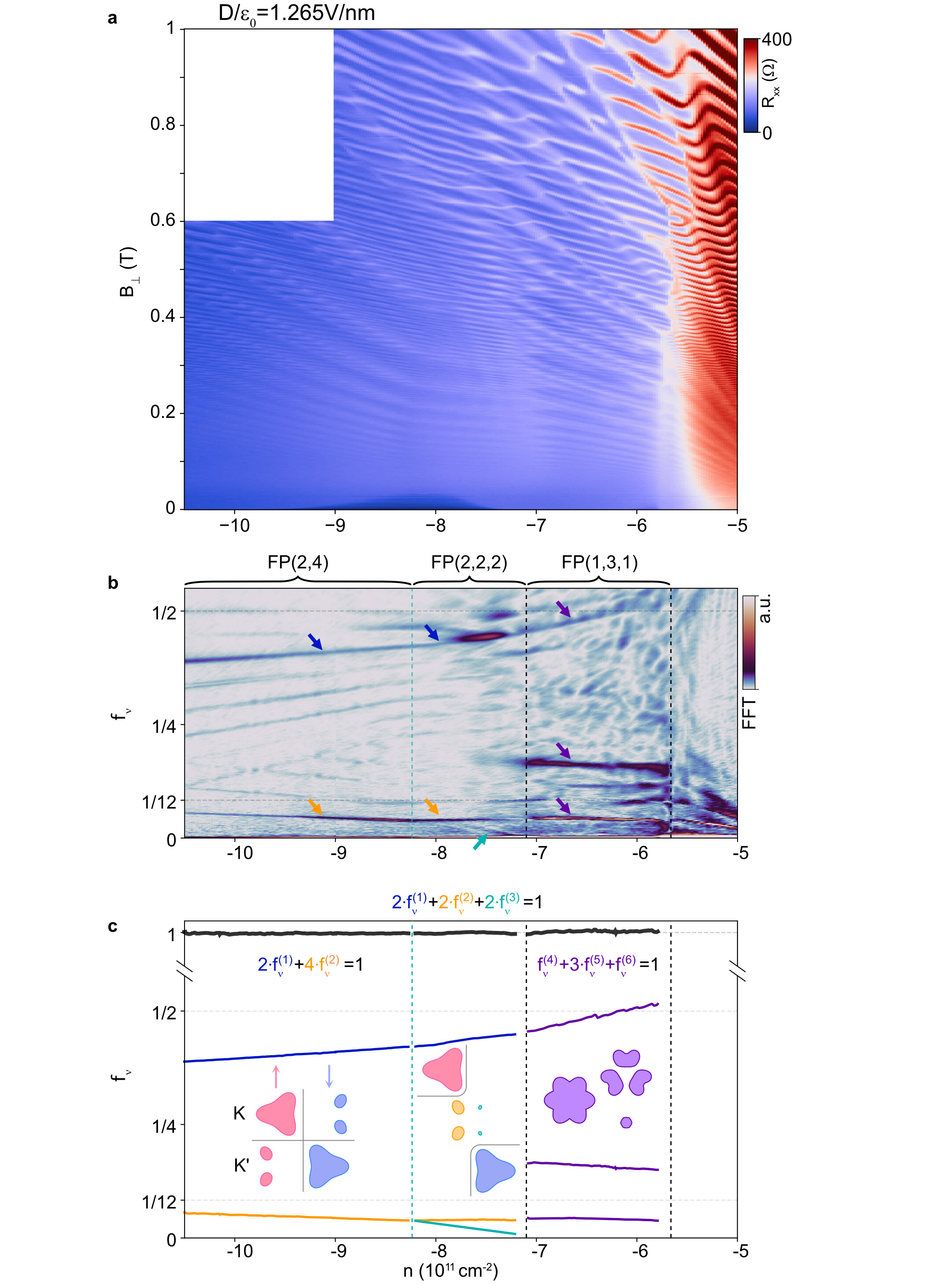}
    \centering
    \caption{{\bf Quantum oscillations and FFT measured at $D/\epsilon_0 = 1.265~\text{V/nm}$.} {\bf a}, $R_{xx}$ versus out-of-plane magnetic field $B_\perp$ and doping density $n$ measured at $D/\epsilon_0 = 1.265~\text{V/nm}$ for a device with $|\lambda_I| \approx 1.5$~meV. {\bf b}, Frequency-normalized Fourier transform of $R_{xx}(1/B_\perp)$ over the same doping density range as in {\bf a}.  {\bf c}, Intensity peaks in $f_\nu$ from {\bf b}.}

\label{exfig:D1p265_Fan}
\end{figure}

\begin{figure}[p]
    \includegraphics[width=14cm]{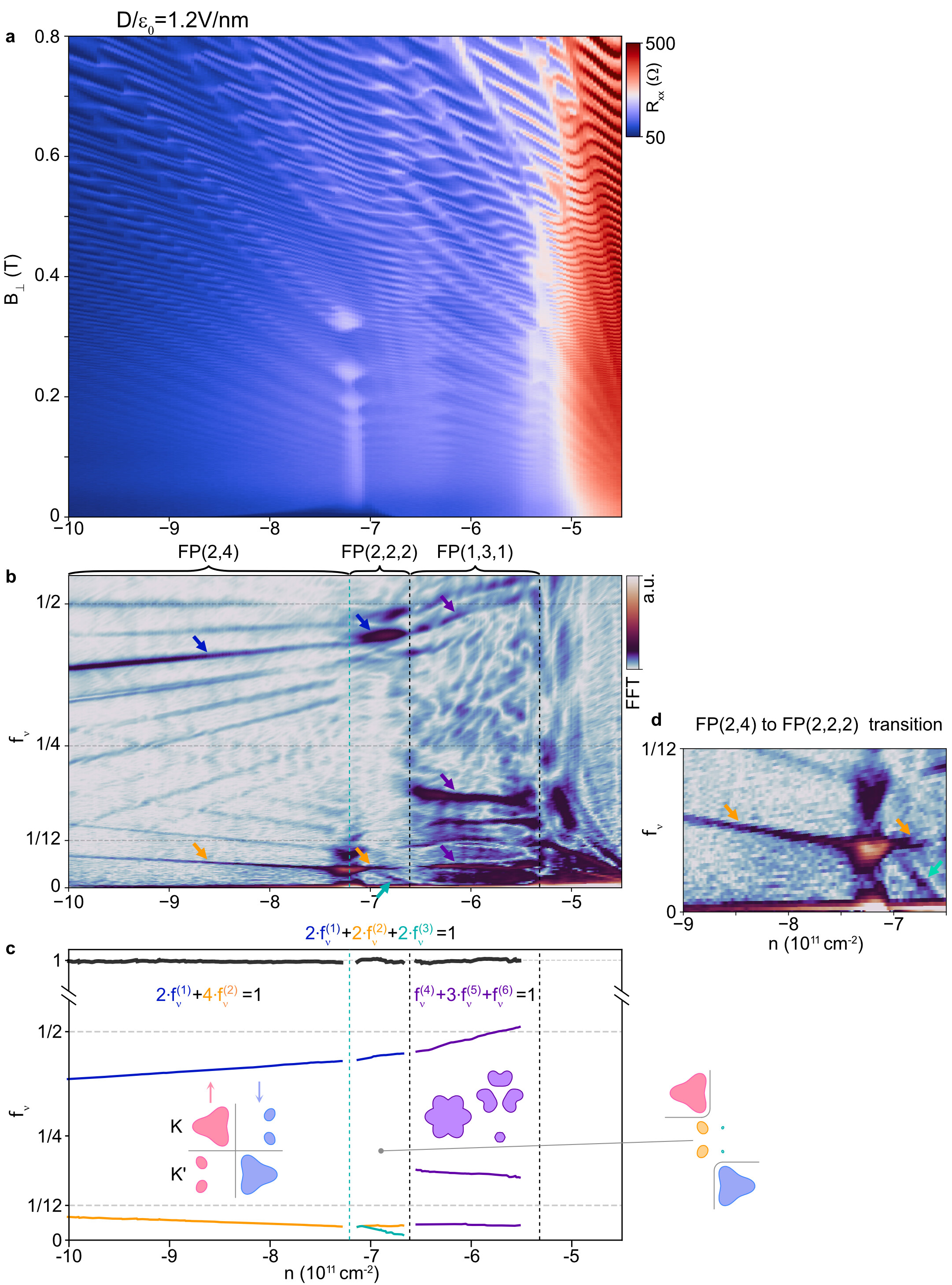}
    \centering
    \caption{{\bf Quantum oscillations and FFT measured at $D/\epsilon_0 = 1.2~\text{V/nm}$.} {\bf a}, $R_{xx}$ versus out-of-plane magnetic field $B_\perp$ and doping density $n$ measured at $D/\epsilon_0 = 1.2~\text{V/nm}$ for a device with $|\lambda_I| \approx 1.5$~meV. {\bf b}, Frequency-normalized Fourier transform of $R_{xx}(1/B_\perp)$ over the same density range as in {\bf a}.  {\bf c}, Intensity peaks in $f_\nu$ from {\bf b}. {\bf d}, zoom-in image at low frequencies from {\bf b}.}

\label{exfig:D1p2_Fan}
\end{figure}

\begin{figure}[p]
    \includegraphics[width=12.5cm]{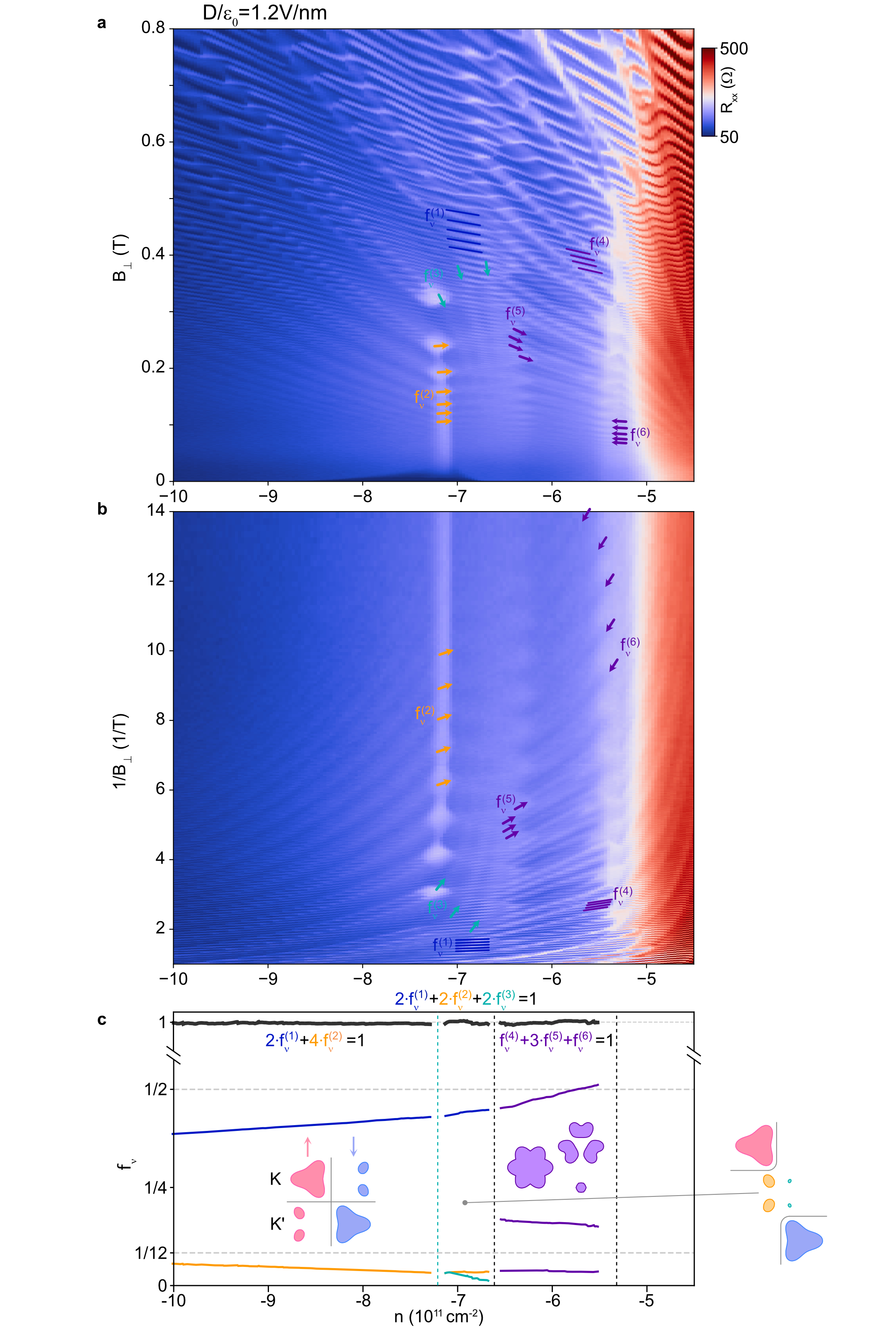}
    \centering
    \caption{
    {\bf Identifying $\mathbf {FP(2,2,2)}$ and $\mathbf{FP(1,3,1)}$ frequencies from the raw data.} {\bf a}, $R_{xx}$ versus out-of-plane magnetic field $B_\perp$ and doping density $n$ measured at $D/\epsilon_0 = 1.2~\text{V/nm}$ for a device with $|\lambda_I| \approx 1.5$~meV. {\bf b}, The same data as in {\bf a}, but plotted as a function of $1/B_\perp$. The corresponding frequencies are marked by colored arrows and lines. {\bf c}, Intensity peaks in $f_\nu$ extracted from the FFT data.
    }

\label{exfig:frequency_raw data}
\end{figure}

\textbf{Identification of $\mathbf{FP(2,2,2)}$ phase:}
Ultrahigh-resolution quantum oscillations at high $D$ fields allow for resolving subtle symmetry-breaking Fermi pockets.
Looking carefully at the FFT frequency $f_\nu^{(2)}$ in \prettyref{exfig:D1p265_Fan}b and \prettyref{exfig:D1p2_Fan}b,d, $f_\nu^{(2)}$ decreases monotonically with lowering doping until reaching $n \approx -8.25\times 10^{11} ~\text{cm}^{-2}$ for $D/\epsilon_0 = 1.265~\text{V/nm}$ ($n \approx -7.2\times 10^{11} ~\text{cm}^{-2}$ for $D/\epsilon_0 = 1.2~\text{V/nm}$), beyond which the dependence of $f_\nu^{(2)}$ is flattened while $f_\nu^{(1)}$ keeps increasing throughout.
This indicates that the sum rule $2 \cdot f_\nu^{(1)} + 4 \cdot f_\nu^{(2)} \approx 1$ of the $\mathrm{FP}(2,4)$ phase at higher doping is no longer satisfied for lower dopings, suggesting an altered Fermi-surface structure.
Indeed, measuring quantum oscillations to higher $B_\perp$ field reveals the emergence of a third very low frequency around the phase boundary as marked by the green arrows in \prettyref{exfig:frequency_raw data}. 
FFT data shown in \prettyref{exfig:D1p265_Fan}b and \prettyref{exfig:D1p2_Fan}b,d clearly reveal the third frequency $f_\nu^{(3)}$ growing rapidly from zero at the phase boundary to a value matching $f_\nu^{(2)}$ at slightly higher doping.
The frequencies obey $2 \cdot f_\nu^{(1)} + 2 \cdot f_\nu^{(2)} + 2 \cdot f_\nu^{(3)} \approx 1$, as discussed in the main text corresponding to an additonal symmetry breaking with trigonally warped pockets of two sizes $f_\nu^{(2)}$ and $f_\nu^{(3)}$.

\textbf{Identification of $\mathbf{FP(1,3,1)}$ phase:}
The identification of $\mathrm{FP}(1,3,1)$ phase is more subtle, involving extensive quantum oscillation measurements.
The raw data (\prettyref{exfig:frequency_raw data}a,b; $n \approx -6.6\times 10^{11} ~\text{cm}^{-2}$ to $-5.3\times 10^{11} ~\text{cm}^{-2}$ taken at $D/\epsilon_0 \approx 1.2~\text{V/nm}$) reveal three oscillation frequencies. 
A high frequency marked $f_\nu^{(4)}$ appears clearly.
At really low $B_\perp \sim 0.05$~T, a relatively low frequency called $f_\nu^{(6)}$ onsets. Further increasing $B_\perp$, each $f_\nu^{(6)}$ period splits into four, giving rise to $f_\nu^{(5)}$ which is indeed roughly four times the frequency $f_\nu^{(6)}$ (\prettyref{exfig:frequency_raw data}c).

It is natural to ask whether  $f_\nu^{(5)}$ is simply a higher (fourth) harmonic of $f_\nu^{(6)}$. This can be answered by 
the measurements at low $D$ fields in the same phase region (\prettyref{exfig:QO_FP(1,3,1)_1} and \ref{exfig:QO_FP(1,3,1)_2}). 
Frequencies are marked by arrows.
When lowering the $D$ fields,  $f_\nu^{(4)}$ gradually increases, while $f_\nu^{(5)}$ and $f_\nu^{(6)}$ gradually decrease. 
At $D/\epsilon_0 = 1~\text{V/nm}$ (\prettyref{exfig:QO_FP(1,3,1)_1}b,d,f), $f_\nu^{(5)}$ already deviates from being four times the value of $f_\nu^{(6)}$. 
Eventually at $D/\epsilon_0 = 0.85~\text{V/nm}$ (\prettyref{exfig:QO_FP(1,3,1)_1}a,c,e), the frequency $f_\nu^{(6)}$ completely disappears while $f_\nu^{(5)}$ independently survives. This $D$ evolution of the two frequencies ($f_\nu^{(5)}$ and $f_\nu^{(6)}$) supports their independence.
Meanwhile at $D/\epsilon_0 = 0.85~\text{V/nm}$, the existing two frequencies ($f_\nu^{(4)}$ and $f_\nu^{(5)}$) obey $f_\nu^{(4)} + 3 \cdot f_\nu^{(5)} \approx 1$; we denote the flavor-polarized phase at this $D$ field as $\mathrm{FP}(1,3)$.
The above $D$-field evolution indicates that the $\mathrm{FP}(1,3,1)$ phase at high $D$ develops from the $\mathrm{FP}(1,3)$ phase at low $D$ as $D$ is increased.

After establishing the existence of three frequencies, we comment on the number of pockets for each type. This relies on the correct identification of intrinsic Fermi-surface frequencies and their harmonics.
At slightly higher $B_\perp$, magnetic breakdown\cite{cohenMagneticBreakdownCrystals1961} kicks in as $B_\perp$-assisted electron tunneling between different Fermi surfaces.
Consequently, the pronounced frequencies might be a sum (or difference) of two base frequencies instead of the intrinsic ones.
This is the case for $\mathrm{FP}(1,3)$ and $\mathrm{FP}(1,3,1)$ phase. 
\prettyref{exfig:QO_FP(1,3,1)_1}c shows the normalized FFT from quantum oscillations going up to $B_\perp = 0.45$~T. The frequency peak at $f_\nu \sim 0.75$ is stronger than the one at  $f_\nu \sim 0.6$ (marked by the arrow).
However, the relative intensity changes by reducing the $B_\perp$ range to 0.23T (\prettyref{exfig:QO_FP(1,3,1)_lowfield}a). At this condition, the peak at $f_\nu \sim 0.6$ is stronger than the one at $f_\nu \sim 0.75$, suggesting that the one marked by the arrow ($f_\nu \sim 0.6$ in \prettyref{exfig:QO_FP(1,3,1)_lowfield}a) is the intrinsic frequency; the one at $f_\nu \sim 0.75$ is instead a sum harmonic $f_\nu^{(4)} + f_\nu^{(5)}$. 
By identifying the intrinsic high frequency $f_\nu^{(4)}$, one obtains $f_\nu^{(4)} + 3 \cdot f_\nu^{(5)} \approx 1$, indicating one large and three small Fermi surfaces, i.e., $\mathrm{FP}(1,3)$. A similar situation holds for the other $D$ fields, such as at $D/\epsilon_0 = 1.2~\text{V/nm}$ (see \prettyref{exfig:QO_FP(1,3,1)_2}c and \prettyref{exfig:QO_FP(1,3,1)_lowfield}c), where we find $f_\nu^{(4)} + 3 \cdot f_\nu^{(5)} +f_\nu^{(6)} \approx 1$. %

\textbf{Inter-valley coherence:}
The occurrence of superconducting state SC$_3$ in the symmetry-breaking 
state $\mathrm{FP}(1,3,1)$ strongly indicates the inter-valley coherent 
nature of $\mathrm{FP}(1,3,1)$. Focusing on the single largest Fermi pocket ($f_\nu^{(4)}$)
that is non-degenerate, there are two options: it is either
valley-polarized and therefore breaks time-reversal symmetry, or it is 
inter-valley coherent. Coherence between the $K$ and $K'$ valleys would restore time 
reversal symmetry for the largest Fermi pocket, naturally more susceptible to pairing. Note that in moir\'e graphene, it is established that superconductivity originates from an inter-valley-coherent 
order\cite{kimImagingIntervalleyCoherent2023, nuckollsQuantumTexturesManybody2023a}.   

Independent evidence for inter-valley coherence comes from analyzing the evolution of 
phase boundaries as a function of $B_\perp$. An out-of-plane magnetic field $B_\perp$ favors valley-polarized states that are characterized by large orbital moments. As $B_\perp$ field 
is increased, valley-polarized states with large orbital moments are expected to take over more of the 
phase space compared to valley-balanced states\cite{arpIntervalleyCoherenceIntrinsic2024}. The evolution of the phase boundaries with $B_\perp$ can be clearly identified from quantum oscillations (\prettyref{exfig:phase boundary moving}). Here, the lowest doping density range ($n > -3\times 10^{11} ~\text{cm}^{-2}$) corresponds to a spin-valley locked $\mathrm{FP}(6)$ phase\cite{zhangEnhancedSuperconductivitySpin2023}. Within this phase, the $K$ and $K'$ valleys are equally populated with opposite spins, resulting in zero net orbital moment.
At the doping density $-3 > n > -4.3\times 10^{11} ~\text{cm}^{-2}$, the oscillation frequency peaks at $f_\nu = 1$ indicating $\mathrm{FP}(1)$ phase. The phase space shows a rich evolution: a phase transition develops with increasing $B_\perp$, consistent with a spin-valley polarized $\mathrm{FP}(1)$ (red line in \prettyref{exfig:phase boundary moving}b) emerges when $B_\perp$  is applied. Importantly, at the lowest $B_\perp$ ($B_\perp \sim 0$~T), the phase boundary between $\mathrm{FP}(1)$ and $\mathrm{FP}(6)$ (black dashed line at $n \approx -3\times 10^{11} ~\text{cm}^{-2}$) doesn't move with $B_\perp$, suggesting that the $\mathrm{FP}(1)$ at $B_\perp \sim 0$~T is characterized by coherence between the two valleys
(over the spin-valley polarized phase), so that the orbital moments cancel.
The same logic applies to the other symmetry-breaking phases  at slightly higher doping. A large FFT frequency dominates at $f_\nu >1/2$ while the phase boundaries persist in doping without moving when $B_\perp$ is applied, suggesting the existence of one large Fermi pocket with 
diminished or no orbital moments and hence inter-valley coherence for $\mathrm{FP}(1,3,1)$.

\textbf{Sample alignment with in-plane magnetic field:} In-plane-field
measurements were performed by mounting the sample vertically with a homemade
frame to access $B_{\parallel}>1$~T. A small $B_{\perp}$ component is inevitably  introduced  when 
$B_{\parallel}$ is applied due to imperfect vertical sample
alignment. 
The $B_x$ and $B_y$ directions of the vector magnet are used to compensate the $B_{\perp}$ component. 
The compensation is crucial for the measurements of large Pauli-limit violation of SC$_3$, because at $B_\parallel = 7$~T or so, an out-of-plane component $B_{\perp} \approx 0.1-0.2$~mT almost completely suppresses the superconductivity. 
Additionally, the cancellation of $B_{\perp}$ gives an accurate $R_{xx}$ dependence on $B_\parallel$ and temperature so that the extraction of $\alpha$ in \prettyref{fig:fig4} is reliable.

\textbf{Predictions of conventional depairing mechanisms for SC$_1$, SC$_2$, and SC$_3$:} Here we summarize the main findings resulting from the conventional 
depairing mechanisms in the context of  SC$_1$, SC$_2$, and SC$_3$,  
 within the Ising superconductor framework\cite{frigeriSuperconductivityInversionSymmetry2004,luEvidenceTwodimensionalIsing2015,
saitoSuperconductivityProtectedSpinvalley2016,zhangEnhancedSuperconductivitySpin2023}. For SC$_2$, we find that experimentally determined values of $\alpha$ are not 
well reproduced by realistic values of $\lambda_R \lesssim 10$ meV (see SI, section~\ref{sec: conventional}). More precisely, 
conventional depairing theory would predict higher resilience (lower 
$\alpha$ by almost an order of magnitude) to the in-plane magnetic field. For the case of 
SC$_1$, we find that the conventional depairing theory applies well, explaining the large 
PVR. 
As such, another mechanism, like the interband-interaction scenario in the main text, must account for the relatively 
minor PVR  in SC$_2$. 
The direction of $B_\parallel$ can further shed light on the depairing mechanism. Theoretically, SC$_1$ is not sensitive
to the in-plane field direction, as reported previously in Ref.~\citenum{holleisNematicityOrbitalDepairing2024}. 
For SC$_2$, because both $\mathrm{FP}(2,4)$ and 
$\mathrm{FP}(2,2,2)$  break the $C_3$ symmetry, we anticipate a dependence of $T_c$ (or 
$\alpha$) as the direction of $B_\parallel$ is rotated. Experimentally, however, SC$_2$ shows no significant variation in $T_c$ as the magnetic field is rotated (\prettyref{exfig:SC2_inplane_angle}), 
suggesting that this simple depairing picture relying on rigid occupancy of the trigonal-warping pockets should be revisited with a self-consistent Hartree-fock (HF) pairing analysis, where the Fermi surfaces adjust in response to the magnetic field. Finally, we also highlight 
that the mechanism discussed in the main text,  the modification of interband and 
intraband interactions with an in-plane magnetic field, 
is entirely independent of the field direction. This is in accordance with the experimental observations, yet contrasts with the ``conventional'' depairing analysis.

\noindent {\bf Acknowledgments:} We thank Jason Alicea, Étienne Lantagne-Hurtubise, Zhiyu Dong, Alex Thomson, Dmitry V. Chichinadze, Andrea Young, and Erez Berg for fruitful discussions.

{\bf Funding:} This work has 
been primarily supported by the Office of Naval Research (grant no. N142112635).
S.N.-P. and D.H. also acknowledge the support of the Institute for 
Quantum Information and Matter, an NSF Physics Frontiers 
Center (PHY-2317110). 
G.S. acknowledges support from the Walter Burke Institute for Theoretical Physics at Caltech, and from the Yad Hanadiv Foundation through the Rothschild fellowship.
H.M. and C.L. were supported by start-up funds from Florida State University and the National High Magnetic Field Laboratory. The National High Magnetic Field Laboratory is supported by the National Science Foundation through NSF/DMR-2128556 and the State of Florida. Y.O. and F.v.O.  acknowledge suppport by Deutsche Forschungsgemeinschaft through CRC 183 (project C02). F.v.O was further supported by Deutsche Forschungsgemeinschaft  through a joint
ANR-DFG project (TWISTGRAPH). 

\noindent {\bf Author Contribution:} Y.Z. and S.N.-P. designed the experiment.
Y.Z. fabricated the devices, performed the measurements, and
analyzed the data. Y.H. and D.H. performed the SHG measurements. G.S., H.M., C.L., F.v.O., and Y.O. developed theoretical models and
performed calculations. K.W. and T.T. provided hBN crystals.
S.N.-P. supervised the project. Y.Z., G.S., C.L., F.v.O., Y.O, and S.N.-P. wrote the manuscript with the input of other authors.
 
\noindent{\bf Competing interests:} The authors declare no competing interests.

\noindent {\bf Data availability:} The data supporting the findings of this
study are available from the corresponding authors on reasonable request.
 
\noindent {\bf Code availability:} All code used in modeling in this study is
available from the corresponding authors on reasonable request.
 
\clearpage

\clearpage
\beginsupplement

\clearpage
\clearpage

\clearpage

\clearpage

\clearpage
\clearpage
\clearpage

\begin{figure}[p]
    \includegraphics[width=16cm]{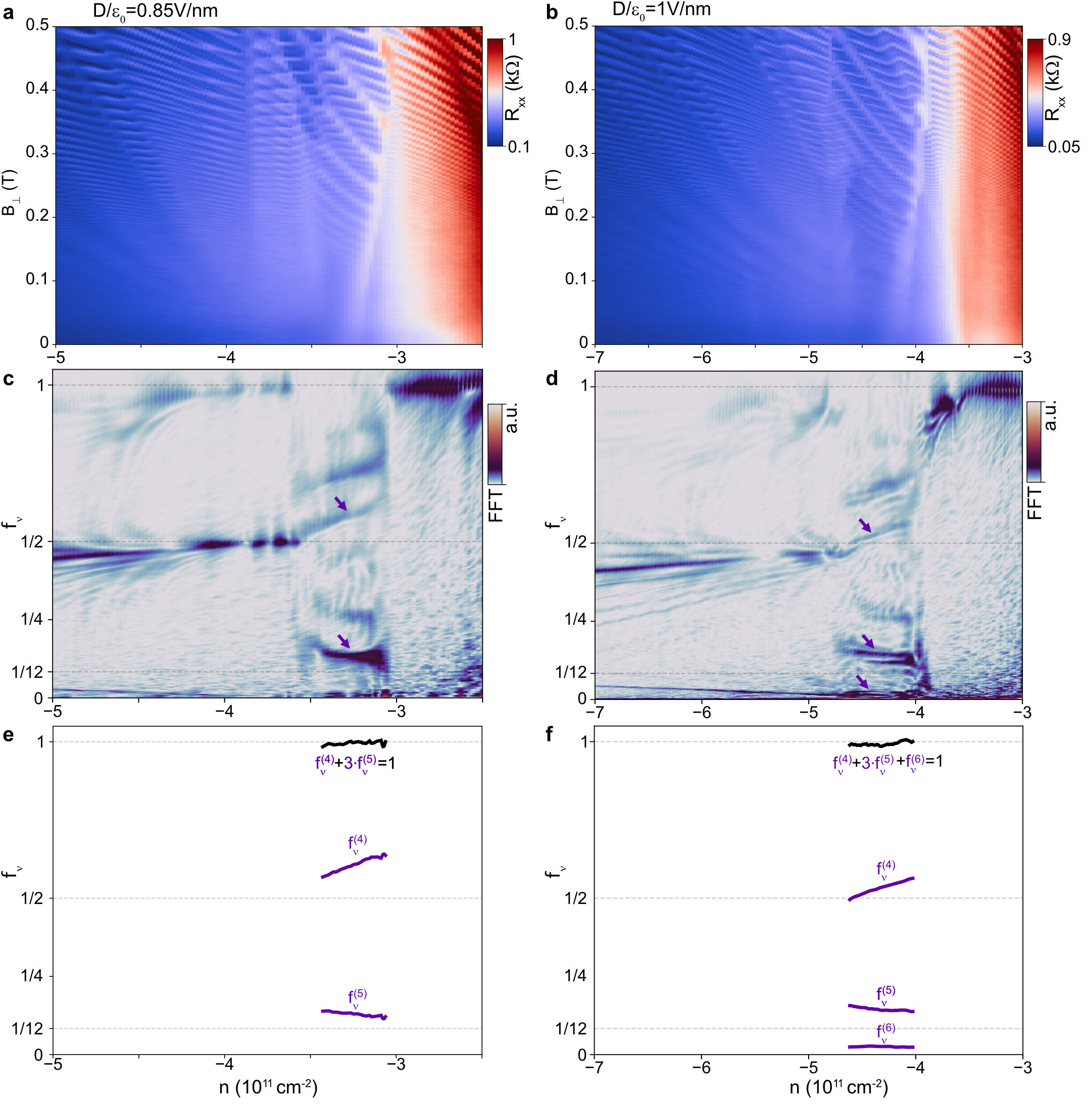}
    \centering
    \caption{{\bf $\mathbf{FP(1,3)}$ and $\mathbf{FP(1,3,1)}$ at $D/\epsilon_0 = 0.85~\text{V/nm}$ and $1~\text{V/nm}$, respectively.} {\bf a},{\bf b}, $R_{xx}$ versus out-of-plane magnetic field $B_\perp$ and doping density $n$ measured at $D/\epsilon_0 = 0.85~\text{V/nm}$ ({\bf a}) and $1~\text{V/nm}$ ({\bf b}), respectively. {\bf c},{\bf d}, Frequency-normalized Fourier transform of $R_{xx}(1/B_\perp)$ at $D/\epsilon_0 = 0.85~\text{V/nm}$ ({\bf c}) and $1~\text{V/nm}$ ({\bf d}), respectively. {\bf e},{\bf f}, Intensity peaks in $f_\nu$ extracted from the FFT data in {\bf c} and {\bf d}.}

\label{exfig:QO_FP(1,3,1)_1}
\end{figure}
\clearpage
\begin{figure}[p]
    \includegraphics[width=16cm]{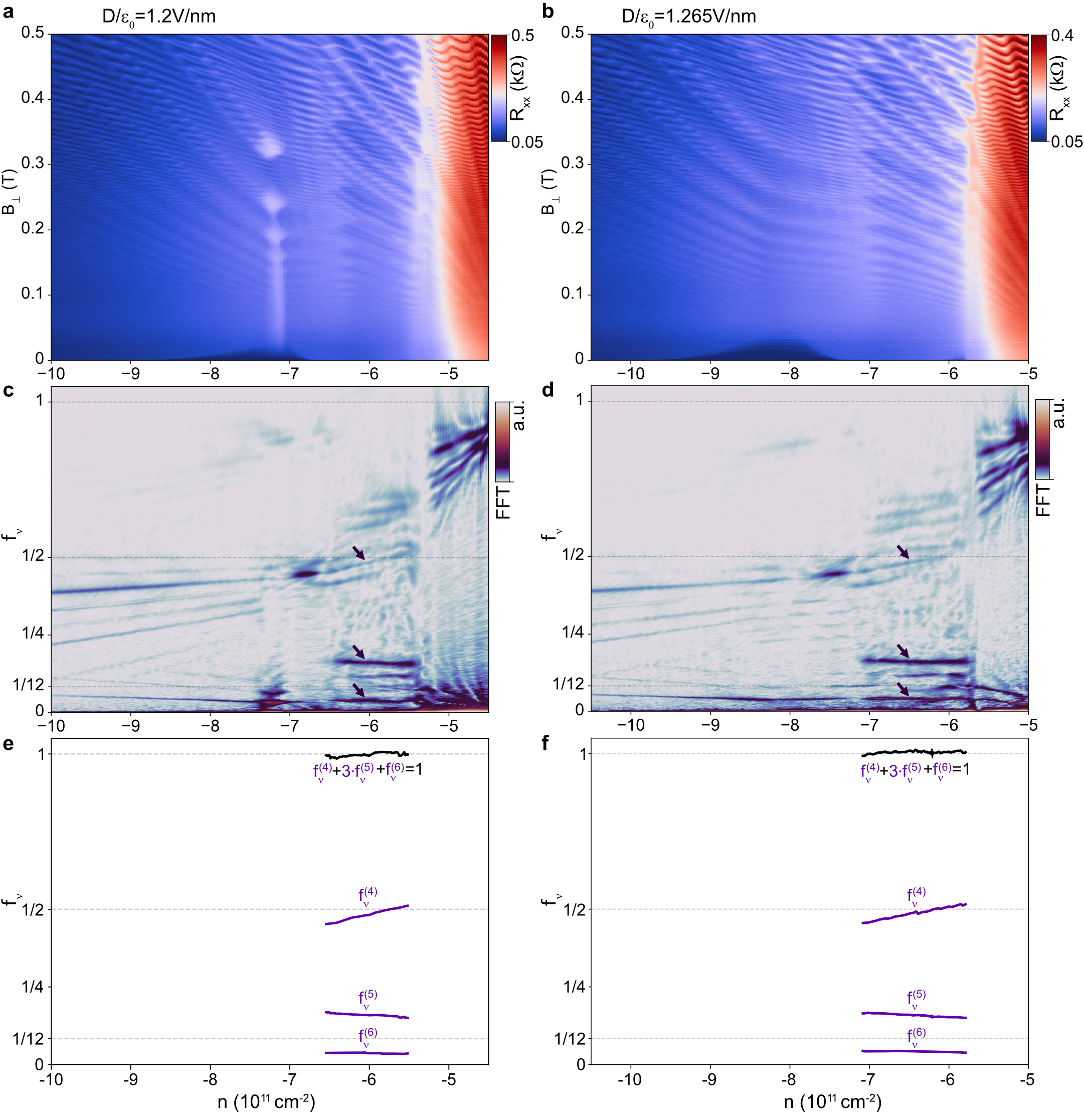}
    \centering
    \caption{{\bf $\mathbf{FP(1,3,1)}$ at $D/\epsilon_0 = 1.2~\text{V/nm}$ and $1.265~\text{V/nm}$.} {\bf a},{\bf b}, $R_{xx}$ versus out-of-plane magnetic field $B_\perp$ and doping density $n$ measured at $D/\epsilon_0 = 1.2~\text{V/nm}$ ({\bf a}) and $1.265~\text{V/nm}$ ({\bf b}), respectively. {\bf c},{\bf d}, Frequency-normalized Fourier transform of $R_{xx}(1/B_\perp)$ at $D/\epsilon_0 = 1.2~\text{V/nm}$ ({\bf c}) and $1.265~\text{V/nm}$ ({\bf d}), respectively. {\bf e},{\bf f}, Intensity peaks in $f_\nu$ extracted from the FFT data in {\bf c} and {\bf d}.}

\label{exfig:QO_FP(1,3,1)_2}
\end{figure}
\clearpage

\begin{figure}[p]
    \includegraphics[width=16cm]{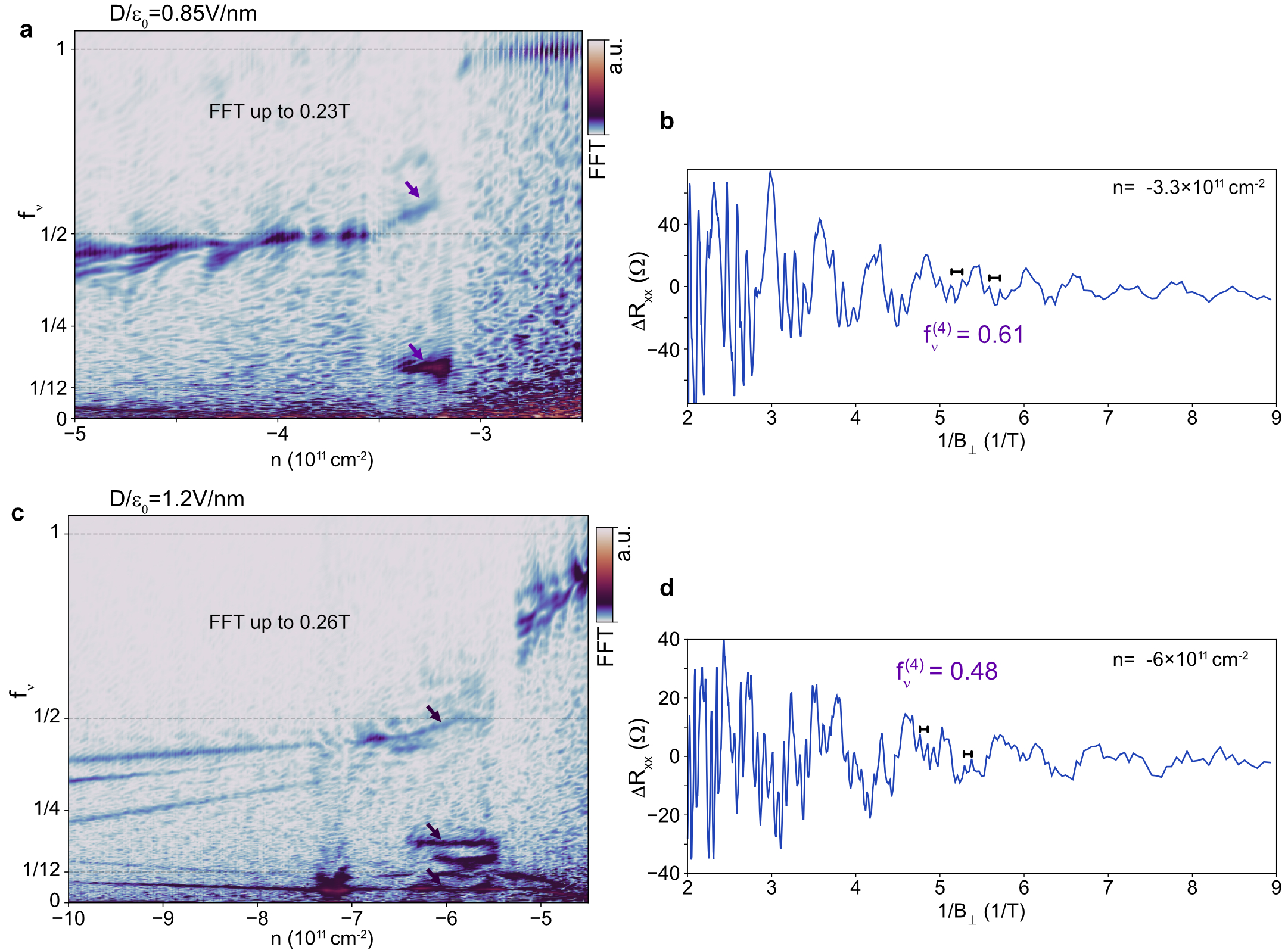}
    \centering
    \caption{{\bf FFT of $\mathbf{FP(1,3)}$ and $\mathbf{FP(1,3,1)}$ with data at lower magnetic field.} {\bf a},{\bf c}, Frequency-normalized Fourier transform of $R_{xx}(1/B_\perp)$ at $D/\epsilon_0 = 0.85~\text{V/nm}$ ({\bf a}) and $1.2~\text{V/nm}$ ({\bf c}), respectively. The $R_{xx}$ data are used up to 0.23~T and 0.26~T respectively. {\bf b},{\bf d}, $R_{xx}$ variation $\Delta R_{xx}$ as a function $1/B_{\perp}$  measured at $n = -3.3\times 10^{11} ~\text{cm}^{-2}$, $D/\epsilon_0 = 0.85~\text{V/nm}$ ({\bf b}) and  $n = -6\times 10^{11} ~\text{cm}^{-2}$, $D/\epsilon_0 = 1.2~\text{V/nm}$ ({\bf d}), respectively.}

\label{exfig:QO_FP(1,3,1)_lowfield}
\end{figure}
\clearpage
\begin{figure}[p]
    \includegraphics[width=14cm]{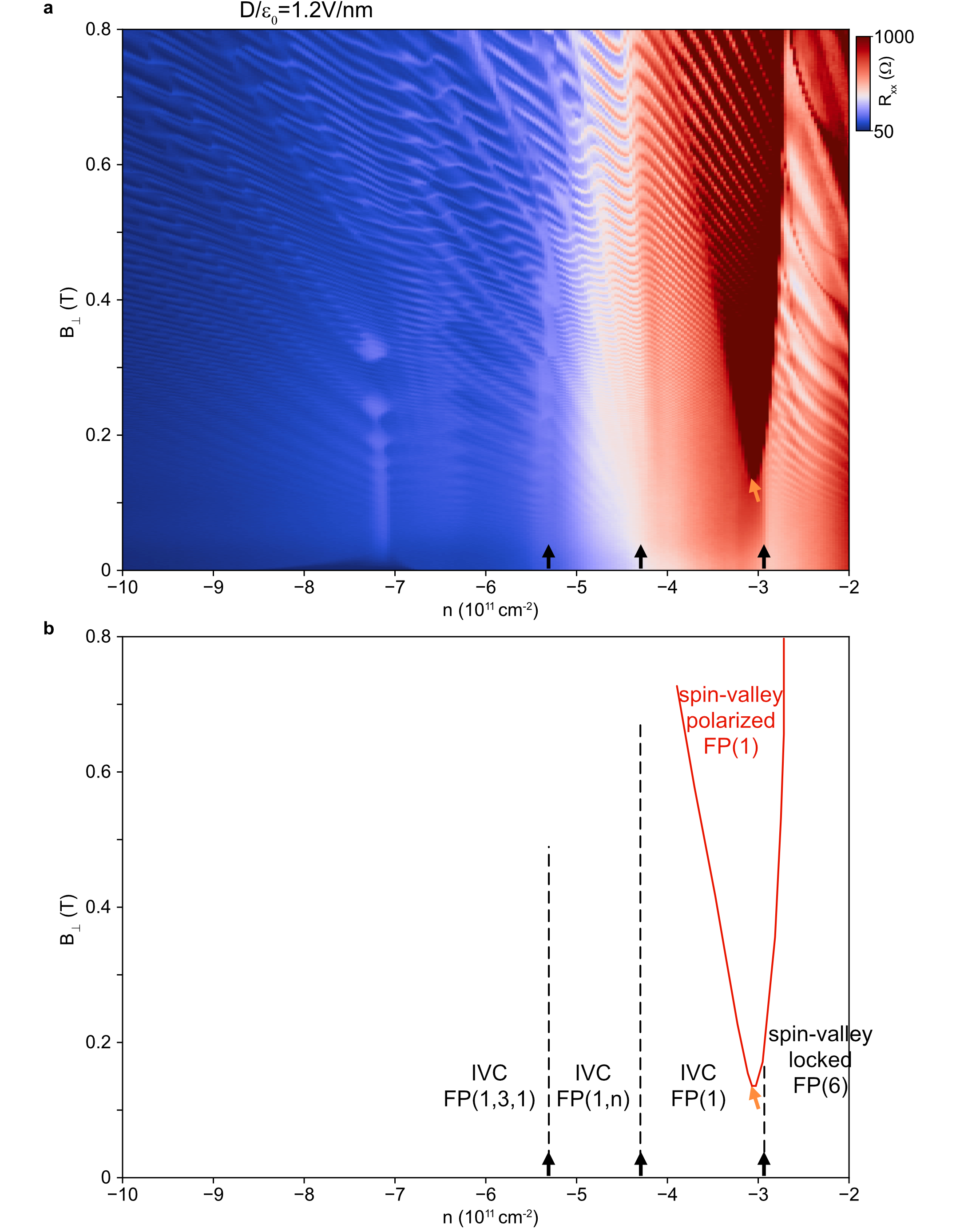}
    \centering
    \caption{
    {\bf Evolution of phase boundaries as a function of $B_\perp$.} {\bf a}, $R_{xx}$ versus out-of-plane magnetic field $B_\perp$ and doping density $n$ measured at $D/\epsilon_0 = 1.2~\text{V/nm}$ for a device with $|\lambda_I| \approx 1.5$~meV. Phase boundaries are marked out in {\bf b}.  The black arrows and dashed lines mark the phase boundaries that are not sensitive to $B_\perp$, suggestive of inter-valley coherence with little or no net orbital moments. The red line draws the phase boundary of the spin-valley polarized $\mathrm{FP}(1)$; the boundary grows (orange arrow) with $B_\perp$ due to large orbital moments.
    }
\label{exfig:phase boundary moving}
\end{figure}
\clearpage

\begin{figure}[p]
    \includegraphics[width=16cm]{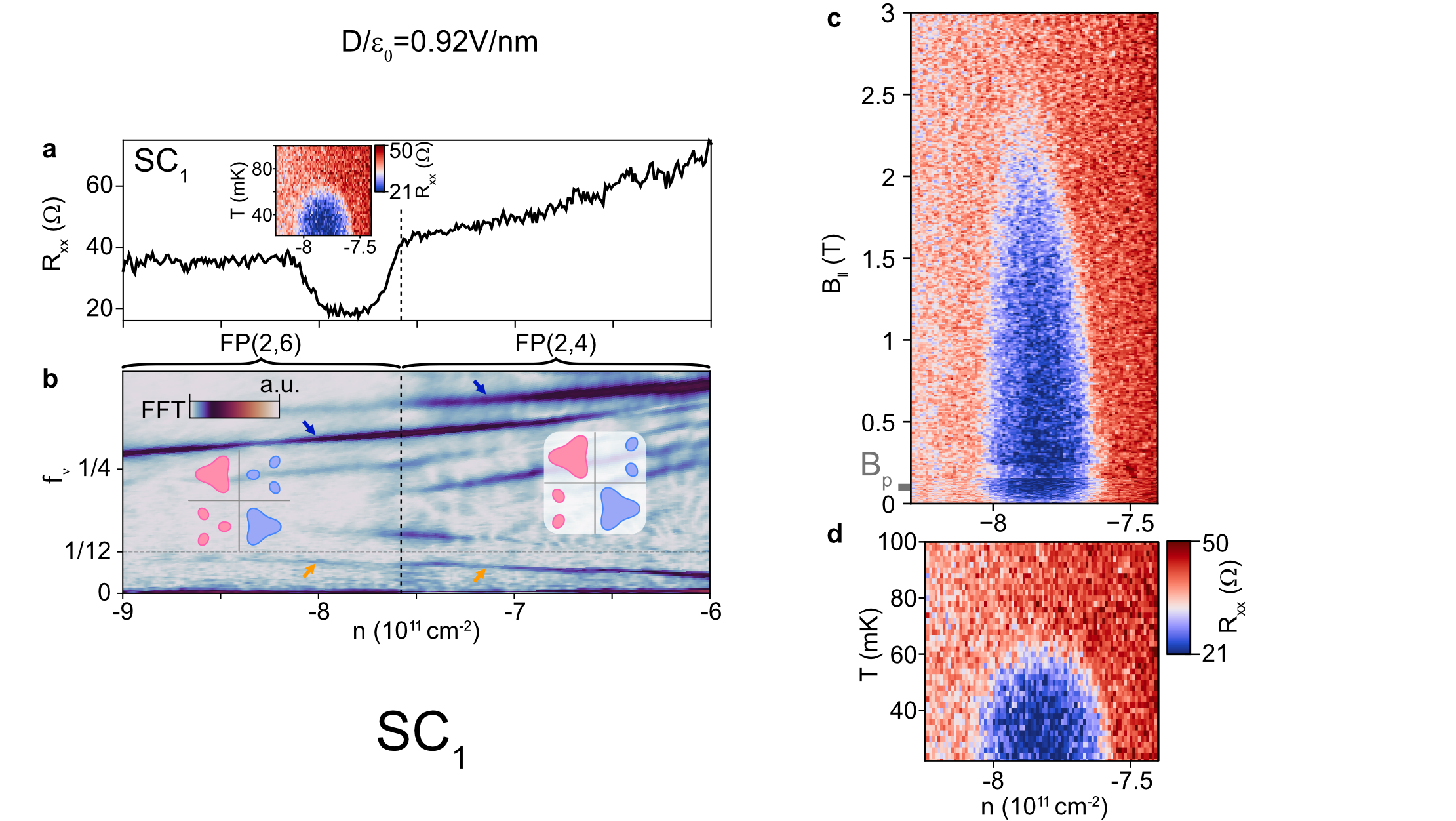}
    \centering
    \caption{{\bf $B_\parallel$ dependence of SC$_1$.}  
    {\bf a}, $R_{xx}$ versus $n$ measured at $D/\epsilon_0 = 0.92~\text{V/nm}$. Inset shows $R_{xx}$ versus $n$ and temperature for the superconducting dome SC$_1$. {\bf b}, Frequency-normalized FFT of $R_{xx}(1/B_\perp)$ over the same doping range as in {\bf a}.
    {\bf c},{\bf d}, $n$-dependent $R_{xx}$ versus in-plane magnetic field ({\bf c}) or versus temperature ({\bf d}), showing the disappearance of SC$_1$. The grey bar marks the Pauli limit $B_p$. 
    } 

\label{exfig:SC1_inplane_field}
\end{figure}
\clearpage

\begin{figure}[p]
    \includegraphics[width=16cm]{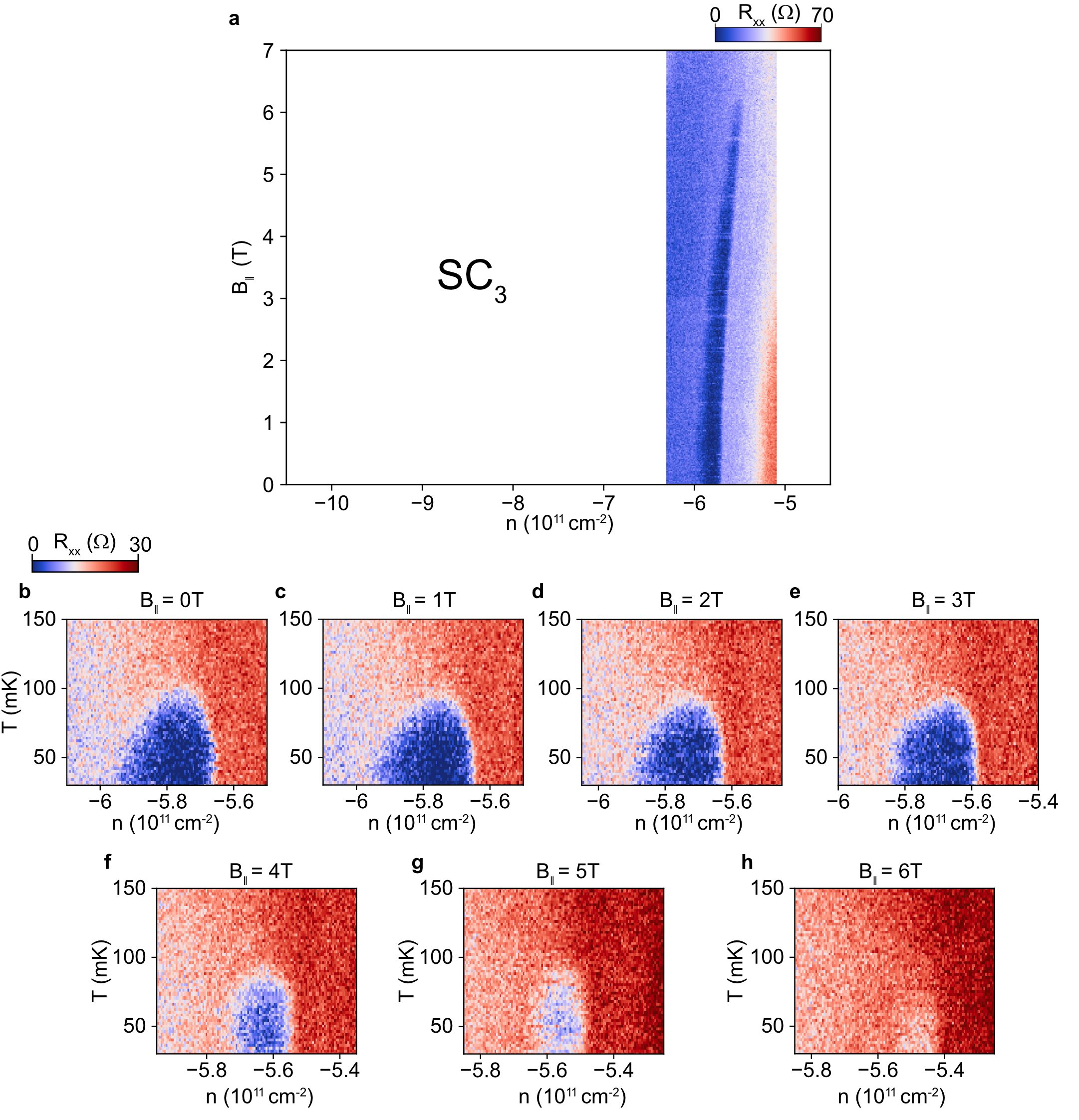}
    \centering
    \caption{{\bf $B_\parallel$ dependence of SC$_3$.}  {\bf a}, $R_{xx}$ versus doping density $n$ and in-plane magnetic field $B_{\parallel}$ showing the evolution of SC$_3$. {\bf b}-{\bf h}, $R_{xx}$ versus doping density $n$ and temperature measured from $B_\parallel = 0$~T ({\bf b}) to $6$~T ({\bf h}), $1$~T increment step.} 

\label{exfig:Inplane_T}
\end{figure}
\clearpage

\begin{figure}[p]
    \includegraphics[width=16cm]{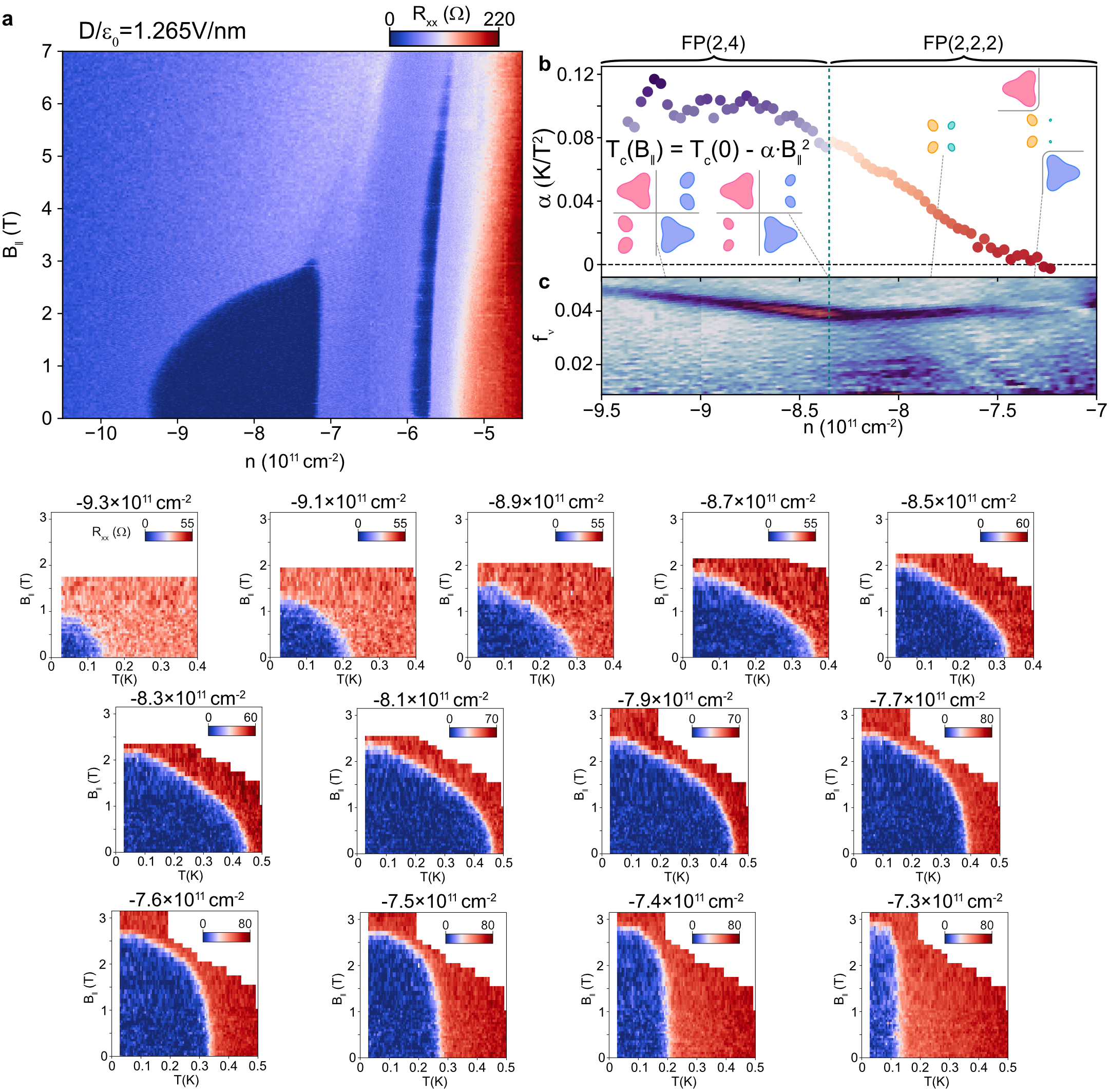}
    \centering
    \caption{{\bf $B_\parallel$ dependence of SC$_2$ at $D/\epsilon_0 = 1.265~\text{V/nm}$.}
    {\bf a}, $R_{xx}$ versus doping and $B_\parallel$ focusing around SC$_2$ at $D/\epsilon_0 = 1.265~\text{V/nm}$. {\bf b}, Fitting coefficient $\alpha$ versus doping density $n$ for SC$_2$ at the same $D$. {\bf c}, Frequency-normalized Fourier transform of $R_{xx}(1/B_\perp)$ over the same doping range as in {\bf b}, focusing around low frequencies representing the two types of  trigonal-warping pockets. Bottom panels show $R_{xx}$ versus temperature and $B_\parallel$ at different doping for $D/\epsilon_0 = 1.265~\text{V/nm}$.
    } 

\label{exfig:Inplane_SC2_D1p265}
\end{figure}
\clearpage
\begin{figure}[p]
    \includegraphics[width=16cm]{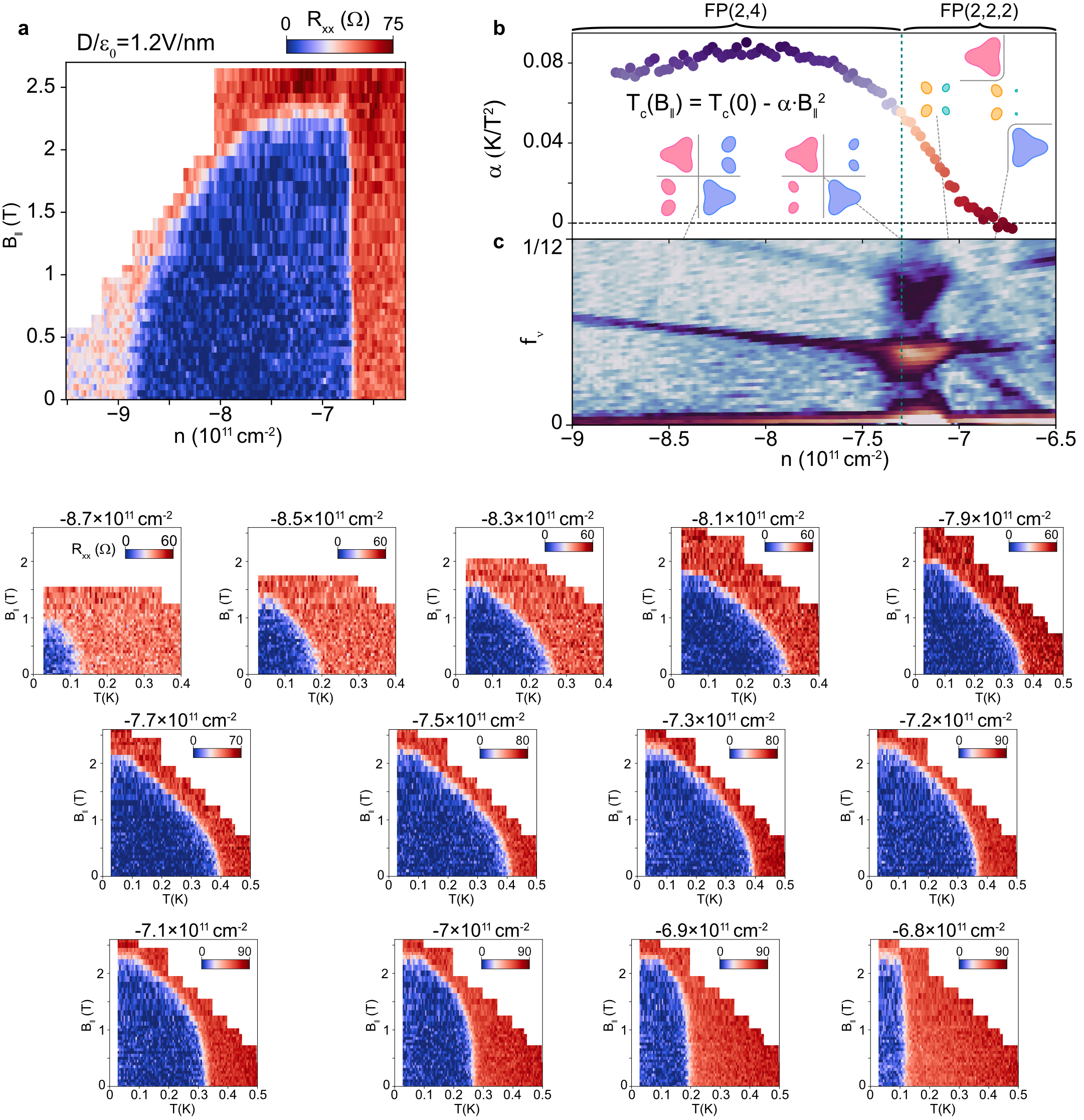}
    \centering
    \caption{{\bf $B_\parallel$ dependence of SC$_2$ at $D/\epsilon_0 = 1.2~\text{V/nm}$.}
    {\bf a}, $R_{xx}$ versus doping and $B_\parallel$ focusing around SC$_2$ at $D/\epsilon_0 = 1.2~\text{V/nm}$. {\bf b}, Fitting coefficient $\alpha$ versus doping density $n$ for SC$_2$ at the same $D$. {\bf c}, Frequency-normalized Fourier transform of $R_{xx}(1/B_\perp)$ over the same doping range as in {\bf b}, focusing around low frequencies representing the two types of  trigonal-warping pockets. Bottom panels show $R_{xx}$ versus temperature and $B_\parallel$ at different doping for $D/\epsilon_0 = 1.2~\text{V/nm}$.} 

\label{exfig:Inplane_SC2_D1p2}
\end{figure}
\clearpage
\begin{figure}[p]
    \includegraphics[width=14cm]{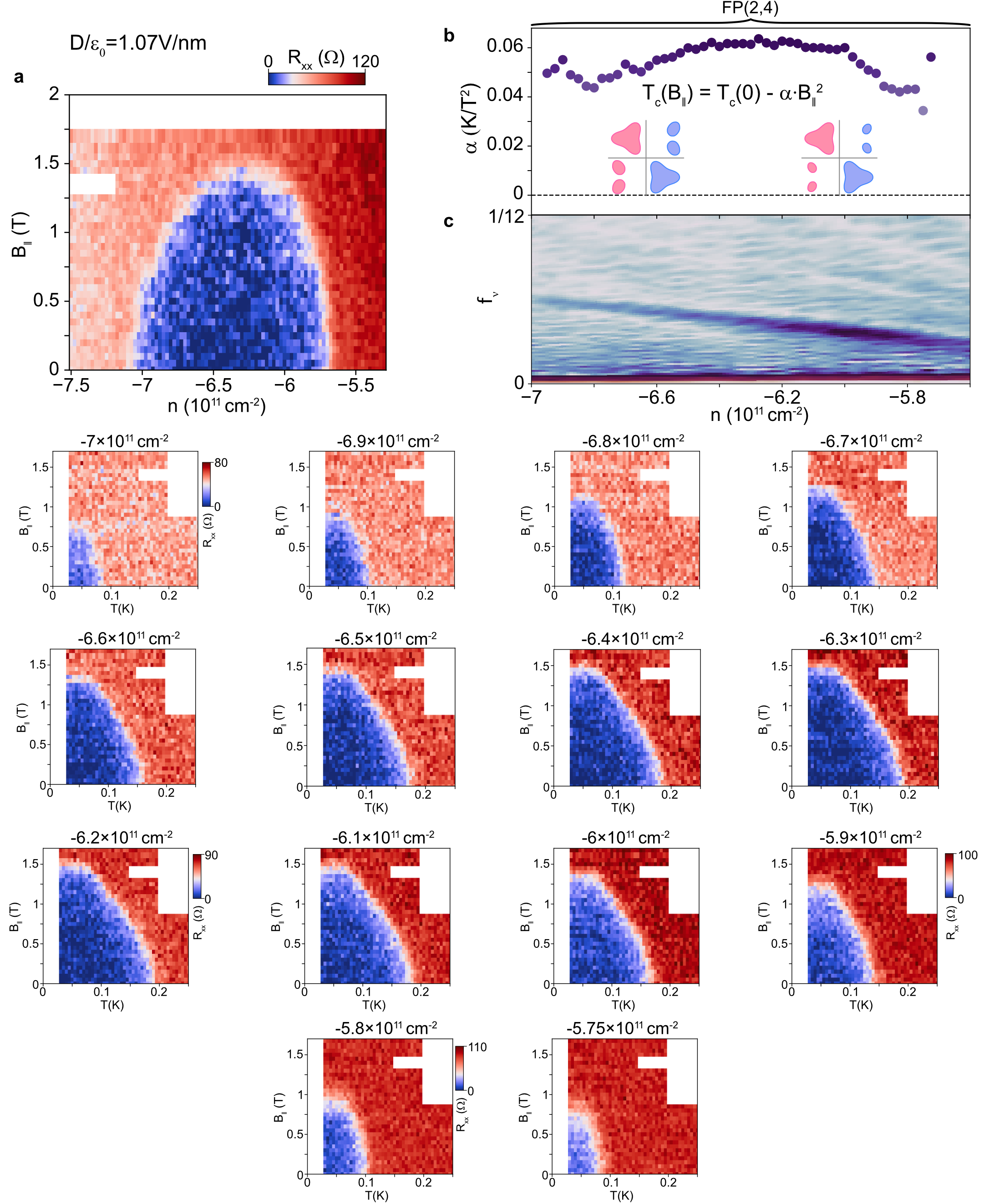}
    \centering
    \caption{{\bf $B_\parallel$ dependence of SC$_2$ at $D/\epsilon_0 = 1.07~\text{V/nm}$.}
    {\bf a}, $R_{xx}$ versus doping and $B_\parallel$ focusing around SC$_2$ at $D/\epsilon_0 = 1.07~\text{V/nm}$. {\bf b}, Fitting coefficient $\alpha$ versus doping density $n$ for SC$_2$ at the same $D$. {\bf c}, Frequency-normalized Fourier transform of $R_{xx}(1/B_\perp)$ over the same doping range as in {\bf b}, focusing around low frequencies representing the single type of trigonal-warping pockets without nematic redistribution of holes. Bottom panels show $R_{xx}$ versus temperature and $B_\parallel$ at different doping at $D/\epsilon_0 = 1.07~\text{V/nm}$. At this $D$ field, SC$_2$ doesn't onset from $\mathrm{FP}(2,2,2)$. The rapidly changing $\alpha$ with diminished values are accordingly absent.} 

\label{exfig:Inplane_SC2_D1p07}
\end{figure}
\clearpage

\begin{figure}[p]
    \includegraphics[width=15cm]{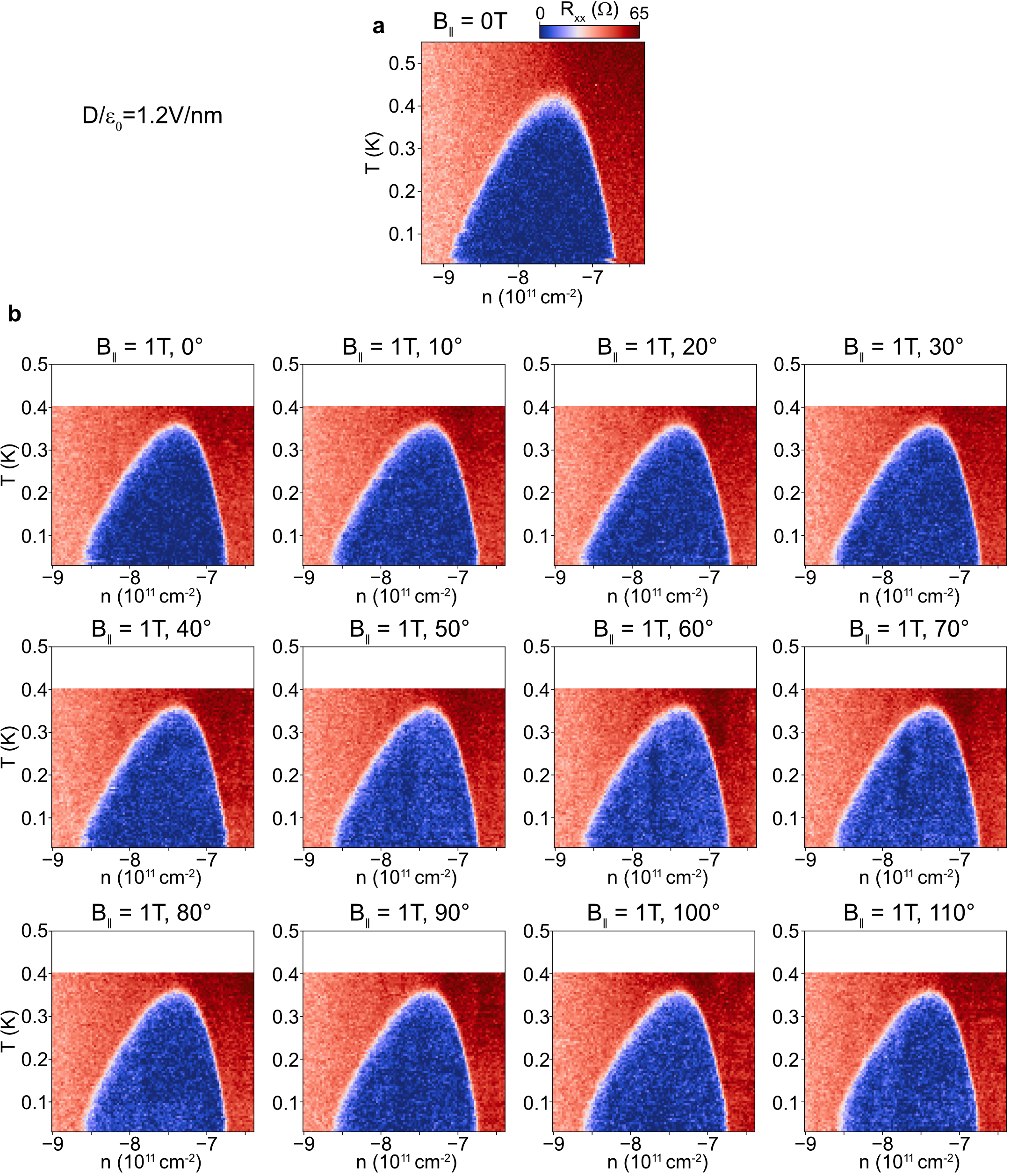}
    \centering
    \caption{{\bf Isotropic $B_\parallel$ dependence of SC$_2$.}
    {\bf a}, $R_{xx}$ versus $n$ and temperature showing SC$_2$ at $D/\epsilon_0 = 1.2~\text{V/nm}$. {\bf b}, $R_{xx}$ versus $n$ and temperature showing SC$_2$, measured with an in-plane magnetic field $B_\parallel = 1$~T, at different in-plane angles. The depairing of SC$_2$ is isotropic along different $B_\parallel$ directions. } 

\label{exfig:SC2_inplane_angle}
\end{figure}
\clearpage

{\large {\bf Supplementary Information:}} \\ 
{\large Twist-Programmable Superconductivity in Spin-Orbit
Coupled Bilayer Graphene} \\
Yiran Zhang, Gal Shavit, Huiyang Ma, Youngjoon Han, Kenji Watanabe, Takashi Taniguchi, David Hsieh, Cyprian Lewandowski, Felix von Oppen, Yuval Oreg, and Stevan
Nadj-Perge

\setcounter{table}{0}
\setcounter{figure}{0}
\renewcommand{\figurename}{Supplementary Information Fig.}
\renewcommand{\tablename}{Supplementary Information Table}

\section{Importance of interband superconductivity in BLG}
\label{sec:Importance_Interband}
Much of the experimental observations in BLG proximitized by WSe$_2$ may be understood in the context of multiband superconductivity.
Here, the bands correspond to the majority and minority bands are split by the Ising spin-orbit coupling (ISOC), possibly further enhanced by electron-electron interactions.
It is the coupling between these two sectors (majority and minority bands) in the Cooper channel, i.e., pair scattering between the bands, that is crucial to interpreting the experimental results.

 The delayed onset (in terms of higher displacement fields) of superconductivity in the main dome when the spin-orbit coupling is increased is elucidated in Sec.~\ref{sec:SConsetISOC}.
When the density of states (DOS) of both bands is high near the Fermi level, screening of the bare Coulomb repulsion becomes much more efficient, enabling superconductivity mediated by some retarded attraction.
As the bands are further split by the ISOC, \textit{the band-concurrent high DOS region is pushed to higher displacement fields}.
This is illustrated in SI Fig.~\ref{fig:onsetFigure}.

The large variance in the behavior of the different superconducting region when an in-plane magnetic field is introduced, as manifested in varying Pauli-limit-violation ratios, is discussed in Sec.~\ref{sec:ZeemanMultibandEffect}.
We consider the effect of a Zeeman term in \textit{gradually decoupling the majority and minor bands in the Cooper channel}.
Analysing the energetics of spin magnetization in the presence of both ISOC and Hund's coupling, we illustrate how the main superconducting dome, where the imbalance between majority and minority carriers is the greatest, becomes most vulnerable to this effect.
We further demonstrate the experimental trends are well understood within our modeling.
We stress that the Zeeman coupling is not assumed to have any pair-breaking role, the effect we consider is purely due to  ``shifting'' interaction strength from the intraband part of the coupling matrix to the interband part. 
In Sec.~\ref{sec: conventional}, we consider a more direct depairing model and contrast with the results of the proposed mechanism here.

Finally, we note that in our analysis we do not specify or rely on what is the pairing glue which mediates superconductivity.
We remain agnostic with regards to its origin, as well as its intraband/interband nature, which do not modify our conclusions.

\section{Variational Hartree-Fock}\label{sec:varitionalHF}

We begin by calculating the the non-interacting band structure of biased Bernal-stacked bilayer graphene. 
Expanded around the valley $K/K'$ points in momentum space, the Hamiltonian is~\cite{mccannElectronicPropertiesBilayer2013} 
\begin{equation}
H_{BLG}=\sum_{\mathbf{k},\tau,s}c_{\tau s\mathbf{k}}^{\dagger}h_{\tau}\left(\mathbf{k}\right)c_{\tau s\mathbf{k}},\label{eq:BLGhAMILTONIAN}
\end{equation}
with $c_{\tau s\mathbf{k}}=\left(A_{1,\tau s\mathbf{k}},B_{1,\tau s\mathbf{k}},A_{2,\tau s\mathbf{k}},B_{2,\tau s\mathbf{k}}\right)^{T},$
where $X_{i,\tau s\mathbf{k}}$ annihilates an electron on sub-lattice
$X$ in layer $i$, with spin $s$, and momentum $\mathbf{k}$ near
the valley $\tau$. The matrix $h_{\tau}$ is given by
\begin{equation}
h_{\tau}\left(\mathbf{k}\right)=\begin{pmatrix}\frac{U}{2} & v_{0}\Pi_{\tau}^{*} & -v_{4}\Pi_{\tau}^{*} & -v_{3}\Pi_{\tau}\\
v_{0}\Pi_{\tau} & \frac{U}{2}+\Delta' & \gamma_{1} & -v_{4}\Pi_{\tau}^{*}\\
-v_{4}\Pi_{\tau} & \gamma_{1} & -\frac{U}{2}+\Delta' & v_{0}\Pi_{\tau}^{*}\\
-v_{3}\Pi_{\tau}^{*} & -v_{4}\Pi_{\tau} & v_{0}\Pi_{\tau} & -\frac{U}{2}
\end{pmatrix},\label{eq:H04by4basis}
\end{equation}
with $\Pi_{\tau}=\tau k_{x}+ik_{y}$, and the parameters $v_{i}=\frac{\sqrt{3}}{2}a\gamma_{i},$
$a=0.246$ nm, $\gamma_{0}=2.61$ eV, $\gamma_{1}=361$ meV, $\gamma_{3}=283$
meV, $\gamma_{4}=138$ meV, and $\Delta'=15$ meV\cite{jungAccurateTightbindingModels2014}.
The interlayer potential difference $U$ is approximately $U\approx-dD/\epsilon$,
where the interlayer distance is $d\approx0.33$nm, $\epsilon\approx4.3$,
and $D$ is the displacement field. 
We diagonalize $H_{BLG}$ at each momentum, and extract the dispersion relation of the lowest-lying valence band, which we denote by $\epsilon_{\tau,\mathbf{k}}$.

Our analysis proceeds with a phenomenological description of the underlying physics of electrons in this band in the spirit of Ref.~\citenum{shavitInducingSuperconductivityBilayer2023}, described by the 4-spinor $\Psi_{\mathbf k}$ of fermionic annihilation operators at momentum $\mathbf{k}$, with valley and spin degrees of freedom, described by Pauli matrices $\tau_{i}$ and $s_{i}$, respectively. 
We analyze the Hamiltonian,
\begin{equation}
H=H_{0}+H_{{\rm ISOC}}+H_{{\rm int}},\label{eq:normalHamiltonian}
\end{equation}
\begin{equation}
H_{0}=\sum_{\mathbf{k}}\Psi_{\mathbf{k}}^{\dagger}\left(
\epsilon_{+,\mathbf{k}}\frac{1+\tau_z}{2}
+
\epsilon_{-,\mathbf{k}}\frac{1-\tau_z}{2}
\right)\Psi_{\mathbf{k}},\label{eq:normalH0}
\end{equation}
\begin{equation}
H_{{\rm ISOC}}=\sum_{\mathbf{k}}\Psi_{\mathbf{k}}^{\dagger}\lambda_{0}\tau_{z}s_{z}\Psi_{\mathbf{k}},\label{eq:normalHisoc}
\end{equation}
\begin{equation}
H_{{\rm int}}=\frac{1}{\Omega}\sum_{\mathbf{q}}\left(\frac{U_{C}}{2}N_{\mathbf{q}}N_{-{\mathbf{q}}}+U_{V}n_{\mathbf{q}}^{+}n_{\mathbf{-q}}^{-}+J\mathbf{S}_{{\mathbf{q}}}^{+}\cdot\mathbf{S}_{-{\mathbf{q}}}^{-}\right),\label{eq:normalHint}
\end{equation}
where $N_{\mathbf{q}}=\sum_{\mathbf{k}}\Psi_{\mathbf{k+q}}^{\dagger}\Psi_{\mathbf{k}}$,
$n_{\mathbf{q}}^{\pm}=\sum_{\mathbf{k}}\Psi_{\mathbf{k+q}}^{\dagger}\frac{1\pm\tau_{z}}{2}\Psi_{\mathbf{k}}$,
$\mathbf{S}_{{\mathbf{q}}}^{\pm}=\sum_{\mathbf{k}}\Psi_{\mathbf{k+q}}^{\dagger}\frac{1\pm\tau_{z}}{2}\mathbf{s}\Psi_{\mathbf{k}}$,
and $\Omega$ is the system area. 
The term proportional to $\lambda_0$ is the bare Ising-type spin-orbit coupling induced in the valence band (at positive values of the electric displacement field) by the proximity to ${\rm WSe}_2$.
The structure of $H_{{\rm int}}$ is the most general form of short-range
interactions which respect the symmetry of the system: time-reversal,
$SU\!\left(2\right)$ spin symmetry (in the absence of magnetic fields
or spin-orbit coupling), and the $U\!\left(1\right)$ charge and (approximate)
valley symmetries. The interaction term proportional to $U_{C}$ is
a structure-less density-density interaction, which is entirely $SU\!\left(4\right)$
symmetric in valley-spin space, and is considered to be dominant as
compared to the other two terms. The term proportional to $U_{V}$
accounts for possible differences between intravalley and intervalley
density-density interactions and will be set to zero throughout this
work, as it is non-essential for correctly capturing the phenomenology
we aim to study. Finally, $J$ is the intervalley Hund's coupling
between electron spins in opposite valleys.

Although the Ising spin-orbit coupling term $H_{{\rm ISOC}}$ explicitly
breaks the flavor symmetry of the system, i.e., not all spin-valley
flavors are populated equally even in the absence of interactions,
the presence of non-negligible interactions in $H_{{\rm int}}$ may
significantly alter the non-interacting picture. We employ a variational
Hartree-Fock procedure to resolve the different flavor-resolved fillings.
Our analysis thus proceeds as follows. At a given chemical potential
$\mu$, the grand-potential $\Phi=\left\langle H-\mu N_{0}\right\rangle _{{\rm H.F.}}$
is minimized, where $\left\langle \right\rangle _{{\rm H.F.}}$ denotes
the expectation value calculated using the variational wavefunction,

\begin{equation}
|\Psi\rangle_{{\rm H.F.}}=\prod_{\tau,s}\left(\prod_{\substack{\mathbf{k}\\
\epsilon_{\mathbf{k}}>\mu_{\tau,s}
}
}\sum_{s'}{\cal U}\left(\tau\theta\right)_{ss'}\psi_{\mathbf{k},\tau,s'}\right)|{\rm CN}\rangle,\label{eq:variationalPsi}
\end{equation}
where $\psi_{\tau,s,\mathbf{k}}$ annihilates an electron in valley $\tau$ and spin $s$ in the valence band at momentum $\mathbf{k}$ with energy $\epsilon_{\mathbf{k}}$,
$|{\rm CN}\rangle$ is the flavor-symmetric charge-neutral Fermi-sea,
and $\mu_{\tau,s}\leq0$ are the four variational parameters corresponding
to the four spin-valley flavors.
The possibility of canting in the different valleys is captured by the matrix ${\cal U}\left(\theta\right)=\begin{pmatrix}\cos\frac{\theta}{2} & \sin\frac{\theta}{2}\\
-\sin\frac{\theta}{2} & \cos\frac{\theta}{2}
\end{pmatrix}$.
Notice that full canting, i.e., alignment of the majority and minority bands according to their spin polarization corresponds to $\theta=\pi/2$, whereas spin-valley locking corresponds to $\theta=0$ (no canting).

Obtaining the different $\mu_{i}$,
we calculate the flavor resolved densities $\nu_{i}=-\frac{1}{\Omega}\sum_{\substack{\mathbf{k}\\
\mu_{i}<\epsilon_{\mathbf{k}}<0 
}
}$, i.e., (minus) the number of $\bf k$ points in the valence band whose energies are larger than $\mu_i$.
The total density is thus $n=\sum_{i}\nu_{i}$. 
The resultant Hartree-Fock Hamiltonian (up to additional constant contributions) can thus be written as 
\begin{equation}
H_{{\rm HF}}=\sum_{\tau s\mathbf{k}}\psi_{\tau s\mathbf{k}}^{\dagger}\left(\epsilon_{\tau\mathbf{k}}-\mu_{\tau s}\right)\psi_{\tau s\mathbf{k}}.\label{eq:Hartree-FockHamiltonian}
\end{equation}

\section{Multiband superconductivity framework}\label{sec:multibandSC}
The starting point we consider for analyzing superconductivity in this work is an electronic system with two bands, with each band being two-fold degenerate (in the present case due to spin).
We are interested in Cooper-channel interactions, and restrict ourselves to a superconducting gap with trivial symmetry.
In that case, the usual mean-field ansatz for the order parameter assumes a vector form due to the multiband nature of the problem,
\begin{equation}
\boldsymbol{\Delta}=\begin{pmatrix}\Delta_{1}\\
\Delta_{2}
\end{pmatrix}=\begin{pmatrix}g & g_x\\
g_x & g
\end{pmatrix}\left\langle \begin{pmatrix}\frac{1}{\Omega}\sum_{\mathbf{\mathbf{k}}}\psi_{-\mathbf{k}1+}\psi_{\mathbf{k}1-}\\
\frac{1}{\Omega}\sum_{\mathbf{\mathbf{k}}}\psi_{-\mathbf{k}2+}\psi_{\mathbf{k}2-}
\end{pmatrix}\right\rangle \equiv\hat{g}_{{\rm initial}}\left\langle \begin{pmatrix}\frac{1}{\Omega}\sum_{\mathbf{\mathbf{k}}}\psi_{-\mathbf{k}1+}\psi_{\mathbf{k}1-}\\
\frac{1}{\Omega}\sum_{\mathbf{\mathbf{k}}}\psi_{-\mathbf{k}2+}\psi_{\mathbf{k}2-}
\end{pmatrix}\right\rangle ,
\end{equation}
where $\Omega$ is the system volume, $\psi_{{\bf k}, n, p}$ annihilates an electron with momentum $\bf k$, in band $n=1,2$, and internal flavor $p=\pm$.
We assume $\hat{g}_{{\rm initial}}$ is a matrix of repulsive interactions, originating from the intrinsic (presumably Thomas-Fermi screened) Coulomb interaction.
For reasons which are clarified below, we take the intraband repulsion $g$ to be the same in both bands, though a generalization is straightforward within our analysis.

Given the form of $\boldsymbol{\Delta}$, the mean-field BCS Hamiltonian takes the form
\begin{align}
H_{{\rm BCS}} & =\sum_{\mathbf{k},n,p}\xi_{\mathbf{k}np}\psi_{\mathbf{k}np}^{\dagger}\psi_{\mathbf{k}np}
+\sum_{\mathbf{\mathbf{k}}}\left(\Delta_{1}^{*}\psi_{-\mathbf{k}1+}\psi_{\mathbf{k}1-}+\Delta_{2}^{*}\psi_{-\mathbf{k}2+}\psi_{\mathbf{k}2-}+{\rm h.c.}\right)\nonumber \\
 & +\Omega\boldsymbol{\Delta}^{\dagger}\left(\hat{g}_{{\rm initial}}\right)^{-1}\boldsymbol{\Delta}.\label{eq:generalizedBCSHamiltonian},
\end{align}
and $\xi_{\mathbf{k}np}$ represents the normal-state energy of the $n,p$ band relative to the Fermi energy.

Our analysis will generally consist of three steps, detailed below.

\subsection{Tolmachev-Anderson-Morel step}
We introduce an energy scale $\omega^*$, below which some retarded interaction is activated.
Clearly, this approach, following Tolmachev\cite{tolmachevLogarithmicCriterionSuperconductivity1962}, and Morel and Anderson\cite{morelCalculationSuperconductingState1962}, is another significant simplification of what should generally be an energy-dependent interaction in the Cooper channel, yet it is sufficient to illustrate all our key results.
In the path-integral formalism, we integrate out the electrons with energy greater than $\omega^*$~\cite{nagaosaQuantumFieldTheory1999}, obtaining the renormalized interaction matrix,
\begin{equation}
    \hat{g}_{\rm TAM}
    \left(\omega^*\right)
    =\left[\left(\hat{g}_{{\rm initial}}\right)^{-1}
    +\begin{pmatrix}\ell_{1}\\
 & \ell_{2}
 \end{pmatrix}
 \right]^{-1},
\end{equation}
where we have defined
\begin{equation}
    \ell_n = \left\{\int_{\omega^*}^\infty+\int^{-\omega^*}_{-\infty}\right\} 
    d\xi \frac{{\cal N}_n\left(\xi\right)}{\left|\xi\right|},
\end{equation}
where ${\cal N}_n\left(\xi\right)$ is the DOS of the $n$-band at a distance $\xi$ away from the Fermi level.
We may explicitly write the elements of this matrix,
\begin{equation}
    \hat{g}_{\rm TAM}
    \left(\omega^*\right)
    =
    \begin{pmatrix}\frac{g+\ell_{2}\left(g^{2}-g_{x}^{2}\right)}{1+g\left(\ell_{1}+\ell_{2}\right)+\left(g^{2}-g_{x}^{2}\right)\ell_{1}\ell_{2}} & \frac{g_{x}}{1+g\left(\ell_{1}+\ell_{2}\right)+\left(g^{2}-g_{x}^{2}\right)\ell_{1}\ell_{2}}\\
\frac{g_{x}}{1+g\left(\ell_{1}+\ell_{2}\right)+\left(g^{2}-g_{x}^{2}\right)\ell_{1}\ell_{2}} & \frac{g+\ell_{1}\left(g^{2}-g_{x}^{2}\right)}{1+g\left(\ell_{1}+\ell_{2}\right)+\left(g^{2}-g_{x}^{2}\right)\ell_{1}\ell_{2}}
\end{pmatrix}\equiv\begin{pmatrix}u_{1} & u_{x}\\
u_{x} & u_{2}
\end{pmatrix}.
\end{equation}

For the sake of illustration, consider two limiting cases.
If the different bands are decoupled, $g_x=0$, we find
\begin{equation}
    \hat{g}_{\rm TAM}
    \left(\omega^*\right)=\begin{pmatrix}\frac{g}{1+g\ell_{1}}\\
 & \frac{g}{1+g\ell_{2}}
\end{pmatrix},
\end{equation}
which is of course two instances of the single band result.
Conversely, when the interband interaction is just as strong as the intraband one, $g_x = g$, 
\begin{equation}
    \hat{g}_{\rm TAM}
    \left(\omega^*\right)=
    \frac{1}{1+g\left(\ell_{1}+\ell_{2}\right)}
    \hat{g}_{{\rm initial}},
\end{equation}
i.e., the DOS from \textit{both bands} contributes to the 
renormalization and suppression of the repulsion -- a first hint at the inter-band interactions importance.

\subsection{Retarded interaction}
We now introduce the retarded interaction in the Cooper channel.
Generally, one may consider two possible scenarios.
In the first, the interaction corresponds to an \textit{intraband attraction}, so we define
\begin{equation}
    \hat{g} =  \hat{g}_{\rm TAM} \left(\omega^*\right) 
    -\begin{pmatrix}g_{{\rm intra}}\\
 & g_{{\rm intra}}
\end{pmatrix}.
\end{equation}
This is the conventional scenario, to which we refer throughout our calculations.
In the second scenario, the retarded interaction scatters Cooper pairs between the bands, and it is necessary to consider \textit{interband repulsion}, i.e.,
\begin{equation}
    \hat{g} =  \hat{g}_{\rm TAM} \left(\omega^*\right) 
    +\begin{pmatrix} & g_{{\rm inter}}\\
g_{{\rm inter}}
\end{pmatrix}.
\end{equation}

We note that in the case of intraband interaction, one must have attraction to have any hope of finding a BCS instability,
yet for interband interactions this is not the case.
However, as we will demonstrate, \textit{interband repulsion is actually more favorable to superconductivity}.
This is because it adds on top of the interband renormalized repulsion $u_x$, instead of working against it.

\subsection{Finding $ \bf T_c$}
Having recovered the interaction matrix $\hat{g}$, we turn to calculate the susceptibility matrix as a function of temperature,
\begin{equation}
    \hat{\chi}\left(T\right)
    =
    \hat{g}^{-1} + 
    \begin{pmatrix}{\cal D}_{1}\left(T\right)\\
 & {\cal D}_{2}\left(T\right)
\end{pmatrix},
\end{equation}
where the electrons inside the $\omega^*$ shell contribute the factors (diverging at $T \to 0$), 
\begin{equation}
    {\cal D}_n \left(T\right)= \int_{-\omega^*}^{\omega^*}d\xi{\cal N}_{n}\left(\xi\right)\frac{\tanh\frac{\xi}{2T}}{\xi}.
\end{equation}

The superconducting instability is identified by the singularity of the inverse-susceptibility matrix, i.e.,
\begin{equation}
    \det \left[\hat{\chi}^{-1}\left(T_c\right)\right]=0.
\end{equation}
This is a transcendental equation for $T_c$, which generally has two solutions, of which one must choose the higher $T_c$, as it marks the superconducting transition.

We may explicitly derive the form of this self-consistent equation for $T_c$ in the two cases discussed above:
\begin{enumerate}
    \item Intraband attraction --
    \begin{equation}
        u_{x}^{2}{\cal D}_{1}{\cal D}_{2}=\left[1-\left(g_{{\rm intra}}-u_{1}\right){\cal D}_{1}\right]\left[1-\left(g_{{\rm intra}}-u_{2}\right){\cal D}_{2}\right].
    \end{equation}
    \item Interband repulsion --
    \begin{equation}
        \left(u_{x}+g_{{\rm inter}}\right)^2{\cal D}_{1}{\cal D}_{2}=\left[1+u_{1}{\cal D}_{1}\right]\left[1+u_{2}{\cal D}_{2}\right].
    \end{equation}
\end{enumerate}

The four-fold degenerate case may be of some interest (though irrelevant here considering the experimental data).
If ${\cal D}_{1}={\cal D}_{2}\equiv{\cal D}$, and $u_1=u_2\equiv u$, the equation for $T_c$ looks the same for both cases above, and reads
\begin{equation}
    \left(g_{{\rm attraction}}-u+u_{x}\right){\cal D}=1.
\end{equation}
It is now made even clearer, that in either case the interband interaction $u_x$ always ``assists'' the retarded pairing glue.
It is actually even simpler in the fully degenerate case $\ell_1=\ell_2\equiv\ell$,
\begin{equation}
    \left(g_{{\rm attraction}}-\frac{g-g_{x}}{1+\left(g-g_{x}\right)\ell}\right){\cal D}=1,
\end{equation}
so $T_c$ is determined by the difference between the bare intraband and intraband repulsion, renormalized by the TAM mechanism.

\section{Superconductivity onset dependence on Ising spin-orbit coupling}\label{sec:SConsetISOC}

We consider the main superconducting region across the different devices, referred to as ``SC$_2$''.
Consistently, it has been shown by Fermiology measurements that the parent normal-state is mostly two-fold degenerate.
In the non-interacting limit, the state is comprised of two spin-valley-locked sectors, the so-called majority and minority bands, split by an energy $\sim 2\lambda_0$.
Interactions tend to modify the energy separation, and possibly induce canting, which render the spin-valley locking only approximate.

As the Ising spin-orbit increases and the bands become more separated (interactions only enhance this trend), the high-DOS regions in the two bands move further away from each other, as well.
Consequently, it becomes exceedingly difficult to stabilize a regime where the DOS in the vicinity of the Fermi energy is high \textit{in both bands}.
A higher displacement field effectively enhances the high DOS region at energies just below the van-Hove singularity, and thus make this simultaneous-high-DOS regime possible again.

As an illustration of this phenomenon, let us consider the effective Tolmachev-Anderson-Morel pseudo-potential
\begin{equation}
    v^*_{\rm TAM} = 
\sqrt{\det \hat{g}_{\rm TAM}
    \left(\omega^*\right)}
    =
    \sqrt{\det \left[\left(\hat{g}_{{\rm initial}}\right)^{-1}
    +\begin{pmatrix}\ell_{1}\\
 & \ell_{2}
 \end{pmatrix}
 \right]^{-1}}.\label{eq:TAMpseudo}
\end{equation}
The interaction $v^*_{\rm TAM}$ can easily be verified to give the familiar TAM pseudo-potential in the single band case.
Investigating this interaction is motivated by the fact that a superconducting transition is associated with a singularity of the $\hat{g}$ matrix emerging at some energy scale.
We point out that this singularity cannot occur without retardation, unless $\left|g_x\right|>g$ (this encompasses both attractive interactions $g<0$, and cases where interband scattering dominates).

\begin{figure}
    \centering
    \includegraphics[width=18cm]{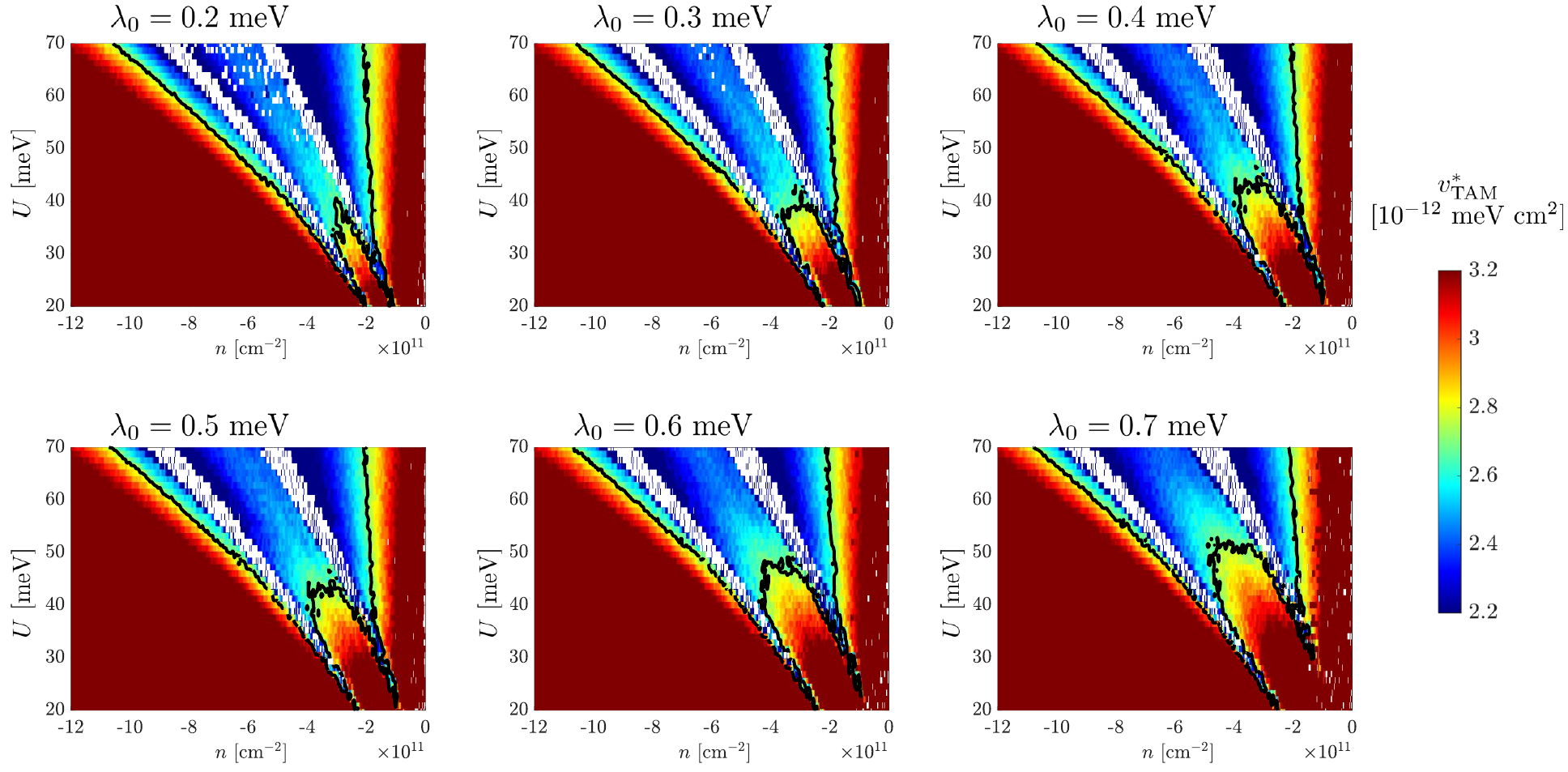}
    \caption{{\bf ISOC dependence of the residual Coulomb repulsion $v^*_{\rm TAM}$.}
    Different panels correspond to calculations of Eq.~\eqref{eq:TAMpseudo} in the presence of different ISOC values, indicated on top of each panel.
    The first and last panels correspond to Fig. 2l in the main text.
    Note that the Ising splitting referred to in the main text is $\lambda_I=2\lambda_0$, where $\lambda_0$ enters the model via Eq.~\eqref{eq:normalHisoc}.
    For emphasis, the black contour line marks the value $v^*_{\rm TAM}=2.7 \times 10^{-12}$ meV cm$^2$.
    The blank white regions mark areas where the system develops valley-polarization, and superconductivity is thus suppressed.
    Here, we use the Thomas-Fermi screening approximation, where $g\approx {\cal N}_{tot.} \left(E_F\right)$, the total Fermi-level DOS, and $g_x=0.5 g$.
    Other parameters used in this plot:
    $\omega^*=0.35$ meV, $U_C=1.4$ eV nm$^2$, $J=U_V=0.2$ eV nm$^2$.
    }
    \label{fig:onsetFigure}
\end{figure}

We plot the interaction strength $v^*_{\rm TAM}$ across the phase diagram for different values of the induced ISOC $\lambda_0$ in SI Fig.~\ref{fig:onsetFigure}.
Focusing our attention to within the blank white regions (those indicate where two-fold degeneracy is broken within our simplified analysis), we discover the sought-after trend.
Namely, the border of low $v^*_{\rm TAM}$, and thus superconductivity threshold, moves to higher displacement fields when the ISOC is gradually enhanced.
Again, this is due to the interband part of the interaction being most efficient when the DOS near the Fermi level is high in both bands at the same time.

\section{Effect of Zeeman magnetic field}\label{sec:ZeemanMultibandEffect}
Let us discuss how Zeeman coupling of the in-plane magnetic field eventually impacts superconductivity itself.

\subsection{Intravalley vs. Intervalley interactions}
Suppose the system we consider is characterized by two interaction scales: low-momentum scattering $V_0$, and intervalley scattering $V_Q$.
Originating in Coulomb repulsion, which scales as $1/q$ ($q$ is the magnitude of momentum scattered), one might assume $V_0\gg V_Q$.
This assumption relies on the Fermi momentum, $k_F\sim \sqrt{n/4} \sim 0.05$ nm$^{-1}$, being much smaller than the intervalley momentum scale, $Q\approx 17$ nm$^{-1}$.
However, as was similarly argued in Refs.~\citenum{chouAcousticphononmediatedSuperconductivityBernal2022,lianTwistedBilayerGraphene2019,shavitInducingSuperconductivityBilayer2023}, $k_F$ is not the relevant momentum here, as significant Thomas-Fermi screening takes place.
The Thomas-Fermi momentum, $q_{TF}=2\pi e^2{\cal N}\left(\bar{\mu}\right)/\epsilon_r$ is of order ${\cal O} \left(10\,{\rm nm}^{-1}\right)$ (considering $\epsilon_r=4$ for hBN and the relevant density ranges).
It is thus entirely possible that $V_0$ and $V_Q$ are comparable in strength, and both are of the order of magnitude of the inverse DOS at the Fermi level (as the Thomas-Fermi approximation suggests).

Considering the two bands or sectors are separated by spin-valley locking (mostly due to the induced Ising spin-orbit coupling), one identifies
\begin{equation}
    g=V_0, \,\,\,\,\, g_x=V_Q,
\end{equation}
where we have assumed the scattering interactions preserve spin.
However, if everything is kept the same, but now a canting angle $\theta\in\left[0,\frac{\pi}{2}\right]$ is introduced, such that $\theta=0$ corresponds to the spin-valley locked phase, and $\theta=\pi/2$ corresponds to two sectors with completely opposite spin polarization within the plane, one finds
\begin{equation}
    g=V_0+V_Q\sin^2\theta, \,\,\,\,\, g_x=V_Q\left(1-\sin^2\theta\right).\label{eq:suppressionbycanting}
\end{equation}
Notice that for complete polarization $g_x$ vanishes -- this process now flips two spins and cannot be facilitated by the Coulomb interactions.
Through this relation, and the self-consistent equation for superconducting $T_c$, \textit{superconductivity becomes directly impacted by the canting angle}.
Importantly, no pair-breaking within the different sectors occurs, and the Zeeman effect (which we will now demonstrate) only acts as a knob on the Cooper channel interaction strength.

\subsection{Canting due to Zeeman in the presence of interactions}
Consider the mean-field free energy of a two-fold symmetric phase, with order parameter $\delta n$, in the presence of short-range repulsion $U_C$, ferromagnetic Hund's coupling $J$, and in-plane Zeeman energy $V_Z$,
\begin{align}
    F\left(\delta n,\phi  \,;\,n\right)
    &=
    2 E_{\rm kin} \left( \frac{n+\delta n}{4}\,;\, \frac{n}{4}\right)
    +2 E_{\rm kin} \left( \frac{n-\delta n}{4}\,;\, \frac{n}{4}\right)
    -\frac{U_C}{8}\delta n^2\nonumber\\
    &-\lambda_0 s_z\tau_z - J \Vec{S}_{+}\cdot\Vec{S}_{-} -V_Z S_x,
\end{align}
where $E_{\rm kin} \left( n_1\,;\, n_2\right)$ is the kinetic energy associated with maintaining the density $n_1$ instead of the paramagnetic density $n_2$.
In terms of the sector imbalance $\delta_n$ and canting angle $\theta$, the magnetic part of the free energy is
\begin{equation}
    \frac{F_{\rm mag}}{\delta n} = -\lambda_0  \cos\theta + \frac{J \delta n}{4} \cos2\theta -V_Z\sin\theta.
\end{equation}
Let us approximate the dependence on $\delta n$ and the canting angle $\theta$ as being roughly separable.
In that case, we may minimize $F_{\rm mag}$ with respect to $\theta$ independently, arriving at the following transcendental equation for extracting $\theta$,
\begin{equation}
    \lambda_0 \tan \theta - J\delta n \sin \theta = V_Z.\label{eq:cantingangle}
\end{equation}
Notice that for $V_Z=0$, one obtains $\theta=\cos^{-1}\frac{{\rm min}\left\{\lambda_0,J \delta n\right\}}{J \delta n}$, and one obtains full spin polarization ($\theta=\pi/2$) in the absence of spin-orbit coupling, as expected.
Importantly, Eq.~\eqref{eq:cantingangle} clearly indicates the role of the majority-minority imbalance $\delta n$ in determining the canting angle, as well as its susceptibility to a Zeeman term.

\subsection{Comparison of different superconducting regions}\label{sec:PVRheirarchy}
The mechanism described in detail above sheds light on the differing behaviors of various superconducting domes when subjected to an in-plane magnetic field.

\begin{figure}
    \centering
    \includegraphics[width=8cm]{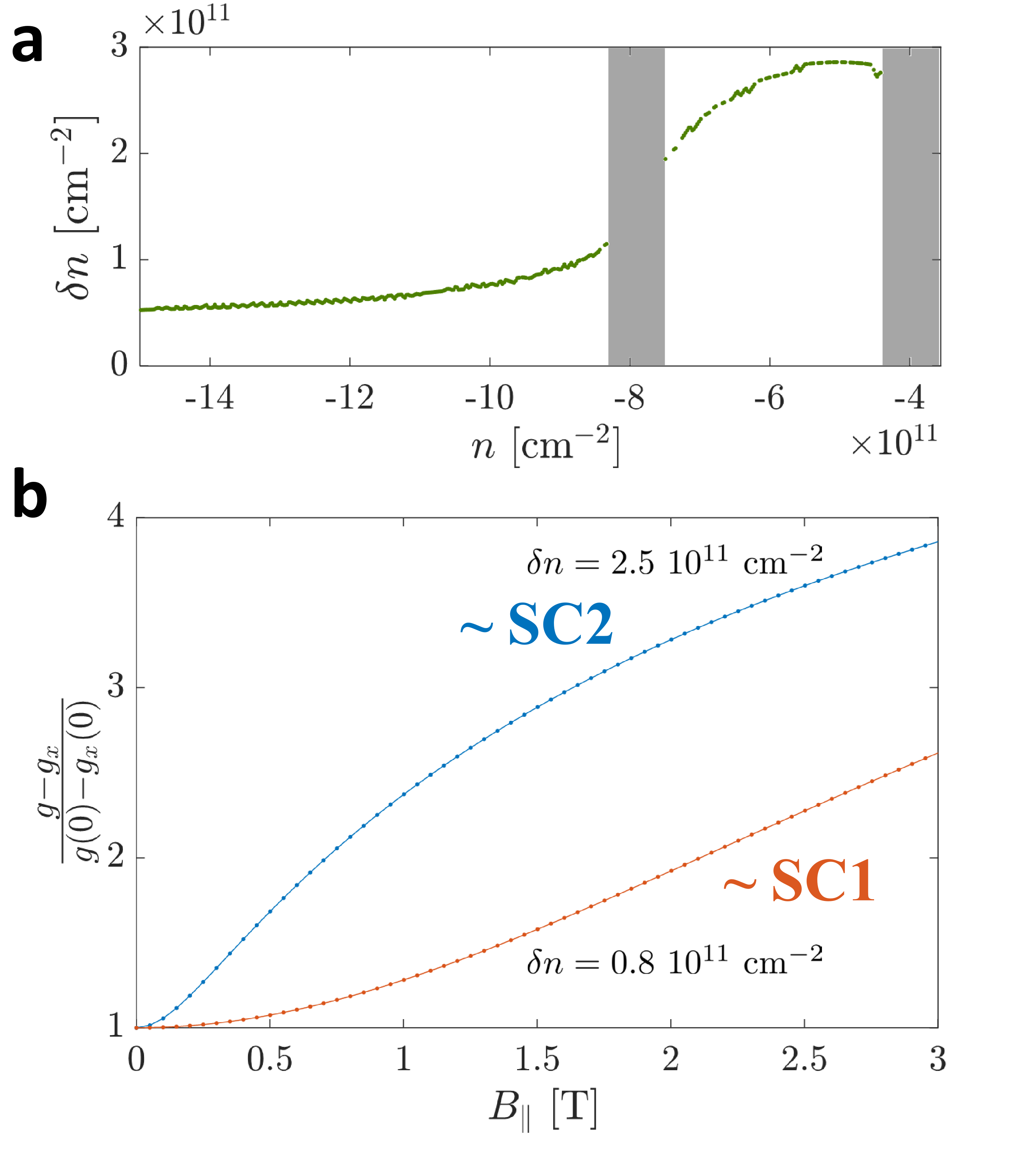}
    \caption{{\bf Understanding the discrepancy of in-plane magnetic field sensitivity between SC$_1$ and SC$_2$.}
    {\bf a}, Majority-minority band population imbalance $\delta n$ as a function of density.
    Here we used $D=0.8$ V/nm, and $\lambda_0=0.6$ meV.
    Gray rectangles mark regions where the two-fold degeneracy is broken within our variational Hartree-Fock analysis.
    The parameters we use to extract the normal state properties are identical to those in SI Fig.~\ref{fig:onsetFigure}.
    {\bf b}, Relative difference between the intraband ($g$) and interband $g_x$ interactions as a function of magnetic field.
    We plot the trend for two representative values of $\delta n$ (indicated in the panel), appropriate for the SC$_1$ (orange) and SC$_2$ regions.
    It is clear that repulsion grows more rapidly with magnetic field in the SC$_2$ region, alluding to the enhanced sensitivity of the corresponding superconductor.
    Here, $ V_0=3.1$ eV nm$^2$, $V_Q=2.2$ eV nm$^2$.
    }\label{fig:sc1sc2comparePVR}
\end{figure}

The particularly relevant regions, considering the flavor occupation picture revealed by quantum oscillations, are SC$_1$ and the higher-hole-density part of SC$_2$ (the main dome).
These two superconductors fit nicely within our framework of interband superconductivity, as they are comprised of two bands, minority and majority, each doubly degenerate.
However, they have one important differentiating feature: the relevant interband polarization $\delta n$ is significantly larger for SC$_2$.
The difference can already be gleaned from the experimental data, see \prettyref{fig:fig3}c of the main text, where the quantum-oscillations feature corresponding to the majority band drops abruptly in the SC$_1$ region.
This indicates a much smaller $\delta n$.
The difference in $\delta n$ is also readily apparent in our variational Hartree-Fock calculation, SI Fig.~\ref{fig:sc1sc2comparePVR}a, where it can be traced to the different DOS in the appropriate filling range.

As such, we would expect [given Eqs.~\eqref{eq:cantingangle} and~\eqref{eq:suppressionbycanting}] SC$_2$ to be much more sensitive to the in-plane field as compared to SC$_1$.
In SI Fig.~\ref{fig:sc1sc2comparePVR}b, we compare the relative strength of intraband and interband interactions as a function of magnetic field, for two representative values of $\delta n$.
The canting angle is extracted via Eq.~\eqref{eq:cantingangle}.
As expected, the interactions corresponding to SC$_2$ are much more sensitive to the in-plane field. 
This is in complete agreement with the experimental data, as the Pauli-limit violation ratio (PVR) for SC$_1$ is $\sim 18$, whereas for SC$_2$ it is noticeably lower, PVR$\sim 4$.

\begin{figure}
    \centering
    \includegraphics[width=15cm]{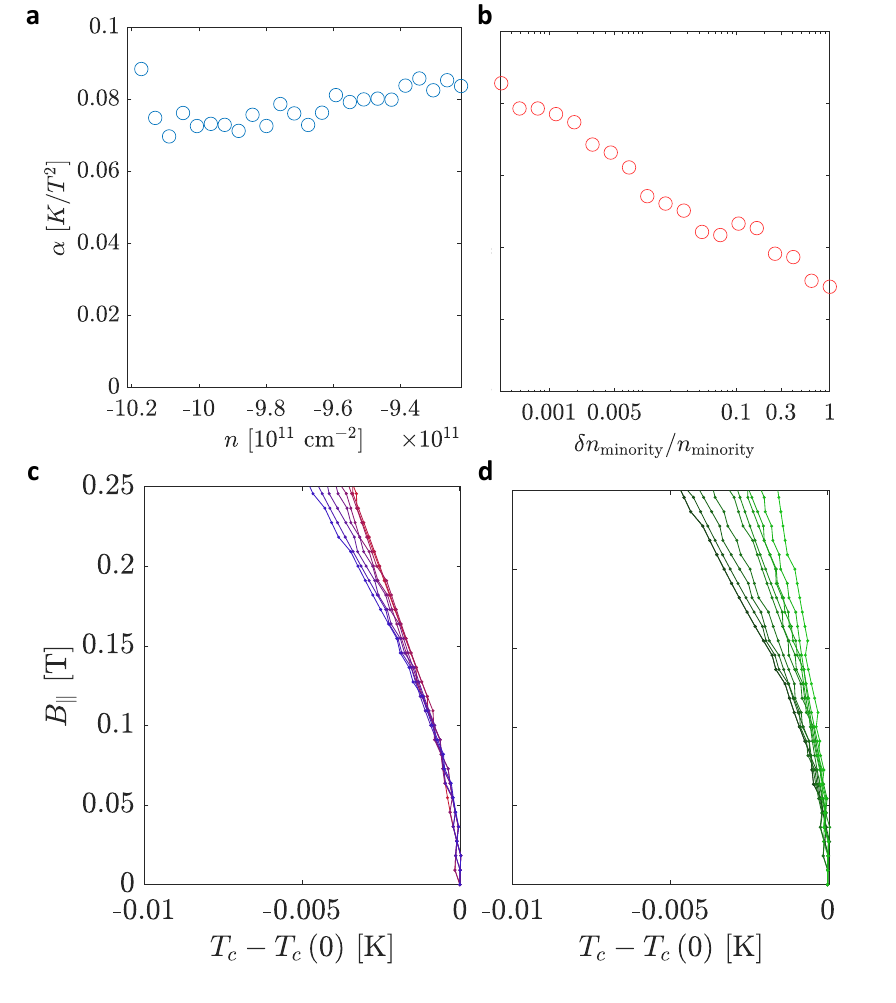}
    \caption{{\bf In-plane field dependence of superconductivity in the SC$_2$ dome.}
    {\bf a}, The coefficient $\alpha$ extracted for a range of densities corresponding to the lower density region of SC$_2$, featuring an approximately constant coefficient.
    {\bf b}, At the highest density of panel {\bf a}, we introduce a polarization within the minority sector. 
    The polarization gradually decouples the minority band from the Cooper channel, making the superconductor less sensitive to the magnetic field.
    {\bf c},{\bf d}, Representative plots used to extract the fit parameter $\alpha$ in panels {\bf a},{\bf b}, respectively.
    Details regarding the methodology and precise parameters used in the calculations can be found in the text of Sec.~\ref{sec:PVRheirarchy}.
    }\label{fig:TcBehaviorBfield}
\end{figure}

Remarkably, the trend in $T_c$ in this part of SC$_2$ is well captured by our model calculations.
For a large portion of this region, the critical temperature at small magnetic fields largely follows the relation
\begin{equation}
    T_c\left(B_\parallel\right) - T_c\left(0\right)
    \approx -\alpha B_\parallel^2.\label{eq:TCb2}
\end{equation}
For example, in the data presented in \prettyref{fig:fig4} in the main text, 
the experimental value remains roughly constant in the left (lower density) part of the SC$_2$ dome, with $\alpha\approx 0.08$ Kelvin/Tesla$^2$.
Our calculations, taking into account the magnetic field effects discussed in this section, are in good agreement both with the $B_\parallel^2$ trend, and with the numerical value of the prefactor $\alpha$.
This is demonstrated in SI Fig.~\ref{fig:TcBehaviorBfield}, where the $T_c$ behavior follows almost perfectly the behavior in Eq.~\eqref{eq:TCb2} with the approximate experimental value.

To perform these calculations, we take as phenomenological input the shape of the Fermi surfaces observed in quantum oscillations.
For the majority bands we take density per flavor in the range $\left[-3.85,-3.35\right]$ 10$^{11}$ cm$^{-2}$, and for the minority bands  $\left[-1.35,-1.25\right]$ 10$^{11}$ cm$^{-2}$.
Moreover, to account for the fact that the minority bands consist of two small pockets (instead of the expected three due to trigonal warping), we introduce a nematic order parameter which breaks the $C_3$ symmetry,
\begin{equation}
    h_\tau\left({\bf k}\right)\to
    h_\tau\left({\bf k}\right)
    +
    \begin{pmatrix} &  &  & \Delta_{{\rm nem}}\\
\\
\\
\Delta_{{\rm nem}}
\end{pmatrix},
\end{equation}
and we used $\Delta_{{\rm nem}}=-15$ meV to match the experimental observation in the desired density range.
Other parameters used:
$\omega^*=0.8$ meV,
$V_0=3.1$ eV nm$^2$,
$V_Q=2.2$ eV nm$^2$,
$J=0.1$ eV nm$^2$,
$g_{\rm attraction}=0.38$ eV nm$^2$,
$\lambda_0=0.6$ meV.

Conversely, there is a marked change in behavior of SC$_2$ when moving closer to the charge neutrality point.
Namely, $T_c$ becomes much less sensitive to magnetic fields and the magnitude of the corresponding $\alpha$ prefactor becomes much smaller.
Again, the quantum oscillation data provides us with important information.
The data is consistent with flavor polarization developing within the minority bands themselves.
Such polarization can be characterized by an order parameter 
$\delta n_{\rm minority}$, such that in the absence of polarization $\delta n_{\rm minority}=0$, and when only one flavor (of the two minority ones) is occupied $\delta n_{\rm minority}=n_{\rm minority}$.
Under the reasonable assumption of zero-momentum Cooper pairing, and considering the fact that the newly polarized flavors correspond to opposite valleys, one expects the minority bands to gradually decouple from the Cooper channel as this polarization increases.

To concretely illustrate this effect, we extract the appropriate Fermi energies in the minority sector, which correspond to a given $\delta n_{\rm minority}$,
\begin{equation}
    2\int_0^{E_+} d\epsilon {\cal N}_2\left(\epsilon\right)=n_{\rm minority}+\delta n_{\rm minority},
    \,\,\,\,\,
    2\int_0^{E_-} d\epsilon {\cal N}_2\left(\epsilon\right)=n_{\rm minority}-\delta n_{\rm minority}.
\end{equation}
Taking $E_0=E_\pm\left(\delta n_{\rm minority}=0\right)$, we define
$\xi_\pm=E_\pm-E_0$.
Thus, in the presence of this secondary polarization the appropriate Cooper integrals in the minority sector are altered,
\begin{equation}
    \ell_2 = \left\{\int_{\max\left\{\omega^*,\xi_+\right\}}^\infty+\int^{\min\left\{-\omega^*,\xi_-\right\}}_{-\infty}\right\} 
    d\xi \frac{{\cal N}_2\left(\xi\right)}{\left|\xi\right|},
\end{equation}
\begin{equation}
    {\cal D}_2 =
    \left\{\int^{\max\left\{\omega^*,\xi_+\right\}}_{\xi_+}
    +\int_{\min\left\{-\omega^*,\xi_-\right\}}^{\xi_-}\right\}
    d\xi{\cal N}_{2}\left(\xi\right)\frac{\tanh\frac{\xi}{2T}}{\xi}.
\end{equation}
We have implicitly assumed here that the relevant temperature regime, i.e., the vicinity of $T_c$, is much smaller than $\xi_\pm/k_B$.
Notice that in the absence of polarization
$\xi_+=\xi_-=0$, and we recover the previous expressions.

As we show in SI Fig.~\ref{fig:TcBehaviorBfield}, we indeed recover the experimental trend.
The underlying reason is that the magnetic field behavior we consider relates solely to the interband interactions in the Cooper channel between majority and minority bands.
Once one of these sectors stops ``pulling its own weight'', eventually only an effective single-band (only majority) superconductor remains, and the magnetic field may only impact it through other effects we do not consider, e.g., an orbital effect (See also discussion of Sec. ~\ref{sec: conventional}).

In this context, it is worth mentioning that in the normal state of SC$_3$ the quantum oscillation data is consistent with single band superconductivity, with significant time-reversal symmetry breaking in at least one of the sectors.
As a result, our analysis would suggest reduced sensitivity to the  $B_{\parallel}$ for the same reason we just discussed.
Remarkably, it appears indeed that SC$_3$ is the superconducting pocket which is most robust in the presence of the magnetic field, and has by far the largest Pauli limit violation in the measured devices.

In summary, the behavior of the different superconducting pockets as a response to an in-plane magnetic field strongly suggests that \textit{interband interactions in the Cooper channel play a crucial role in BLG superconductivity}.
Recognizing this prominent role, we may comment on the disparity between superconductivity in BLG with or without WSe$_2$.
Namely, in the former superconductivity is much stronger and more robust than the latter.
In the two-fold polarized regime (majority and minority sectors are well separated in energy), the system is spin-polarized without WSe$_2$ due to intervalley Hund's coupling.
Thus, interband interactions are entirely absent in the Cooper channel due to spin conservation.
Introduction of Ising spin orbit coupling ``re-shuffles'' the polarization cascade, and favors spin-valley locking as the two-fold phase even at moderate values of the induced spin orbit coupling.
This ``shifts'' interaction strength from the intraband part of the coupling matrix to the interband part, Eq.~\eqref{eq:suppressionbycanting}.
As we have demonstrated above, such a shift towards stronger interband interactions may have a \textit{great positive impact on superconductivity}.

\section{Analysis of SOC, orbital and Zeeman depairing mechanisms}
\label{sec: conventional}

We now discuss SOC, orbital, and Zeeman depairing mechanisms in Ising superconductors. We 
follow the treatment first developed in
Ref.~\citenum{frigeriSuperconductivityInversionSymmetry2004} to compute the
response to a Zeeman field of non-centrosymmetric superconductors with Rashba
SOC, later generalized in Refs.~\citenum{luEvidenceTwodimensionalIsing2015,
saitoSuperconductivityProtectedSpinvalley2016} to systems with a coexistence 
of Ising and Rashba SOC. Parts of the discussion present in this supplement 
mirror our earlier analysis of Ref. \citenum{zhangEnhancedSuperconductivitySpin2023}. 
The analysis of this section complements the previous section, intending 
to provide contrast and further strengthen the interpretation above.
Namely, the trends observed in the experiment as a function of 
in-plane magnetic field likely points to the prominence of majority-minority 
interband interactions. 

We assume that as a result of flavor symmetry-breaking transition (cascade), the carriers 
in the parent states of SC$_1$, SC$_2$, and SC$_3$ occupy the relevant 
number of trigonal-warping pockets as discussed in the main text. Each trigonal-warping pocket we model as two electronic bands centered around
trigonal-warping loci $\vec{T}$ that respectively originate from the $K$
and $K'$ valleys. We also assume that the
selected pockets in the two valleys are time-reversed partners of each other to allow for the formation of zero-momentum pairing naturally.

As mentioned in the text, we stress that the quantum oscillation data do not 
directly reveal nematicity but rather only inform about the number and relative 
sizes of Fermi pockets. If the Fermi pockets are not the trigonal warping pockets 
but rather some other Fermi pockets resulting from the interaction-driven 
reconstruction of the electronic spectrum, then the following
modeling could still apply, provided that time reversal 
symmetry $\mathcal{T}$, which relates the two remaining small
pockets is preserved. The specific relation between the effective 
Ising, Rashba, and orbital coupling, however, in such a case, would 
not be related to the microscopic parameters of the continuum model that we will derive in the following subsection.

\subsection{Effective Hamiltonian in the trigonal warping pockets}

Here we derive an effective low-energy $2\times 2$ Hamiltonian describing two 
SOC split bands at each trigonal warping loci. This Hamiltonian serves as a 
starting point for the following analysis as it allows us to connect microscopic 
model parameters (Ising, Rashba, Orbital, displacement field) with the low-energy 
theory. The purpose of this exercise is to provide analytic intuition about 
what energy and momentum scales in the problem control the depairing. Moreover, 
this procedure allows us to reduce the number of fitting parameters to just 
Rashba coupling as Ising SOC is determined experimentally, and orbital 
coupling is a property of the system.

The starting point of our analysis is the valley-dependent ($\tau=\pm 1$) 
Hamiltonian in spin, layer, and sublattice basis:
\begin{equation}
    h_{full, \tau} = h_{0,\tau} + h_{I,\tau} + h_{R,\tau} + h_{orb,\tau} + h_{Z}\,,
\end{equation}
where $h_{0,\tau} = s_0 \otimes h_{\tau} $ with $h_\tau$ given by Eq.~\eqref{eq:H04by4basis} (we set $\Delta'=0$ for simplicity) and describes the non-interacting Hamiltonian of bilayer graphene in the absence of SOC and magnetic field. We use the shorthand notation, $s_0, s_x, s_y, s_z$, for Pauli matrices in the spin space. The other entries in the above expression correspond to Ising SOC ($h_{I,\tau}$), Rashba SOC($h_{R,\tau}$), orbital coupling ($h_{orb,\tau}$), and Zeeman energy ($h_{Z}$). They are given by:
\begin{align}
  h_{I,\tau} &=\frac{1}{2} \tau \lambda_I \left[ s_z  \otimes\left(
\begin{array}{cccc}
 1 & 0 & 0 & 0 \\
 0 & 1 & 0 & 0 \\
 0 & 0 & 0 & 0 \\
 0 & 0 & 0 & 0 \\
\end{array}
 \right)\right]\\
h_{R,\tau}&=\frac{1}{2} \lambda_R \left[\tau s_y \otimes \left(
\begin{array}{cccc}
 0 & 1 & 0 & 0 \\
 1 & 0 & 0 & 0 \\
 0 & 0 & 0 & 0 \\
 0 & 0 & 0 & 0 \\
\end{array}
\right)-s_x \otimes \left(
\begin{array}{cccc}
 0 & -i & 0 & 0 \\
 i & 0 & 0 & 0 \\
 0 & 0 & 0 & 0 \\
 0 & 0 & 0 & 0 \\
\end{array}
\right)\right]\\
h_{Z} &=\mu_B B \left[\left(\cos\theta_B s_x + \sin\theta_B s_y\right)\otimes \left(
\begin{array}{cccc}
 1 & 0 & 0 & 0 \\
 0 & 1 & 0 & 0 \\
 0 & 0 & 1 & 0 \\
 0 & 0 & 0 & 1 \\
\end{array}
\right)\right]\\
h_{orb,\tau} &=\lambda_{orb}\left[\tau sin\theta_B s_0 \otimes
\left(
\begin{array}{cccc}
 0 & -1 & 0 & 0 \\
 -1 & 0 & 0 & 0 \\
 0 & 0 & 0 & 1 \\
 0 & 0 & 1 & 0 \\
\end{array}
\right)+\cos\theta_B \otimes
\left(
\begin{array}{cccc}
 0 & -i & 0 & 0 \\
 i & 0 & 0 & 0 \\
 0 & 0 & 0 & i \\
 0 & 0 & -i & 0 \\
\end{array}
\right)\right]
\end{align}
In the above expressions, we explicitly write the sublattice and layer structure of the matrices to highlight that the Ising and Rashba SOC are induced only on one layer, which is in proximity to the TMD. The angle $\theta_B$ determines the direction of the in-plane magnetic field $B$ with $\theta_B=0$ corresponding to the x-direction in the continuum model. The orbital coupling enters via minimal coupling (See Ref.\citenum{kheirabadiMagneticRatchetEffect2016} for derivation). Its magnitude (See also Ref. \citenum{zhangEnhancedSuperconductivitySpin2023}):
\begin{equation}
    \lambda_{orb} = \frac{v_0 e d}{2 \hbar} = \frac{\sqrt{3} \gamma_0 a_0 e d }{4 \hbar} \approx 0.14 \frac{\rm meV}{\rm Tesla} .
\end{equation}
yields a stronger coupling than the spin Zeeman term $h_{Z}$  with the Bohr magneton $\mu_B \sim 0.06$ meV/Tesla hinting at the role of orbital coupling in depairing. In the above Hamiltonian, in principle, all parameters are known or determined via experiment except $\lambda_R$.

We now follow the standard perturbation theory approach discussed, for example, in Ref. \citenum{mccannElectronicPropertiesBilayer2013,zaletelGatetunableStrongFragile2019} to arrive at a low-energy theory description of the trigonal warping pockets. The difference in our analysis compared to that of Ref. \citenum{mccannElectronicPropertiesBilayer2013,zaletelGatetunableStrongFragile2019} is that we keep both the trigonal warping term, $v_3$, which is necessary to account for the three pockets as well as the displacement field $U$ necessary to polarize charges to one side. Details and further discussion of the calculation are provided in Ref. \citenum{maUpperCriticalInplane2024}. Carrying out the analysis, we arrive at an effective Hamiltonian $h_{eff, \tau}$ in each trigonal warping pocket (in the spin up and down basis):
\begin{equation}
\label{eq:eff_Hamiltonian}
h_{eff, \tau} = \tilde{h}_{0,\tau} + \tilde{h}_{I,\tau} + \tilde{h}_{R,\tau} + \tilde{h}_{orb,\tau} + \tilde{h}_{Z}\,,
\end{equation}
where the individual terms are:
\begin{align}
\tilde{h}_{0,\tau} &\approx \epsilon^0(\vec{p}) s_0\\
\tilde{h}_{I,\tau} &\approx \frac{1}{2} \tau \lambda_I  s_z\\
\tilde{h}_{orb,\tau} &\approx g_{orb}(\vec{p}) s_0\\
\tilde{h}_{R,\tau} &\approx \frac{1}{2} \left(g_{R,x}(\vec{p}) s_y-g_{R,y}(\vec{p}) s_x\right)\label{eq:effective_rashba}\\
\tilde{h}_{Z} &\approx \mu_B B \left(\cos\theta_B s_x + \sin\theta_B s_y\right)\,.
\end{align}
Here we expanded around a trigonal warping center located along the $x$-axis focusing on leading order terms where $\gamma_0$ ($v$) and $\gamma_1$ were the large energy scales in the problem (Analogously to Ref. \citenum{mccannElectronicPropertiesBilayer2013,zaletelGatetunableStrongFragile2019}). The dispersion of other pockets can be analogously derived by the rotating location of the $\vec{T}$. Momentum $\vec{p}$ is measured with respect to the trigonal warping pocket center $\vec{T} = T\hat{x}$:
\begin{equation}
    T = \frac{\sqrt{\gamma_1^2 v_3^2+4 U^2 v^2}+3 \gamma_1 v_3}{4 v^2}
\end{equation}
The parameters $\gamma_1, v_3, v$ are the continuum model parameters defined in Eq.~\eqref{eq:H04by4basis}. $U$ is the applied displacement field in units of energy difference across layers. As the displacement field grows, the trigonal warping centers move away from the $K, K'$ points with $U v$ comparable to $\gamma_1 v_3$ for the relevant $U$ for the experiment.

In the above equation the effective orbital coupling scale $g_{orb}$ is
\begin{equation}
\label{eq:g_orb}
g_{orb} \approx \lambda_{orb} \frac{ U v (p_y \cos\theta_B-(\tau T+p_x) \sin\theta_B)}{\gamma_1^2}.
\end{equation}
We see that as $U$ increases, the role of orbital coupling becomes more pronounced. 
We also see that the effective orbital coupling has two characteristic momentum 
scales:  ($i$) one sensitive to the filling of the small pockets (i.e., momentum $\vec{p}$ 
dependent) and ($ii$) one set by the location of the trigonal warping pocket $T$. 
Depending on the orientation of the magnetic field ($\theta_B$), the effective 
orbital coupling can change drastically, hinting at a potential sensitivity to 
in-plane field orientation.

Lastly the effective Rashba couplings $g_{R,x}(\vec{p})$ and $g_{R,y}(\vec{p})$ are 
given by:
\begin{align}
g_{R,x}(\vec{p}) &\approx \lambda_R \frac{p_y v \left(14 \gamma_1 T v_3+U^2\right)}{4 \gamma_1^2 U}\\
g_{R,y}(\vec{p}) &\approx \lambda_R \left[ \frac{\tau T U v}{4 \gamma_1^2} +\frac{p_x  v \left(10 \gamma_1 T v_3+3 U^2\right)}{4 \gamma_1^2 U}\right]\,.
\end{align}
Just as in the case of the effective orbital coupling, we find that there are two 
contributions to the effective Rashba coupling: ($i$) one set by the location of the 
trigonal warping pocket $T$ and ($ii$) one controlled by the doping of each pocket 
and hence momentum $p_x,p_y$. Physically, the two origins of the terms make 
sense: ($i$) controls the winding of the Rashba spin texture around the $K, K'$ points, 
and ($ii$) controls the winding of the Rashba spin texture around the trigonal warping 
center. In the above expression, we focused on the leading order terms for typical 
numerical values (See Ref. \citenum{maUpperCriticalInplane2024} for further discussion).

Crucially, from the above result, we find that the effective Ising in each trigonal 
pocket is, to leading order, unmodified from the microscopic Ising value by the 
projection to each pocket. However, the effective orbital and Rashba contributions 
are different. Both orbital and Rashba contributions feature dependence on two characteristic momenta: ($i$) momentum location of the trigonal warping pocket 
from the $K, K'$ (which is displacement field ($U$) dependent) and ($ii$) a momentum $p$ 
measured with respect to each pocket center. Both terms are significant in 
determining the magnitude of the effective orbital and Rashba parameters. 

\subsection{Depairing analysis for $\lambda_I \ll E_F$}

We now proceed to analyze the depairing mechanisms using the practical model of Eq. \eqref{eq:eff_Hamiltonian} as the normal state Hamiltonian. As mentioned before, we will follow the treatment first developed in
Ref.~\citenum{frigeriSuperconductivityInversionSymmetry2004} to compute the
response to a Zeeman field of non-centrosymmetric superconductors with Rashba
SOC, later generalized in Refs.~\citenum{luEvidenceTwodimensionalIsing2015,
saitoSuperconductivityProtectedSpinvalley2016} to systems with a coexistence of
Ising and Rashba SOC. The approach mirrors that carried out by us in Ref. \citenum{zhangEnhancedSuperconductivitySpin2023}, however here we focus on analytical understanding of the $T_C\to T_{C,0}$ limit (here $T_{C,0}$ is the depairing temperature in the absence of in-plane field). We will also use the results of the previous subsection to constrain the estimates with realistic microscopic parameters.
 
To connect to the previous work of Ref. \citenum{zhangEnhancedSuperconductivitySpin2023}, we further rewrite the normal-state Hamiltonian as
\begin{equation}
    h_{eff, \tau}(\vec{p}) = \xi(\vec{p}) + \frac{1}{2} \tau \lambda_I s^z + \frac{1}{2} (\vec{s}\times\vec{g_{R}(\vec{p})})\cdot \vec{z} + \vb  \cdot  \vs + \left( \vb \times \vec{g}_{orb}(\vec{p})  \right) \cdot \vec{z}    ,
    \label{eq:toy_model}
\end{equation}
On the right side $\xi({\vk})$ 
is the spin-orbit-free normal state band structure, which we linearize near the 
Fermi surface as $\xi({\vk}) \approx v_F (p-p_F)$ ($p_F$ denotes the Fermi 
momentum measured from the center $\vec{T}$ of the pocket). The next two terms 
incorporate Ising and Rashba SOC. Although the Rashba SOC, when projected to the trigonal warping pocket, is not isotropic with respect to the pocket center (as the pocket is not isotropic either), see Eq. \eqref{eq:effective_rashba}, for simplicity of the analysis, we will approximate the Rashba as
\begin{equation}
\vec{g}_{R}(\vec{p}) \approx g_{R,0}+ g_{R,1}(p_y,p_x)\,\quad g_{R,0}=\lambda_R  \frac{\tau T U v}{4 \gamma_1^2}\, g_{R,1}=\lambda_R \frac{ v \left(10 \gamma_1 T v_3+3 U^2\right)}{4 \gamma_1^2 U}.
\end{equation}
By comparison with Eq.\eqref{eq:effective_rashba}, this overestimates the Rashba coupling, but for the purpose of this analysis, it is sufficient. For the orbital coupling, we focus on the in-plane field oriented along $x$-axis (we leave the discussion of in-plane direction dependence to the next section), giving
\begin{equation}
\vec{g}_{orb}(\vec{p}) = \lambda_{orb} \frac{ U v p_y}{\gamma_1^2} \left(0,1\right)
\end{equation}
The term $\vb \cdot \vs$ is simply the Zeeman energy, which for an in-plane magnetic
field $B$ along $x$-axis takes the form $\vb = (\mu_B B, 0, 0)$ 
with $\mu_B$ the Bohr magneton. We pause here to highlight that unlike the analysis of Ref. \citenum{zhangEnhancedSuperconductivitySpin2023}, here, in principle, there is only one undermined parameter $\lambda_R$ with the remainder of the Hamiltonian known.

To analyze depairing, we then consider a local (momentum-independent) spin-singlet pairing term that 
gives rise to superconductivity with a critical temperature $T_c^0$ at zero 
magnetic field. In the presence of an in-plane magnetic field $B$, the 
superconductivity is weakened through a combination of spin and orbital 
effects, with $T_c < T_c^0$ given by the solution of a self-consistent gap 
equation linearized near the second-order transition at $T_c$ where the pairing
amplitude $\Delta \rightarrow 0$,\cite{saint-jamesTypeIISuperconductivity1969, 
frigeriSuperconductivityInversionSymmetry2004, 
luEvidenceTwodimensionalIsing2015, 
saitoSuperconductivityProtectedSpinvalley2016, 
zwicknaglCriticalMagneticField2017}
\begin{equation}
\label{eq:gap_equation}
    \ln \left( \frac{T_c}{T_c^0} \right) =  \frac{T_c}{2} \sum_{\omega_n}\left(\left\langle \int d \xi_{\vec{p}} \mathrm{Tr}\left\{s_y G_{0,\tau}(\vec{p},i\omega_n) s_y G_{0,-\tau}^*(-\vec{k}, i\omega_n)\right\}\right>_{\rm FS}-\frac{\pi}{|\omega_n|}\right) .
\end{equation}
Here $\sum_{\omega_n}$ denotes a summation over Matsubara frequencies 
$i\omega_n$, $\langle\cdots\rangle_{\rm FS}$ denotes a Fermi surface average and
$\mathrm{Tr}\{ \cdots \}$ is a trace over spin Pauli matrices, and 
$G_{0,\tau}(\vec{p}, i\omega_n)$ is the normal-state Green's function given of the Hamiltonian from Eq.\eqref{eq:toy_model}. The above equation can be further simplified (as done previously and in literature\cite{frigeriSuperconductivityInversionSymmetry2004,luEvidenceTwodimensionalIsing2015,
saitoSuperconductivityProtectedSpinvalley2016}) to yield a non-linear equation connecting critical depairing magnetic field with the relevant temperature $T_C$. The pairing glue strength is parametrized here by the $T_{C,0}$ - the temperature without a magnetic field. We also note that in arriving at the above equation, we used the condition that chemical potential $E_F$ is larger than any SOC scale. Thus, we also allow for an interband pairing in the above expression, however the pairing glue is not renormalized as discussed in the mechanism from the previous section. In the next subsection, we focus on the other limit of $E_F \ll \lambda_I$.

To make an explicit connection to the experimental fitting of $T_C=T_{C,0}-\alpha B^2$, we focus on the $T_C\to T_{C,0}$ limit of Eq.\eqref{eq:gap_equation}. In such cases, we can arrive at (See Ref. \citenum{maUpperCriticalInplane2024} for further discussion and details of the derivation) at a microscopic expression for $\alpha$
\begin{align}
\label{eq:alpha_two_band}
    \alpha &=\alpha_0+\alpha_{orb}\\
    \alpha_0 &\approx T_{C,0}\mu_B^2\frac{2\lambda^{2}_{I}+\tilde{g}^2_R}{(\lambda^{2}_{I}+\tilde{g}^2_R)^{2}}\Phi(\rho)+\frac{\alpha_C \mu_B^2}{2 k_B^2  T_{c0}}\frac{\tilde{g}^2_R}{\lambda^{2}_{I}+\tilde{g}^2_R}\\
    \alpha_{orb} &\approx  \alpha_C \frac{ \tilde{g}_{orb}^2}{k_B^2 T_{c0}}\,, \quad \alpha_C = \frac{\Psi''(1/2)}{8\pi^{2}}\,
    \end{align}
    where we separated $\alpha$ into two parts orbital independent ($\alpha_0$) and orbital dependent part($\alpha_{orb}$). Here, the effective Fermi surface averaged Rashba and orbital couplings are given by:
    \begin{equation}
\tilde{g}_{orb} \approx \lambda_{orb} \frac{ U v p_F}{\gamma_1^2}\,,\quad\tilde{g}_{R}\approx \lambda_R  \frac{\tau T U v}{4 \gamma_1^2} +\lambda_R \frac{ v \left(10 \gamma_1 T v_3+3 U^2\right)}{4 \gamma_1^2 U} p_F.
\end{equation}
The special functions are defined as
\begin{equation}
    \Phi(\rho)\equiv\frac{1}{2}\{Re\left[\Psi\left(\frac{1+i\rho}{2}\right)-\Psi\left(\frac{1}{2}\right)\right]\,, \rho = \frac{\sqrt{\lambda^{2}_{I}+\tilde{g}^2_R}}{\pi k_B T_{C,0}}
\end{equation}
and $\Psi(z)$, $\Psi''(z)$ is the DiGamma function and its second order derivative.

\begin{figure}
\includegraphics[width=\textwidth]{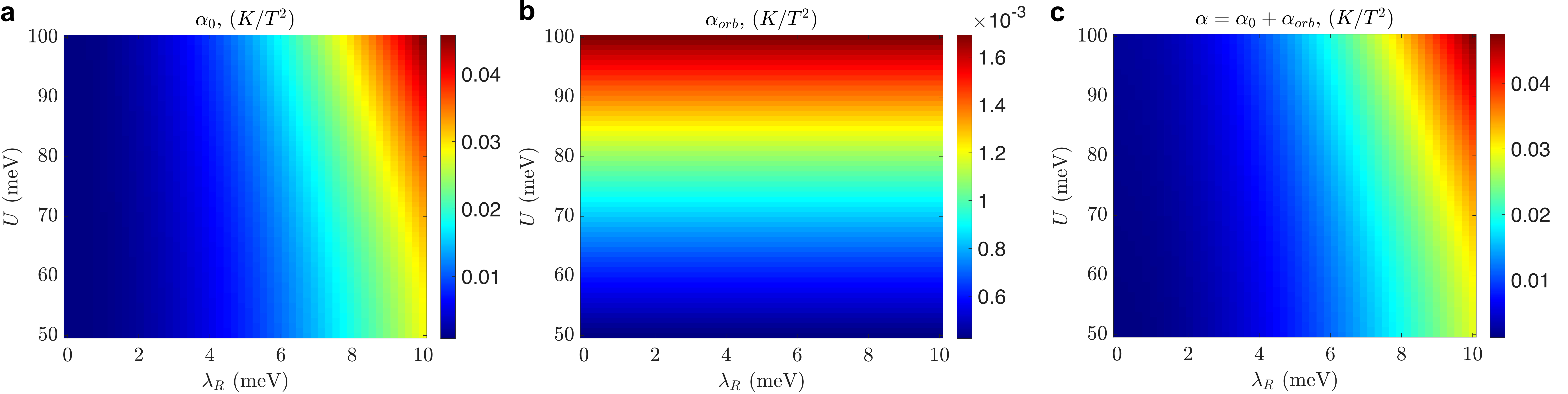}
\caption{{\bf Coefficient $\alpha$ in $T_C=T_{C,0}-\alpha B^2$ in the $E_F\gg \lambda_I$ limit.} Dependence of the coefficents $\alpha_0$ ({\bf a}), $\alpha_{orb}$ ({\bf b}), and $\alpha=\alpha_0+\alpha_{orb}$ ({\bf c}) from Eq.\eqref{eq:alpha_two_band} on layer potential $U$ and Rashba coupling $\lambda_R$. Here we use $p_F\sim \sqrt{n/4} \sim 0.05$ nm$^{-1}$ at $T_{C,0}\approx 100 $~mK and $\lambda_I=1.5$~meV.}
\label{sfig:cyprian_two_band}
\end{figure}

Let us now analyze the consequences of the result in Eq.\eqref{eq:alpha_two_band}. As expected, when Ising SOC increases, $\alpha$ decreases, in line with the expected Ising superconductor behavior (i.e., superconductor becomes more resilient to depairing). Rashba SOC and orbital coupling contribute to depairing, with the contributions being additive. SI Fig. \ref{sfig:cyprian_two_band}a-c shows plots of the two parts of the $\alpha$ coefficient separately (SI Fig. \ref{sfig:cyprian_two_band}a,b) and combined (SI Fig. \ref{sfig:cyprian_two_band}c) as a function of layer potential $U$ and microscopic $\lambda_R$ for $p_F\sim \sqrt{n/4} \sim 0.05$ nm$^{-1}$ at $T_{C,0}\approx 100 mK$ and $\lambda_I=1.5$ meV. We find that no realistic value of $\lambda_R < 10$ meV reproduces the experimental values of $\alpha{\sim} 0.1 K/T^2$ seen in SC$_2$ (we are about an order of magnitude too small). This conclusion also holds if the direction of the $B_\parallel$ were rotated to increase the orbital coupling, c.f. Eq.\eqref{eq:g_orb}. For the SC$_1$ case however where the superconductor is more robust to depairing with the experimental $\alpha\lesssim 0.01$ $K/T^2$ the theoretically predicted $\alpha$ is in agreement for reasonable $\lambda_R < 2$ meV values.

To phrase it differently, using realistic microscopic SOC and orbital coupling values is insufficient to account for the suppression of superconductivity with an in-plane field in the SC$_2$ phase. This apparent robustness is a consequence of approximating the intra and interband pairing as having constant magnitude. As argued in Sec.~\ref{sec:ZeemanMultibandEffect}, however, the interband and intraband interactions are modified in the presence of a magnetic field due to canting effects, thus reducing the robustness to the in-plane field and increasing $\alpha$.

\subsection{Depairing analysis for $\lambda_I \gg E_F$}

In the analysis of the previous section, we were working in the conventional limit used for the Ising superconductors of $E_F$ being the largest energy scale in the problem. As a result, the microscopic dispersion dependence $\xi_\vec{p}$ effectively dropped out of the calculation. However, given that in the large $U$ limit, the trigonal warping pockets become flat promoting interactions, it is also insightful to consider the other limit of $\lambda_I \gg E_F$. In this limit, we will focus solely on an intraband pairing between the small trigonal warping pockets (the minority carriers) in each valley, as the interband pairing should be subleading.

We begin by writing the linearized intraband gap equation for the pairing temperature
\begin{equation}
\label{eq:A}
\frac{2}{g}=\int \frac{d^2 \vec{p}}{(2\pi)^2} \frac{\tanh\left(\frac{\epsilon_{\tau,\vec{p}}}{2 k_B T_C}\right)+\tanh\left(\frac{\epsilon_{-\tau,-\vec{p}}}{2 k_B T_C}\right)}{\epsilon_{\tau,\vec{p}}+\epsilon_{-\tau,-\vec{p}}}\,,
\end{equation}
where $\epsilon_{\tau,\vec{p}}=\epsilon^0_{\tau,\vec{p}}+s_{\tau,\vec{p}}$ is the dispersion of carriers. Here $\epsilon^0_{\tau,\vec{p}}$ is the dispersion of the carriers in the absence of SOC and magnetic field (See Eq.\eqref{eq:eff_Hamiltonian}) and the $s_{\tau,\vec{p}}$ is the SOC, orbital and Zeeman field dependent part. Crucially $\epsilon^0_{\tau,\vec{p}}=\epsilon^0_{-\tau,-vec{p}}$, but $s_{\tau,\vec{p}}\neq s_{-\tau,-\vec{p}}$ giving rise to depairing.

To eliminate the pairing constant $g$ we can use the analogous $T_{C,0}$ expression
\begin{equation}
\label{eq:B}
\frac{2}{g}=\int \frac{d^2 \vec{p}}{(2\pi)^2} \frac{\tanh\left(\frac{\tilde{\epsilon}_{\tau,\vec{p}}}{2 k_B T_{C,0}}\right)+\tanh\left(\frac{\tilde{\epsilon}_{-\tau,-\vec{p}}}{2 k_B T_{C,0}}\right)}{\tilde{\epsilon}_{\tau,\vec{p}}+\tilde{\epsilon}_{-\tau,-\vec{p}}}\,,
\end{equation}
where $\tilde{\epsilon}_{-\tau,-\vec{p}}$ is the dispersion $\epsilon_{\tau,\vec{p}}=\epsilon^0_{\tau,\vec{p}}+s_{\tau,\vec{p}}$ evaluated at $B\to 0$. Using Eq.\eqref{eq:A} and Eq.\eqref{eq:B}, eliminated $g$ and expanding near $T_C\to T_{C,0}$ we arrive at the desired form
\begin{equation}
T_C = T_{C,0} - \alpha B^2\,,
\end{equation}
where
\begin{align}
\label{eq:alpha_one_band}
\alpha \approx T_{C,0} \frac{F}{G}\,,\quad G \approx \int \frac{d^2 \vec{p}}{(2\pi)^2} \frac{1}{2 k_B T_{C,0}} \mathrm{sech}^2\left(\frac{2 \epsilon^0_{+,\vec{p}}+\lambda_I}{4 k_B T_{C,0}}\right)
\end{align}
and the function $F$ is given by a series expansion of
\begin{equation}
F \approx \frac{1}{B^2} \int \frac{d^2 \vec{p}}{(2\pi)^2} \frac{\tanh \left(\frac{\tilde{\epsilon}_{\tau,\vec{p}}}{2 k_B T_{C,0}}\right)+\tanh \left(\frac{\tilde{\epsilon}_{-\tau,-\vec{p}}}{2 k_B T_{C,0}}\right)}{\tilde{\epsilon}_{\tau,\vec{p}}+\tilde{\epsilon}_{-\tau,-\vec{p}}}  -\frac{\tanh \left(\frac{\epsilon_{\tau,\vec{p}}}{2 k_B T_{C,0}}\right)+\tanh \left(\frac{\epsilon_{-\tau,-\vec{p}}}{2 k_B T_{C,0}}\right)}{\epsilon_{\tau,\vec{p}}+\epsilon_{-\tau,-\vec{p}}}    
\end{equation}
in powers of magnetic field $B$, the leading order of which is independent of $B$ as required. The perturbation can be evaluated in a closed form; however, the resulting expression is not particularly transparent.

\begin{figure}
\includegraphics[width=\linewidth]{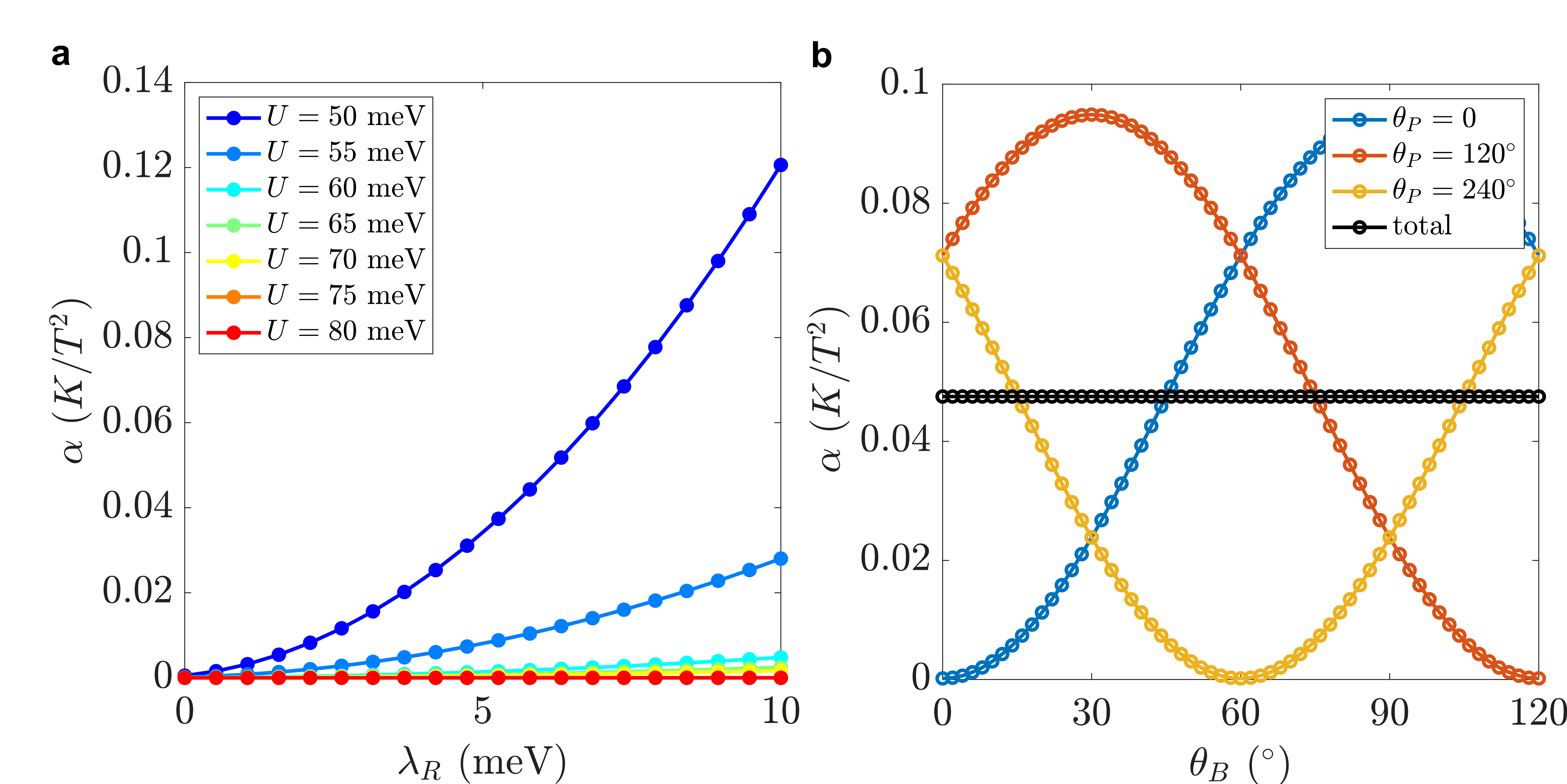}
\caption{{\bf Coefficient $\alpha$ in $T_C=T_{C,0}-\alpha B^2$ in the $\lambda_I\gg E_F$ limit}. {\bf a}, Dependence of the coefficent $\alpha$ from Eq.\eqref{eq:alpha_one_band} for various potentials $U$ and Rashba coupling $\lambda_R$. Here we use $p_F\sim \sqrt{n/4} \sim 0.05$ nm$^{-1}$ at $T_{C,0}\approx 100$~mK and $\lambda_I=1.5$ meV. {\bf b}, Dependence of $\alpha$ on an in-plane B field direction $\theta_B$. Here $U=65$ meV, $\lambda_R=2$ meV. $\theta_P=0,120^\circ,240^\circ$ label the three trigonal warping pockets. See also the discussion in the text.}
\label{sfig:cyprian_one_band}
\end{figure}

Let us now analyze the consequences of the result in Eq.\eqref{eq:alpha_one_band} mirroring the analysis of the previous section.  SI Fig. \ref{sfig:cyprian_one_band}a shows a plot of numerically determined $\alpha$ coefficient from Eq.\eqref{eq:alpha_one_band} as a function of layer potential $U$ and microscopic $\lambda_R$ for $p_F\sim \sqrt{n/4} \sim 0.05$ nm$^{-1}$ at $T_{C,0}\approx 100$~mK and $\lambda_I=1.5$ meV. We again see that no realistic value of $\lambda_R {\lesssim} 10$ meV reproduces the experimental values of $\alpha{\sim} 0.1 K/T^2$ seen in SC$_2$ for large $U$ values. This again supports our conclusions of the previous depairing analysis and points towards the mechanism of Sec.\ref{sec:ZeemanMultibandEffect} as controlling the resilience to the magnetic field of the SC$_2$.

In SI Fig. \ref{sfig:cyprian_one_band}b, we fix $\lambda_R=2$ meV and $U=65$ meV and vary the orientation of in-plane magnetic field $\theta_B$. The three curves ($\theta_B =0,\,120^\circ\, 240^\circ$) correspond to the expected depairing if only one of the three trigonal warping pockets is occupied. When all three pockets are occupied (as in SC$_1$), there is no dependence on an in-plane B field direction in line with results of Ref.~\citenum{holleisNematicityOrbitalDepairing2024}. While the precise value of $\alpha$ is controlled by $T_{C,0},\lambda_R$ and $U$, we notice that if only one pocket were occupied, then one expects a significant variation in $T_C$ as a function of in-plane B field direction. This contrasts the experiment; see \prettyref{exfig:SC2_inplane_angle}, thus suggesting that perhaps a strict ``flocking'' model is not applicable and the Fermi surface could reconstruct itself in response to the applied $B$ field to minimize depairing - a proposal that can be verified in future self-consistent HF works that allow for nematicity.

\end{document}